\documentclass{aa}  
\usepackage{graphicx}
\usepackage{txfonts}
\usepackage{lscape} 
\usepackage{placeins}
\usepackage[draft=False, colorlinks=true, linkcolor=blue, urlcolor=blue, citecolor=blue]{hyperref}
\usepackage[T1]{fontenc}
\usepackage{xspace}
\usepackage{booktabs}
\usepackage{siunitx}
\usepackage{romannum}
\usepackage{float}
\usepackage{xcolor}

\newcommand{\hi}{H~\textsc{i}\xspace}
\newcommand{\hei}{He~\textsc{i}\xspace}
\newcommand{\ha}{\ensuremath{\mathrm{H}\alpha}\xspace}

\def\lmbdrestvac{\ensuremath{\lambda_{\textrm{rest},\,\textrm{vac}}}\xspace}
\def\lmbdobsvac{\ensuremath{\lambda_{\textrm{obs},\,\textrm{vac}}}\xspace}

\newcommand*{\teff}{\ensuremath{T_{\mathrm{eff}}}\xspace}
\newcommand{\mdot}{\ensuremath{\dot{M}_{\mathrm{acc}}}\xspace}
\newcommand{\lline}{\ensuremath{L_{\mathrm{line}}}\xspace}
\newcommand{\lacc}{\ensuremath{L_{\mathrm{acc}}}\xspace}
\newcommand{\msun}{\ensuremath{M_{\odot}}\xspace}

\newcommand{\mj}{\ensuremath{M_{\mathrm{Jup}}}\xspace}

\newcommand{\mjyr}{\ensuremath{\mj~\mathrm{yr}^{-1}}\xspace}
\newcommand{\msyr}{\ensuremath{\msun~\mathrm{yr}^{-1}}\xspace}
\newcommand{\kms}{\ensuremath{\mathrm{km\,s^{-1}}}\xspace}
\newcommand{\lsun}{\ensuremath{L_{\odot}\xspace}}
\newcommand{\cgs}{erg~s$^{-1}$~cm$^{-2}$\xspace}

\begin{document}
\pagenumbering{arabic}
   \title{ExoplaNeT accRetion mOnitoring sPectroscopic surveY\\(ENTROPY)}

   \subtitle{III. Optical He~\textsc{i} line profiles of the accreting super Jupiter Delorme 1 (AB)b}

   \authorrunning{G.\ Viswanath et al.}
   \author{Gayathri~Viswanath\inst{\ref{Stockholm},\ref{Grenoble}}
        \and
        Micka{\"e}l~Bonnefoy\inst{\ref{Grenoble}}
        \and
        Catherine~Dougados\inst{\ref{Grenoble}}
        \and
        Simon~C.~Ringqvist\inst{\ref{Stockholm}}
        \and
        Markus~Janson\inst{\ref{Stockholm}}
        \and 
        Dorian~Demars\inst{\ref{Virginia}}
        \and 
        Aurora~Sicilia-Aguilar\inst{\ref{SUPA}}
        \and         J\'erôme~Bouvier\inst{\ref{Grenoble}}
        \and
        Gabriel-Dominique~Marleau\inst{\ref{duisburg},\ref{mpia}, \ref{bern}}
        \and
        Evelyne Alecian\inst{\ref{Grenoble}}
        \and
        Ga\"el~Chauvin\inst{\ref{nice}}
        }
   \institute{Institutionen f\"{o}r astronomi, Stockholms universitet, AlbaNova universitetscentrum, 106 91,                  Stockholm, Sweden \\\email{gayathri.viswanath@astro.su.se}
   \label{Stockholm}
   \and
    Universit\'e Grenoble Alpes, CNRS, IPAG, 38000 Grenoble, France
    \label{Grenoble}
    \and
    Department of Astronomy, University of Virginia, 530 McCormick Rd, Charlottesville, VA 22904, USA
    \label{Virginia}
    \and
    SUPA, School of Science and Engineering, University of Dundee, Nethergate, DD1 4HN, Dundee, UK
    \label{SUPA}
    \and
    Fakultät für Physik, Universität Duisburg--Essen, Lotharstra\ss{}e 1, 47057 Duisburg, Germany
    \label{duisburg}
    \and
    Max-Planck-Institut für Astronomie, Königstuhl 17, 69117 Heidelberg, Germany
    \label{mpia}
    \and
    Division of Space Research \& Planetary Sciences, Physics Institute, University of Bern, Sidlerstr.~5, 3012 Bern, Switzerland
    \label{bern}
    \and 
    Laboratoire Lagrange, Universit\'e C\^ote d'Azur, CNRS, Observatoire de la C\^ote d'Azur, 06304 Nice, France
    \label{nice}
    }

   \date{Received ...; }

    \abstract{Observations of helium emission lines from classical T Tauri stars at high resolution ($R_{\lambda}>10,000$) offer great potential, showing distinct profile characteristics that help probe regions within the accretion geometry untapped by hydrogen lines. Parallel studies in the planetary-mass regime have not been explored.}{We investigate helium line emission from the nearby (47~pc), wide orbit ($\sim84$~au), $\sim13$~\mj, accreting circumbinary companion Delorme 1 (AB)b and analyse the resolved profile characteristics to infer clues to line origin.}{We obtained high signal-to-noise spectra of the target over 33 exposures with VLT/UVES over near-ultraviolet to optical wavelengths at high resolution ($R_{\lambda}\sim50,000$). We studied the helium line profiles in the spectra and compared them to helium emission recorded from both accreting and non-accreting young stellar objects.}{We detected seven neutral helium (\hei) lines $\lambda\lambda3890,4027,4473,4923,5017,5877,6680$ at high confidence ($>5\sigma$), with notable flux variation between epochs. The line profiles of \hei $\lambda\lambda5877,4923,4473,4027$ show clear asymmetry, with a narrow component at $\sim0$~\kms and a broad component redshifted by $\sim15$~\kms. The accretion luminosity ($1.3^{+1.6}_{-0.7}\times10^{-5}$~\lsun) and mass accretion rate ($0.7^{+0.9}_{-0.4}\times10^{-8}$~\mjyr) obtained from median \hei line luminosities using empirical scaling relations from stars are comparable but slightly higher than from the target's ultraviolet excess emission.}{The protoplanet Delorme 1 (AB)b exhibits asymmetric \hei lines similar to classical T Tauri stars, but with much smaller widths for the narrow and broad components. The triplet--singlet line ratio, a strong correlation with ultraviolet excess and the near-zero, redshifted velocities obtained for the narrow component suggest that it originates within the post-shock region, close to the planet surface. The persistent redshift of the broad component, its line width, and velocity correlation with the narrow component imply an origin within the shock structure, closer to the shock front. Emission seems to be dominated by accretion based on the obtained accretion luminosities, but a contribution from chromospheric activity may be present. }


   \maketitle

\defcitealias{demars2026}{ENTROPY~\Romannum{2}}
\defcitealias{beristain2001}{BEK01}
\defcitealias{betti2023}{CASPAR}
\defcitealias{schneeberger1978}{Schneeberger et al. 1978}

\section{Introduction}

Decades of observations of classical T Tauri stars (CTTS) have provided a rich database to study the interplay between young, low-mass stars and the accretion disks that feed their formation. A wealth of accretion signatures have been recorded and studied from these class of objects; a few prominent ones include infrared excess from the accretion disk, ultraviolet (UV) excess emission from the accretion shock on the stellar surface, subsequent strong \ha and higher order Balmer line emission, both neutral (\hei) and ionised helium (He~\textsc{ii}) emission as well as Ca~\textsc{ii} emission. A forest of metallic emission lines from species such as Fe, Ca, Na, and Ti has also been detected, including forbidden lines that trace outflows and jets from the accretion process \citep[see][]{hartmann2016}. These observations have led to the consensus that in young, accreting, solar-type stars, material from the inner disk is fed to the star along magnetic field lines of a strong magnetosphere; this is the so-called magnetospheric accretion paradigm. Analysing the emission line profiles from this phenomenon at high resolution ($R_{\lambda}>10,000$) has helped uncover details of the geometry of magnetospheric accretion as well as the physical conditions within these regions. 

In comparison, the parallel in the planetary-mass regime has been much less explored owing largely to the faintness of the phenomenon. Early observations with seeing-limited echelle spectrograph revealed a diversity of profiles in brown dwarfs down to $\sim$15~\mj \citep{jayawardhana2003,mohanty2003,mohanty2005,muzerolle2005}. More recent observations have begun to exploit the accretion signatures of planetary mass objects (PMOs; $<15$~\mj) but have been mostly limited to hydrogen line emission and conducted at modest resolution ($R_{\lambda}<10,000$) \citep[e.g.][]{boucher2016,santamaria2018,santamaria2019,haffert2019,chinchilla2021,eriksson2020, betti2022,demars2023,aoyama2024,almendros2025, currie2025}. Attempts to predict the physical origin of these emission lines have been made with various accretion models \citep[e.g.][]{aoyama2018,thanathibodee2019,szulagyi2020,aoyama2024,hashimoto2025}. Compared to \hi, helium lines have a much higher excitation potential, requiring high temperatures or close proximity to an ionizing source, and thus serve a stricter confinement of the region of line origin. Resolved helium line profiles catalogued in CTTS \citep[e.g.][]{ulrich1981, muzerolle1998, beristain2001, edwards2003, edwards2006, erkal2022} show both broad and narrow components, as well as absorption features, attributed to composite line origins  that are difficult to probe with hydrogen lines \citep{kwan2007, kwan2024}; prominent narrow components (NC) from accretion shock at the stellar surface, absorption features from inner winds and prominent broad components (BC) whose origin, although determined to be non-shock, is yet unclear. However, high-resolution observations of helium emission from PMOs have not been reported, except in one case \citep{viswanath2024}, despite \hei $\lambda5877$ being a prevalent accretion signature in the higher-mass regime ($>20$~\mj).

Located $47\pm3$~pc away in the Tucana--Horologium stellar association, Delorme 1 (AB)b is a wide-orbit ($\sim84$~au) $\sim$13~\mj companion to an M5.5-type binary system \citep{delorme2013}. \cite{eriksson2020} reported strong \ha emission from this planetary-mass companion (PMC; $<15~\mj$) with the Multi Unit Spectroscopic
Explorer \citep[MUSE;][]{bacon2010} at the Very Large Telescope (VLT) in Chile, followed by detections of Pa$\beta$, Pa$\gamma$, and Br$\gamma$ emission in the near-infrared (NIR) by \cite{betti2022}. Recent observations by \cite{malin2025} with the \textit{James Webb }Space Telescope's Mid-Infrared Instrument \citep[JWST/MIRI;][]{wells2015} confirmed the presence of warm gas in a circumplanetary disk (CPD) around the object, with indications of outflows and an inner disk cavity. These evidences show that this object is actively accreting despite its relatively old age of $\sim40$~Myr.
Having a mass at the deuterium-burning limit, its nearby distance and wide orbital separation (1.8\arcsec) offer a convenient opportunity to study up close the accretion at a PMC using high-resolution spectroscopy.

Previous observations in October 2021 with the Ultraviolet and Visual Echelle Spectrograph \citep[UVES;][]{dekker2000} at VLT, at a high resolution of $R_{\lambda}\sim50,000$, revealed resolved spectral lines of higher-order Balmer transitions H5 to H9 \citep{ringqvist2023} from the target, with complex line profiles consisting of both broad and narrow components similar to CTTS. Recently, follow-up observations were conducted with VLT/UVES, spanning multiple epochs, as part of the ExoplaNeT accRetion mOnitoring sPectroscopic surveY (ENTROPY) program. ENTROPY aims to understand the accretion process in planetary-mass objects ($<20~\mj$) using high-resolution spectroscopic techniques \citep{viswanath2024, demars2026}. 
The second paper of this series (\citealp{demars2026}; hereafter \citetalias{demars2026}) presented a detailed analysis of the hydrogen emission lines in Delorme 1 (AB)b and their variability based on these observations. The complex Balmer line profiles obtained from the target's spectra were decomposed into two static components: a narrow `core' component, centred at $\sim0$~\kms, and a broad `wings' component at $\sim-50$~\kms with an absorption feature at $\sim15$~\kms. While the wings component is reproduced best by modelling emission from accretion columns, the core component is consistent with both predictions from shock emission models and chromospheric activity. The overall line profiles show strong variability of $\sim100\%$ on average on a time scale of weeks. 

Our work here presents the helium emission lines in the same spectra, providing the first focused report on helium lines at high resolution from an accreting PMC. The analysis combines the recently acquired data with the target's observations from 25 October 2021 \citep{ringqvist2023} taken with VLT/UVES with the same set-up. The paper is organised as follows. Section~\ref{obs} provides a summary of the observations and reduction routines employed to get the final spectrum used for the analysis in this work. The main results from the identified lines are outlined in Section \ref{results}. Section~\ref{analysis} provides a detailed analysis of the characteristics of the lines, and its implications are discussed in Section~\ref{discussion}, while Section~\ref{conclusion} summarises the key conclusions of this work.

\section{Observations and data reduction} \label{obs}
Delorme 1 (AB)b was observed with VLT/UVES between 13 October 2022 and 2 January 2023 as part of ENTROPY. The observations covered near-UV to optical wavelengths (3300--6800~\AA) at a resolution of $R_{\lambda}\sim50,000$. The integration time was 1482 sec, providing the same observing set-up as the 2021 epoch \citep{ringqvist2023} for Delorme 1 (AB)b. Including the two previous epochs on 25 October 2021, the new observations yielded 33 exposures over 11 nights. The total on-source integration time is 13.6~hr, yielding a high S/N for the emission lines detected in the median spectrum. 

Data were reduced using the ESO/UVES data handling pipeline \citep{ballester2000} and the extraction of the primary and companion spectra was performed using a custom CLEAN-like procedure \citep{hogbom1974} outlined in detail in \citetalias{demars2026}. The companion spectra were flux calibrated based on the spectrophotometric calibration of the primary spectra using its \textit{Gaia} DR3 \citep{gaiadr3} $G_{BP}$ magnitude (15.690$\pm$0.005). The resulting companion spectra were then corrected for the target's radial velocity \citep[7.3$\pm$2.6 km/s;][]{kraus2014}. After correcting for barycentric motion, a grand median was obtained over all 33 exposures which offered high S/N for the emission lines in the spectrum, optimal for their identification. Additionally, individual exposures from each night of observation were combined to yield a median spectrum per night, which allowed for assessing epoch-to-epoch flux variation in the emission lines identified from the grand median. 

We used the python package STAR-MELT\footnote{\url{https://github.com/justyncw/STAR_MELT}} \citep{white2021} to identify emission lines in the spectra, with a tolerance velocity shift of $\pm8~\kms$ ($\sim3\sigma$ of the target's radial velocity) around the rest wavelengths. We also fit the line profiles using STAR-MELT and defined the S/N of the identified lines based on the peak of the Gaussian fit to its profile and the standard deviation within $\pm150$~\kms around the central wavelength. STAR-MELT has been extensively used for analysing CTTS emission spectra \citep[e.g.][]{arul2023, white2023, white2023b, aguilar2023}. For details on the data reduction procedure, the observing log, and the final spectra over the full range of wavelength, we refer the reader to \citetalias{demars2026}.

\section{Results} \label{results}

\begin{figure*}[!ht]
    \centering
    \includegraphics[width=0.33\linewidth]{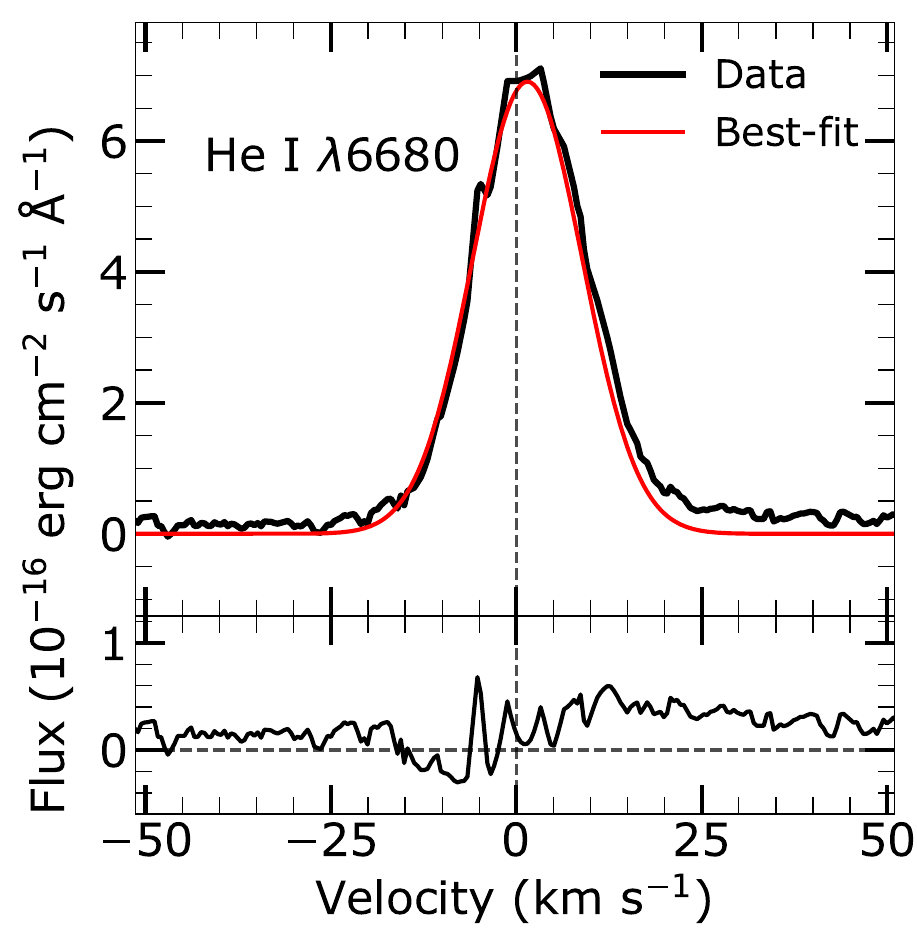}
    \includegraphics[width=0.33\linewidth]{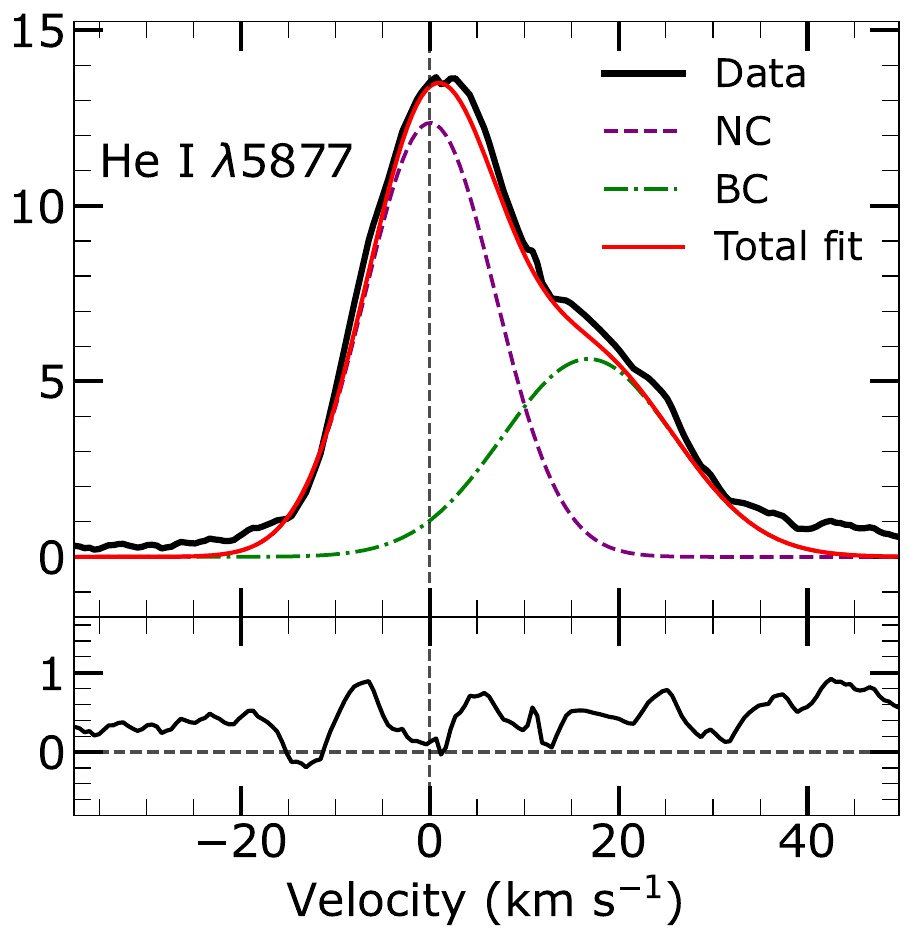}
    \includegraphics[width=0.33\linewidth]{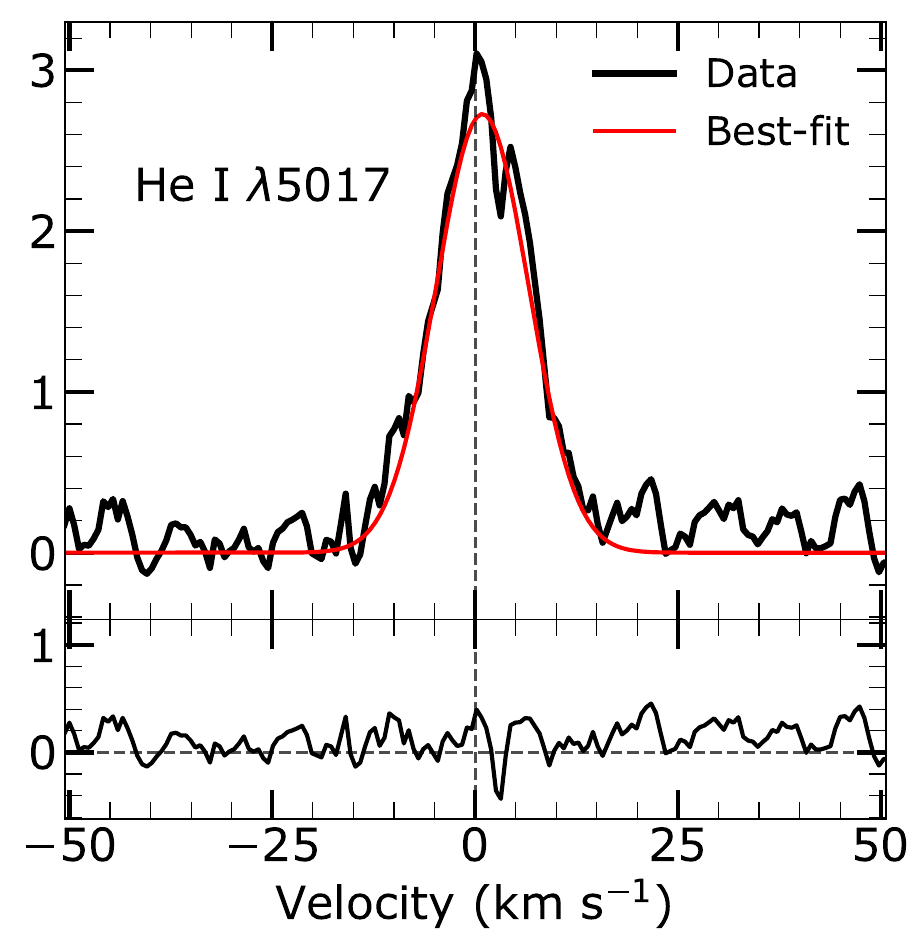}\\
    \includegraphics[width=0.33\linewidth]{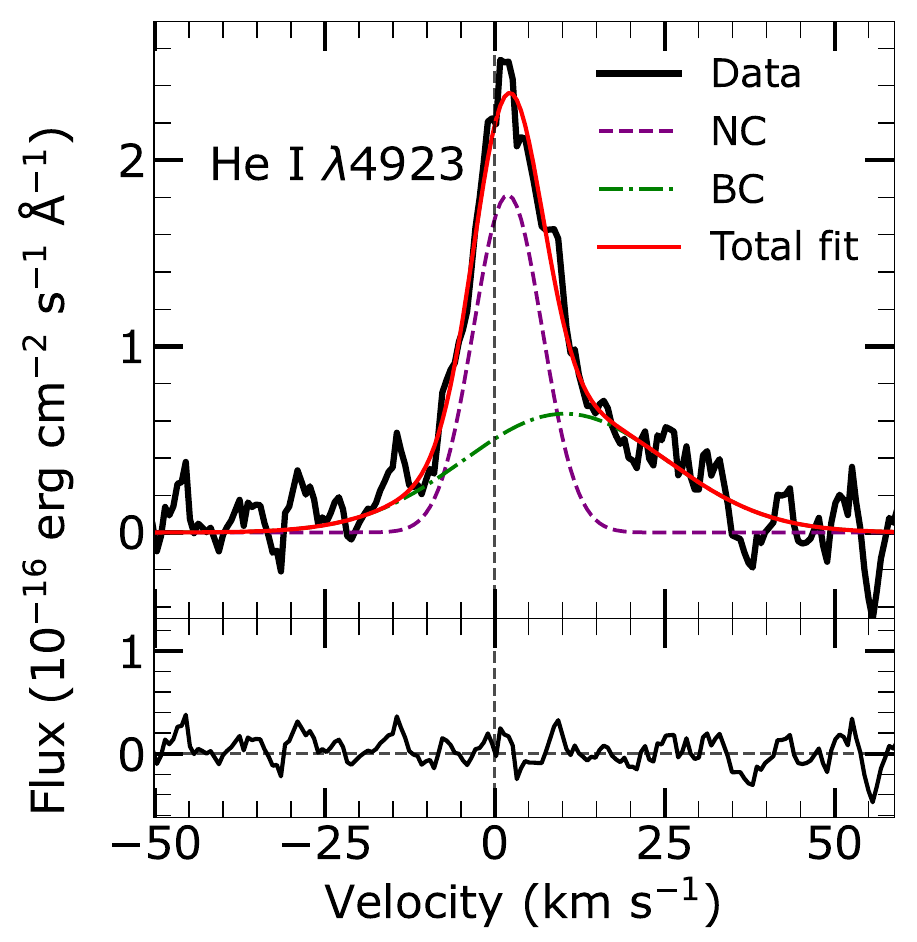}
    \includegraphics[width=0.33\linewidth]{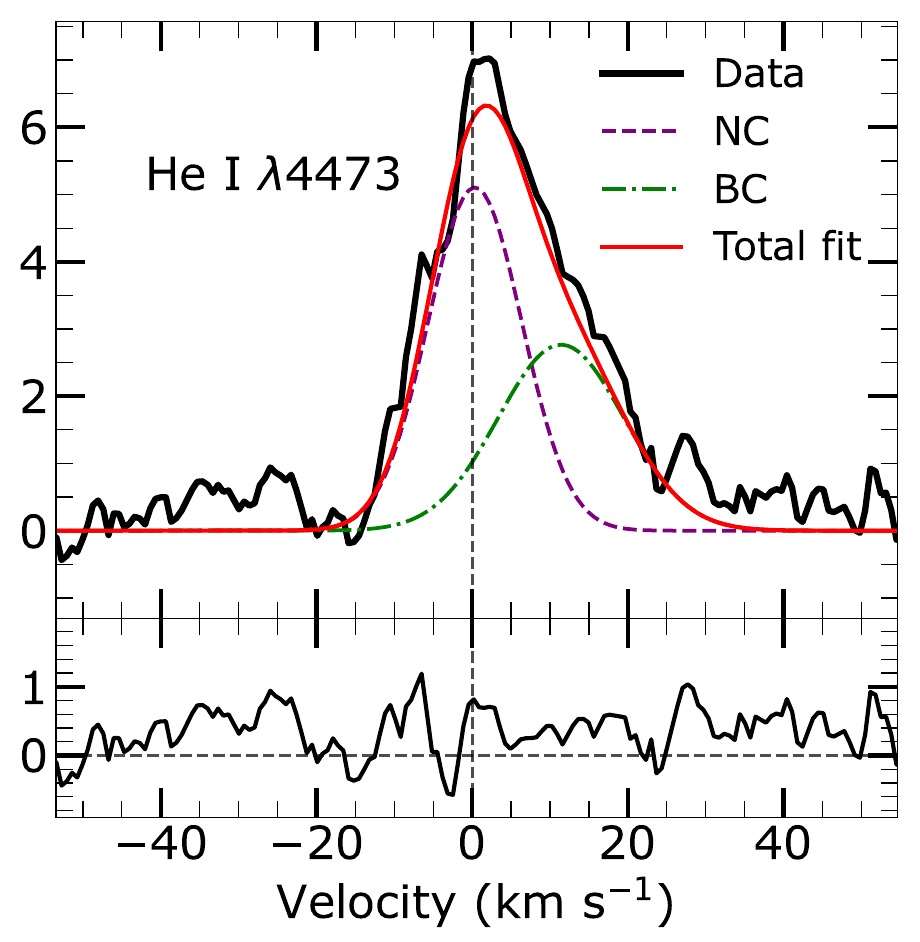}
    \includegraphics[width=0.33\linewidth]{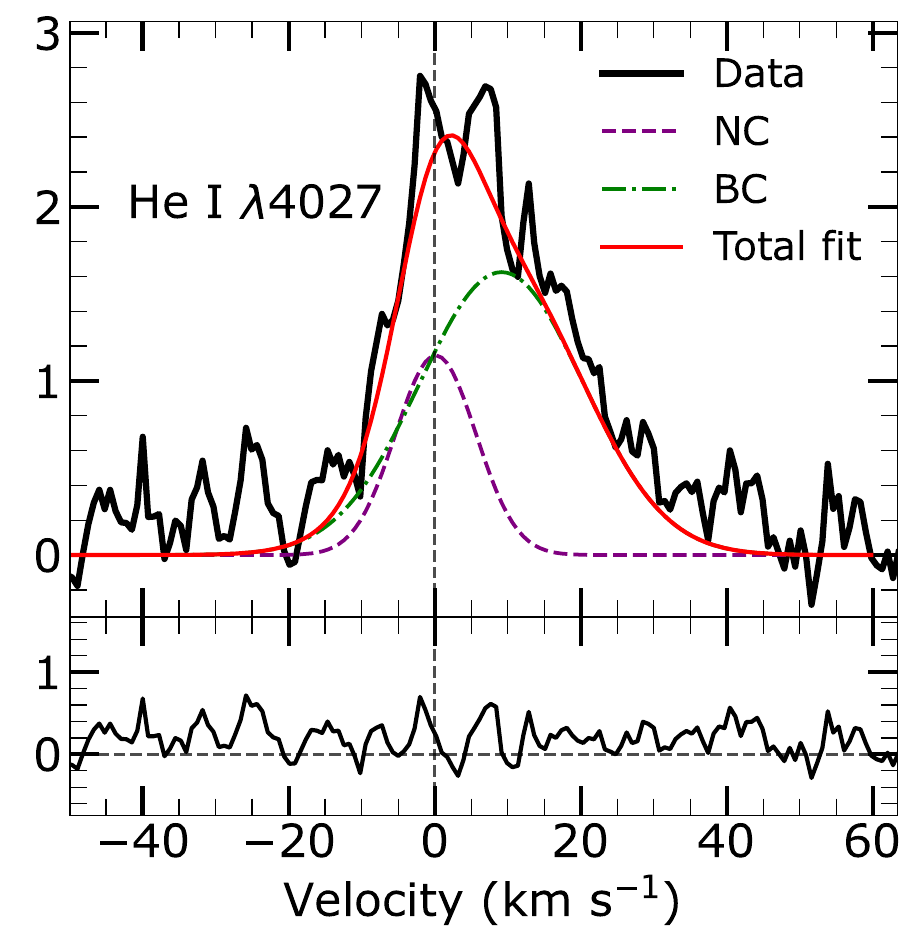}\\
    \includegraphics[width=0.33\linewidth]{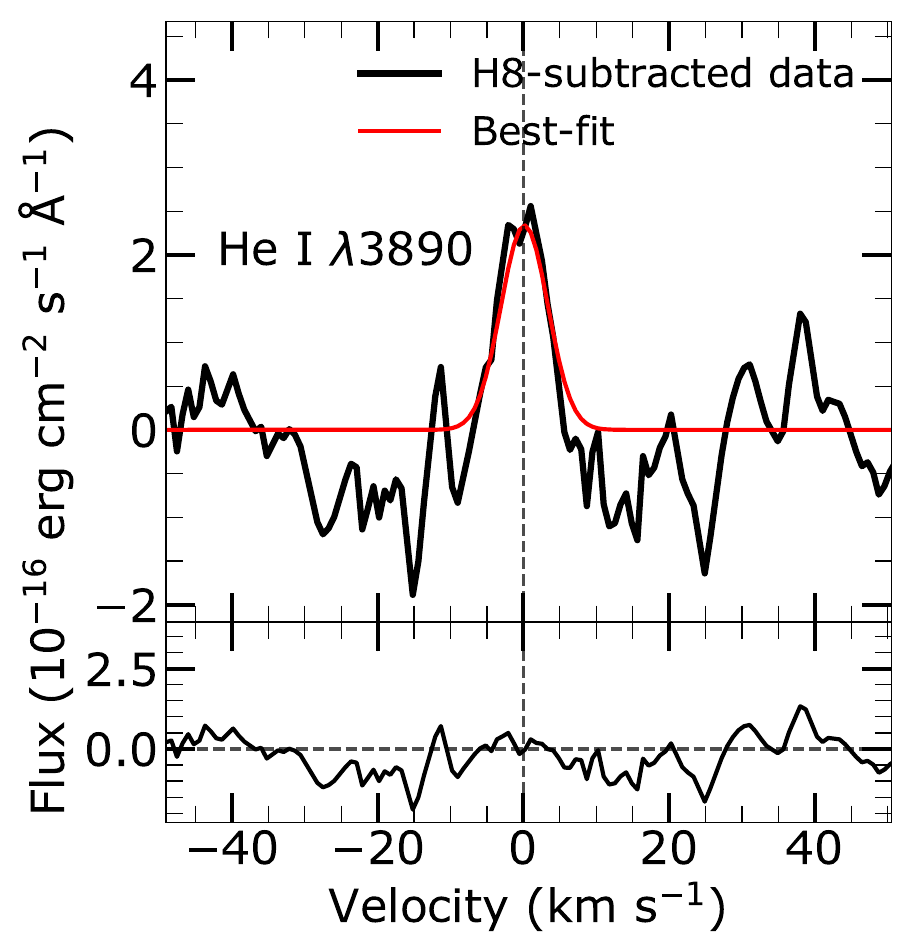}
    \includegraphics[width=0.33\linewidth]{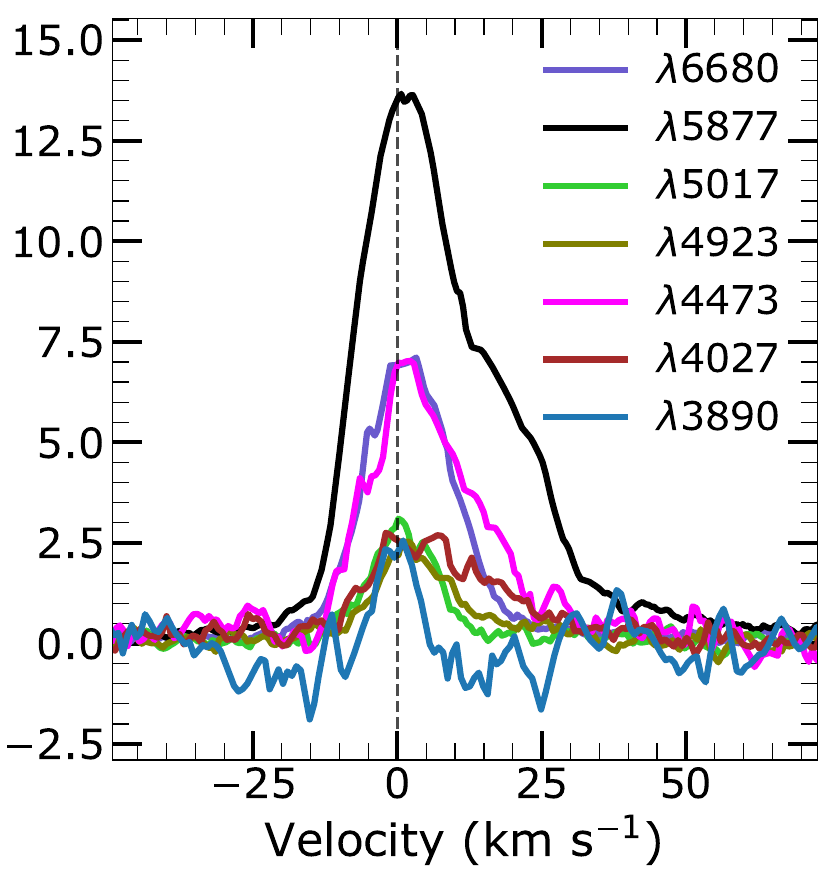}
    \includegraphics[width=0.33\linewidth]{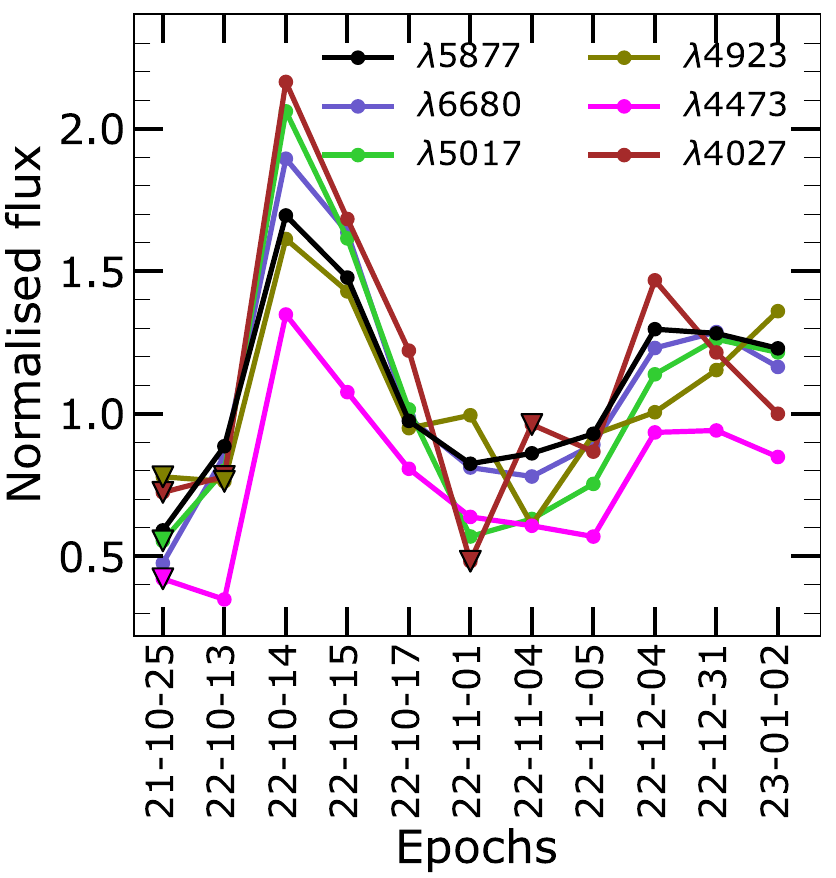}
    \caption{Resolved profiles of \hei emission lines (black) from Delorme 1 (AB)b detected from its grand median spectrum in this work. The red curve shows the least-$\chi^2$ fit to the line profile, composed of either pure NC or with the addition of a BC. In case of the latter, the red curve represents the total profile fit. Residual from the fit is shown in the bottom sub-panel of each line plot. The line profile for $\lambda$3890 (bottom left) is obtained after the Balmer line H8 was modelled and subtracted from the data (see Appendix~\ref{app-d}). The bottom middle panel shows all the \hei lines plotted against respective velocity scales, demonstrating relative strength and asymmetry. The bottom right panel illustrates the variability of their integrated line flux in time, normalised with respect to the corresponding grand median values. Downward triangles represent tentative detections with confidence between $1-3\sigma$. The y-axis for all panels except the bottom right shows the flux in units of $10^{-16}$~erg~cm$^{-2}$~s$^{-1}$~\AA$^{-1}$.}
    \label{fig1}
\end{figure*}

\citetalias{demars2026} showed that the Balmer emission lines in the reduced UVES spectra of Delorme 1 (AB)b display high flux variability ($\sim$100$\%$ on average) over time scales of weeks to months. These Balmer line profiles and the corresponding epochs can be classified into three classes depending on the observed shape of the \ha profile, namely `A', `B', and `C' (see Fig.~\ref{figA1}). Class A profiles exhibit low flux levels as seen on 1, 4, and 5 November 2022, as well as the 2021 epoch. Class B profiles have profile shapes similar to those of Class A but with visibly higher flux intensities, pointing to higher mass accretion rates during these epochs. Class C profiles stand out from those of Class A and B due to the presence of an enhanced blue wing emission during the nights of 14, 15, and 17 October 2022, which could possibly trace an accretion burst or a chromospheric activity burst (see \citetalias{demars2026}).
At wavelengths less than $\sim$3700~\AA, there is substantial UV continuum excess in the spectra, likely originating from accretion shock at the surface of the PMC. The spectra also present a nearly flat continuum throughout the wavelength range, with a gradual increase beyond $\sim$5300~\AA, which becomes more prominent during Class A epochs (see Fig.~\ref{figA2}). Using STAR-MELT, the continuum was fit using a standard Savitzky--Golay polynomial smoothing filter \citep[see][]{white2021, temmink2024, malin2025} within every 12~\AA~wavelength window, with a first-order polynomial. All reported emission lines here were identified from the grand median spectrum after subtracting the above-fit continuum. The line list for line identification was taken from the NIST Atomic Database\footnote{\url{https://www.nist.gov/pml/atomic-spectra-database}} and all wavelengths reported here are in vacuum.  

\subsection{Helium emission from Delorme 1 (AB)b}
We report seven resolved optical helium emission lines from Delorme 1 (AB)b. Using the grand median over all epochs, we confirm, with high confidence, the tentative \hei $\lambda\lambda$3890, 4027, 4473 emission lines reported in \cite{ringqvist2023} from the 2021 epoch. In addition, we also detect four strong \hei emission lines $\lambda\lambda$4923, 5017, 5877, and 6680 in the grand median spectrum. No He~\textsc{ii} emission was detected in the spectra. The strongest \hei emission in the spectra is the triplet line $\lambda5877$, followed by the singlet line $\lambda6680$, both of which are prevalent in CTTS spectra. Table~\ref{tab1} lists the confirmed ($>5\sigma$) \hei emission lines from the grand median spectra of Delorme 1 (AB)b and the characteristics of the corresponding line profiles. Since the major contribution to the flux uncertainty is from continuum subtraction, error bars on the integrated line flux values are estimated conservatively as the rms of the local continuum. Figure~\ref{fig1} illustrates the respective resolved line profiles in the grand median spectra and their least-$\chi^2$ Gaussian fits. 

The upper energy levels ($E_k$) of the detected \hei line transitions are between 23--24~eV, with a maximum $E_k$ of 24.04 eV for the line transition at 4027.33~\AA. The \hei transition at 3820.70~\AA~($E_k=24.21$~eV) is tentatively detected in emission at an S/N of 2$\sigma$ (see Appendix~\ref{app-c}). For \hei to be excited to such high energy levels of $\gtrsim 20~eV$, it requires either strong UV flux causing photoionisation, followed by recombination and cascade at lower temperatures of 8000--15,000~K, or collisional excitation resulting from high temperatures of 25,000--90,000~K in the emitting region \citep[][hereafter \citetalias{beristain2001}]{beristain2001}. Section~\ref{discussion} provides further discussion of the origin of these \hei lines from Delorme 1 (AB)b.

\renewcommand{\arraystretch}{1.2}
\begin{table*}[!ht]
\caption{Characteristics of \hei emission lines detected in this work.}
\centering
\begin{tabular}[c]{c c c c c c c c c c}
\hline\hline
\lmbdrestvac & $E_k$ & Transition\tablefootmark{a}  & $A_{ki}$\tablefootmark{b} & Fit\tablefootmark{c} & \lmbdobsvac & FWHM & Velocity & Flux & S/N \\
(\AA) & (eV) & & ($10^6$\,s$^{-1}$) & & (\AA) & (\kms) & (\kms) & ($10^{-17}$ erg~s$^{-1}$~cm$^{-2}$) & ($\sigma$) \\
\hline
3889.75 & 23.01 & T 3p--2s & 9.5 & \textit{s} & 3889.75 & $\phantom{1}7.2\pm0.8$ & $-0.2\pm0.3$ & $\phantom{1}3.0\pm3.6$ & 6 \\
4027.33 & 24.04 & T 5d--2p & 4.8 & \textit{s} & 4027.40 & $23.5\pm1.0$ & $\phantom{1}5.4\pm0.3$ & $\phantom{1}7.9\pm2.5$ & 9 \\
     &         &         &   & \begin{tabular}{@{}c@{}}\textit{d}, NC \\ \textit{d}, BC\end{tabular} & \begin{tabular}{@{}c@{}}4027.33 \\ 4027.45\end{tabular} & \begin{tabular}{@{}c@{}}$13.0\pm2.2$ \\ $26.4\pm1.7$\end{tabular} & \begin{tabular}{@{}c@{}}$\phantom{1}0.1\pm0.6$ \\ $\phantom{1}9.2\pm1.6$\end{tabular} & $\phantom{1}8.3\pm2.5$ & 10 \\
4472.73 & 23.74 & T 4d--2p & 24.6 & \textit{s} & 4472.78 & $20.1\pm0.5$ & $\phantom{1}3.7\pm0.2$ & $19.5\pm3.2$ & 18 \\
     &         &         & & \begin{tabular}{@{}c@{}}\textit{d}, NC \\ \textit{d}, BC\end{tabular}  & \begin{tabular}{@{}c@{}}4472.74 \\ 4472.89\end{tabular}& \begin{tabular}{@{}c@{}}$14.4\pm2.4$ \\ $22.0\pm6.3$\end{tabular}& \begin{tabular}{@{}c@{}}$\phantom{1}0.3\pm1.1$ \\ $10.4\pm7.0$\end{tabular}& $20.9\pm3.2$ & 20 \\
4923.31 & 23.74 & S 4d--2p & 19.9 & \textit{s} & 4923.36 & $16.2\pm0.4$ & $\phantom{1}2.9\pm0.2$ & $\phantom{1}6.1\pm1.6$ &  13 \\
     &         &         & & \begin{tabular}{@{}c@{}}\textit{d}, NC \\ \textit{d}, BC\end{tabular} & \begin{tabular}{@{}c@{}}4923.34 \\ 4923.47\end{tabular} & \begin{tabular}{@{}c@{}}$11.8\pm0.5$ \\ $33.3\pm2.1$\end{tabular} & \begin{tabular}{@{}c@{}}$\phantom{1}1.9\pm0.2$ \\ $\phantom{1}9.8\pm1.2$\end{tabular}& $\phantom{1}7.5\pm1.6$ & 14 \\
5017.08 & 23.09 & S 3p--2s & 13.4 &\textit{s } & 5017.10 & $13.4\pm0.3$ & $\phantom{1}0.9\pm0.1$ & $\phantom{1}6.5\pm1.7$ & 20 \\
5877.25 & 23.07 & T 3d--2p & 53.0 & \textit{s } & 5877.32 & $23.5\pm0.3$ & $\phantom{1}3.7\pm0.2$ & $61.6\pm1.5$ & 132 \\
     &         &         & & \begin{tabular}{@{}c@{}}\textit{d}, NC \\ \textit{d}, BC\end{tabular} & \begin{tabular}{@{}c@{}}5877.25 \\ 5877.56\end{tabular} & \begin{tabular}{@{}c@{}}$16.0\pm0.2$ \\ $23.4\pm0.6$\end{tabular} & \begin{tabular}{@{}c@{}}$-0.2\pm0.1$ \\ $16.0\pm0.5$\end{tabular} & $68.4\pm1.5$ &147 \\
6679.99 & 23.07 & S 3d--2p & 63.7 & \textit{s} & 6680.03 & $17.4\pm0.1$ & $\phantom{1}1.5\pm0.0$ & $28.6\pm1.8$ & 77 \\\hline
\end{tabular}
\tablefoot{\tablefoottext{a}{Triplet ($S=1$) and singlet ($S=0$) line transitions of \hei, denoted respectively by T and S, are identified from \citet{delzanna2020}.}\\ 
\tablefoottext{b}{$A_{ki}$ denotes the Einstein coefficient for the transition.}\\
\tablefoottext{c}{\textit{s} refers to a single Gaussian fit and \textit{d} refers to a double Gaussian fit. Line characteristics for the narrow and broad components for the latter are indicated as NC and BC respectively.}}
\label{tab1}
\end{table*}

\subsection{Asymmetry in the \hei line profiles}
\hei $\lambda5877$ (\hei~D3) is one of the strongest and most prevalent permitted emission lines in CTTS optical spectra (\citetalias{beristain2001}). It is the strongest emission line in the Delorme 1 (AB)b spectra after the Balmer lines. The grand median profile of the line shows clear asymmetry in its red wing emission, with the least-$\chi^2$ fit yielding two Gaussians (see  Fig.~\ref{fig1}, top middle panel), one centred at near-zero velocity with full width at half maximum (FWHM$_{\mathrm{NC}}$) $16.0\pm0.2$~\kms, and a second, broader component (FWHM$_{\mathrm{BC}}=23.4\pm0.6$~\kms) redshifted by 16~\kms. Similar asymmetry in the red wing emission is also seen in the blue \hei triplet lines $\lambda\lambda4473,4027$, whose grand median profiles can also be decomposed into a narrow and broad component as shown in Fig.~\ref{fig1}. These profiles show BCs with a slightly lower redshift ($\sim10$~\kms) than \hei~D3 but with an NC that remains at near-zero velocities.   

The \hei triplet lines are composed of two equal-intensity transitions at similar wavelengths and a third less intense transition separated by $\sim$15~\kms. The second broader component in the $\chi^2-$fit for \hei $\lambda\lambda5877,4473,4027$ line profiles is centred very close to the respective third transition of these triplet lines, suggesting the possibility that the high resolution of the data could have partially de-blended the lines leading to the asymmetry. However, the BC/NC line ratio is much higher than expected from the ratio of relative intensities of the individual transitions listed in the NIST database. More importantly, the singlet line \hei $\lambda4923$ also displays a similar asymmetry in its grand median profile, as illustrated in the first middle panel of Fig.~\ref{fig1}. Its BC is redshifted similarly at $\sim$10~\kms, but with a slightly broader profile  ($\sim$33~\kms) than those of the triplet lines. A possible explanation for this asymmetry in the singlet line could be blending with an Fe~\textsc{ii} transition at 4923.56~\AA, but the absence of any other Fe~\textsc{ii} line detections ($>3\sigma$) in the grand median spectrum renders this unlikely.

The singlet line \hei $\lambda$6680 also displays a small but broad flux increase in its red wing (see the fit-residual in the first panel of Fig.~\ref{fig1}), which can be subsequently fit with a very shallow BC at a redshift of $\sim$15~\kms. However, the significantly lower intensity of the BC relative to the NC renders this fit tentative, so we only consider a single Gaussian fit to its profile for the line characteristics used in subsequent analysis. However, we note that the line profile is not very symmetric for \hei~$\lambda6680$, with more emission on the red side of the line centre than on the blue side, although not as pronounced as in \hei~D3. The velocity and FWHM of NCs and BCs of all the asymmetric \hei lines in the grand median spectrum are listed in Table~\ref{tab1}. The NC widths measured for all the asymmetric lines are broader than the spectral point spread  function (PSF; 6~\kms)\footnote{\url{https://www.eso.org/sci/facilities/paranal/instruments/uves/doc/ESO_514367_User_Manual_P115.pdf}} of UVES, which implies that these components are spectrally resolved by the instrument. 

\subsection{Time variability of the \hei line profiles}\label{variability}
All \hei lines detected in this work demonstrate notable flux variation between individual epochs, similar to the Balmer lines shown in \citetalias{demars2026}. The bottom right panel of Fig.~\ref{fig1} illustrates the integrated flux of these \hei emission lines (see Table~\ref{tabg1} for values) varying in a similar pattern within a time scale of days, offering clear indications of an accretion origin. Line emissions are strongest on 14 October. The flux of the strongest lines \hei $\lambda\lambda5877,6680$ varies by a factor of 3--4, similar to accretion signatures in CTTS. 
\hei~D3 remains asymmetric in all epochs (see Fig.~\ref{figG2}), with the intensities and widths of the respective NC and BC evolving in an increasing pattern from Class A to Class C epochs. While the centroid velocity of NC remains mostly within $\pm1$~\kms, the velocity of BC is slightly more variable with time (see Fig.~\ref{figE} and Table~\ref{tabe2}). The contribution of the BC to the total line strength also varies through the epochs; the line profile remains mostly NC-dominated but the BC/NC flux ratio varies from 0.31 (1 November 2022, Class A) to 0.76 (14 October 2022, Class C). The only exception is on 13 October 2022 when the \hei~D3 profile becomes BC-dominated with BC/NC flux ratio of 1.65 (see Fig.~\ref{figE2}). This epoch, which is right before the peak outburst on 14 October, stands out from the rest due to significantly lower UV excess emission in the spectrum, strong core \ha component, and strong \hei~D3 BC. In general, the relative BC contribution (i.e. the BC/NC flux ratio) in \hei~D3 line profile increases in the epochs of Class B and Class C compared to Class A (see Fig.~\ref{figE}). This variation in the BC/NC flux ratio with time is another strong argument against the asymmetry in the line profiles originating from resolved triplet components.

Compared to \hei~D3, the asymmetric \hei lines at shorter wavelengths $\lambda\lambda4923,4473,4027$ are much lower in S/N. \hei $\lambda4473$ is detected only at $2.6\sigma$ on 25 October 2021, and its profile is asymmetric in nine of the ten remaining epochs with S/N$>3\sigma$. \hei $\lambda4923$ remains asymmetric in eight of the nine epochs where its S/N is above $3\sigma$. The emission at $\lambda4027$ is relatively weaker with the detection falling below $3\sigma$ in four of the nights, for which a two-component Gaussian fit was not warranted. The relative BC contributions in the \hei~$\lambda4473$ and $\lambda$4923 line profiles differ in their behaviour from that of \hei~D3, with both increasing on average in Class A epochs (see Fig.~\ref{figE}) compared to Class C. For the relatively weaker $\lambda4027$ emission, the BC contribution is quite diverse, ranging from pure-NC profiles to nearly pure-BC profiles, but we refrain from a detailed analysis because the line is noisier.

\subsection{Kinematics of NC and BC}\label{kinematics}
\begin{figure}[!ht]
    \centering
    \includegraphics[width=0.9\linewidth]{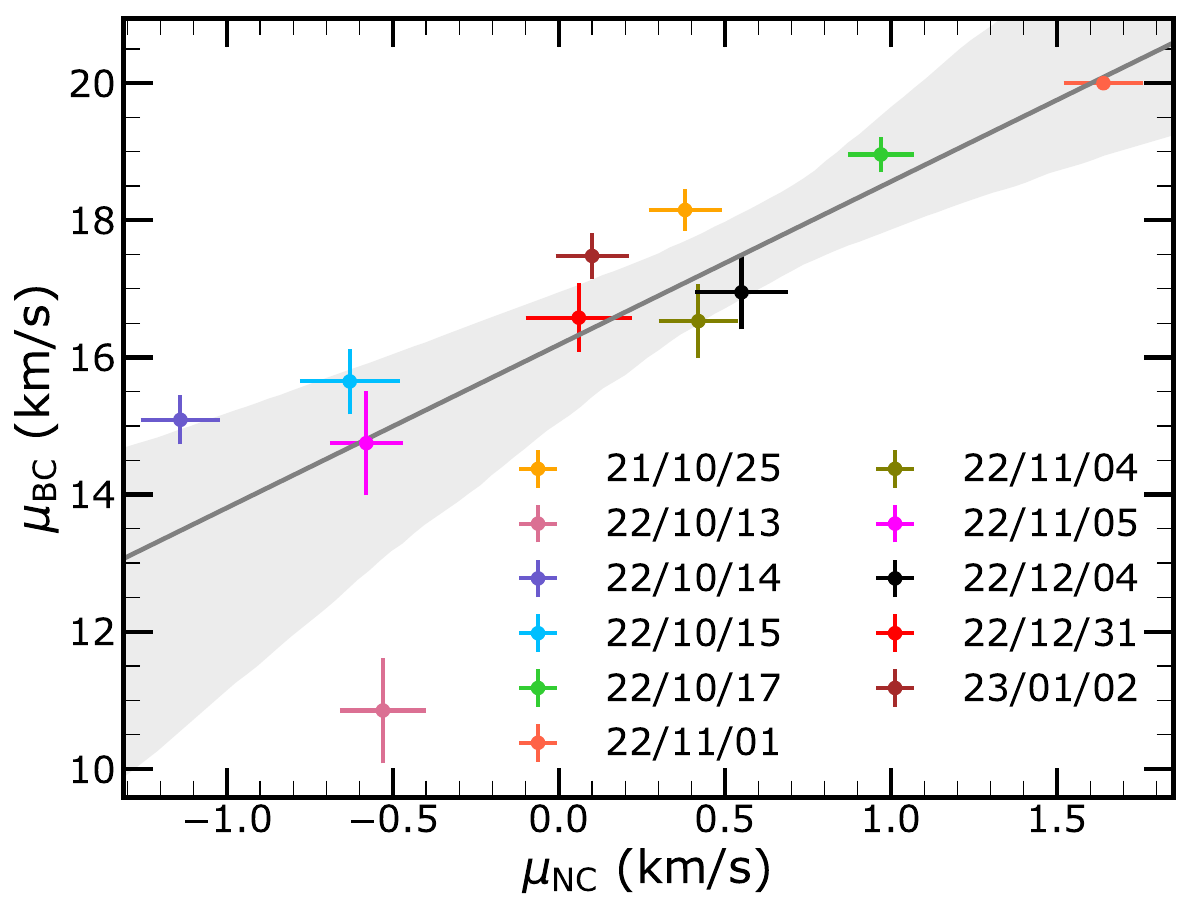}
    \caption{Variation in the centroid velocity, $\mu$, of NC and BC in \hei~D3 through the epochs. The grey line denotes the linear fit between NC and BC velocities, which are strongly correlated.}
    \label{vel-variation}
\end{figure}

The variation in the centroid velocity ($\mu$) of NC and BC in \hei~D3 throughout the epochs is shown in Fig.~\ref{vel-variation} (see Table~\ref{tabe2} for the values). Its NC velocity seems to vary between positive and negative values over time, but remains mostly near-zero with a mean value of 0.1$\pm$0.8~\kms. The NC velocities of the lower S/N \hei lines from the target also vary with time. The mean of these velocities from individual epochs is given in Table~\ref{tab3} for each of the detected \hei lines of Delorme 1 (AB)b. The respective FWHM measurements are also given in the Table, with the width distribution relatively uniform among the NCs. In general, the mean NC velocities of these \hei lines are slightly redshifted, in line with the origin in the immediate post-shock region at the base of the accretion column, as predicted in magnetospheric accretion scenarios \citep[][\citetalias{beristain2001}]{hartmann2016}. 

 \renewcommand{\arraystretch}{1.2}
\begin{table}[!ht]
\caption{Mean kinematics of \hei lines, taken over individual fits from each epoch.}
\centering
\begin{tabular}[c]{c c c c c}
\hline\hline
\lmbdrestvac\tablefootmark{a} & $\mu_{\mathrm{NC}}$ &  FWHM$_{\mathrm{NC}}$ & $\mu_{\mathrm{BC}}$ &  FWHM$_{\mathrm{BC}}$\\
(\AA) & (\kms) & (\kms) & (\kms) & (\kms) \\
\hline
3880.75  & $0.5\pm0.6$ & $\phantom{1}8\pm2$\tablefootmark{b} & -- & -- \\
4027.33 & $2.1\pm3.8$ & $15\pm9$ & $13.5\pm6.5$ & $27\pm6$ \\
4472.73 & $1.6\pm1.6$ & $15\pm4$ & $15.0\pm3.7$ & $18\pm8$ \\
4923.31 & $2.2\pm0.5$ & $14\pm2$ & $16.6\pm5.0$ & $\phantom{1}30\pm17$ \\
5017.08 & $0.6\pm0.6$ & $14\pm1$ & -- & --\\
5877.25 & $0.1\pm0.8$ & $16\pm1$ & $16.5\pm2.3$ & $21\pm4$ \\
6679.99 & $1.4\pm0.4$ & $16\pm1$ & -- & --\\ \hline
\end{tabular}
\tablefoot{\\ \tablefoottext{a}{\hei $\lambda\lambda3890, 5017, 6680$ lines are fit with a single Gaussian profile. }\\
\tablefoottext{b}{Line width of \hei $\lambda3890$ emission is dependent on the efficiency of H8 modelling, and hence should be interpreted with caution. }
}
\label{tab3}
\end{table}

The centroid velocities of BC in \hei~D3 also varies in a pattern similar to that of its NC through the epochs. From Fig.~\ref{vel-variation}, we see that the centroid velocities of the two components are strongly correlated in time with a Pearson's correlation coefficient, $r=0.8$, at $p=0.005$\footnote{The $p$-value in the Pearson's correlation test indicates the statistical significance of the correlation, and the obtained correlation is considered not relevant if $p>0.05$.}, which implies that NC and BC are likely formed via the same mechanism. The mean value of the BC velocities of individual epochs is given in Table~\ref{tab3} for \hei~D3 and the other asymmetric \hei lines. On average, the centroid velocities of BC in all asymmetric \hei lines from the target seem redshifted with respect to NC, with the individual epochs having BC velocities ranging from $\sim$10--22~\kms. The BC width distribution is more diverse between the \hei lines than for NC, with mean values between 16--30~\kms. For the strongest line \hei~D3, the width of BC has a negative correlation in time with that of its NC ($r=-0.7$, $p=0.02$; see Fig.~\ref{figE2} right panel), resulting in the overall profile shape of the line being conserved over epochs.

\section{Analysis}  \label{analysis}

\subsection{Comparison to \hei emission in CTTS}\label{ctts-comparison}
\subsubsection{Line profiles}
\citetalias{beristain2001} analysed \hei profiles in the high-resolution ($R_{\lambda}\sim20,000$) spectra of 31 CTTS down to spectral type M5 using the KPNO 4~m telescope, detecting the \hei lines $\lambda\lambda4473, 5877, 6680$ in several of the targets in their sample. Similarly to Delorme 1 (AB)b, the strongest \hei emission in the sample was at \hei~D3, followed by \hei $\lambda6680$. No \hei emission was reported from the sample at $\lambda\lambda3890,4027,4923$, and 5017, while we detect these lines at high S/N from Delorme 1 (AB)b.

Among the \hei~D3 profiles in the \citetalias{beristain2001} sample the morphology is quite diverse, ranging from BC-dominated profiles (flux ratio\footnote{Line strengths in \citetalias{beristain2001} were measured in terms of veiling-corrected equivalent widths since the spectra were not flux-calibrated.} BC/NC$>5$) to NC-dominated profiles (BC/NC$<1$). The composite \hei~D3 profile of Delorme 1 (AB)b remains NC-dominated in all epochs, except on 13 October 2022 (see Section~\ref{variability}). A similar asymmetry in the red wing, as seen in the \hei emission lines in Delorme 1 (AB)b, was also displayed in the \hei $\lambda5877$ and $\lambda4473$ emission line profiles of the M0.5 type TW Hya \citep[$\mdot\approx10^{-9}$~\msyr; see][]{herczeg2023}.

The line profiles of \hei $\lambda6680$ emission reported from the  \citetalias{beristain2001} sample also range from NC-dominated to BC-dominated profiles, while in  Delorme 1 (AB)b the line is clearly NC-dominated, with only a tentative detection of a BC. \citetalias{beristain2001} also reported composite profiles for the \hei $\lambda$4473 emission, but excluded it from their analysis due to the suspected blending of the line with the strong BC in the Ti~\textsc{ii} emission at 208~\kms. 
For Delorme 1 (AB)b, the BC at $\sim$13~\kms in \hei $\lambda4473$ can be confirmed since at the higher resolution of the spectra, any blending from $\sim$200~\kms can be confidently dismissed. Additionally, no Ti~\textsc{ii} lines are identified in the whole spectrum.

\subsubsection{Kinematics of \hei~D3}
\begin{figure*}
    \centering
    \includegraphics[width=0.45\linewidth]{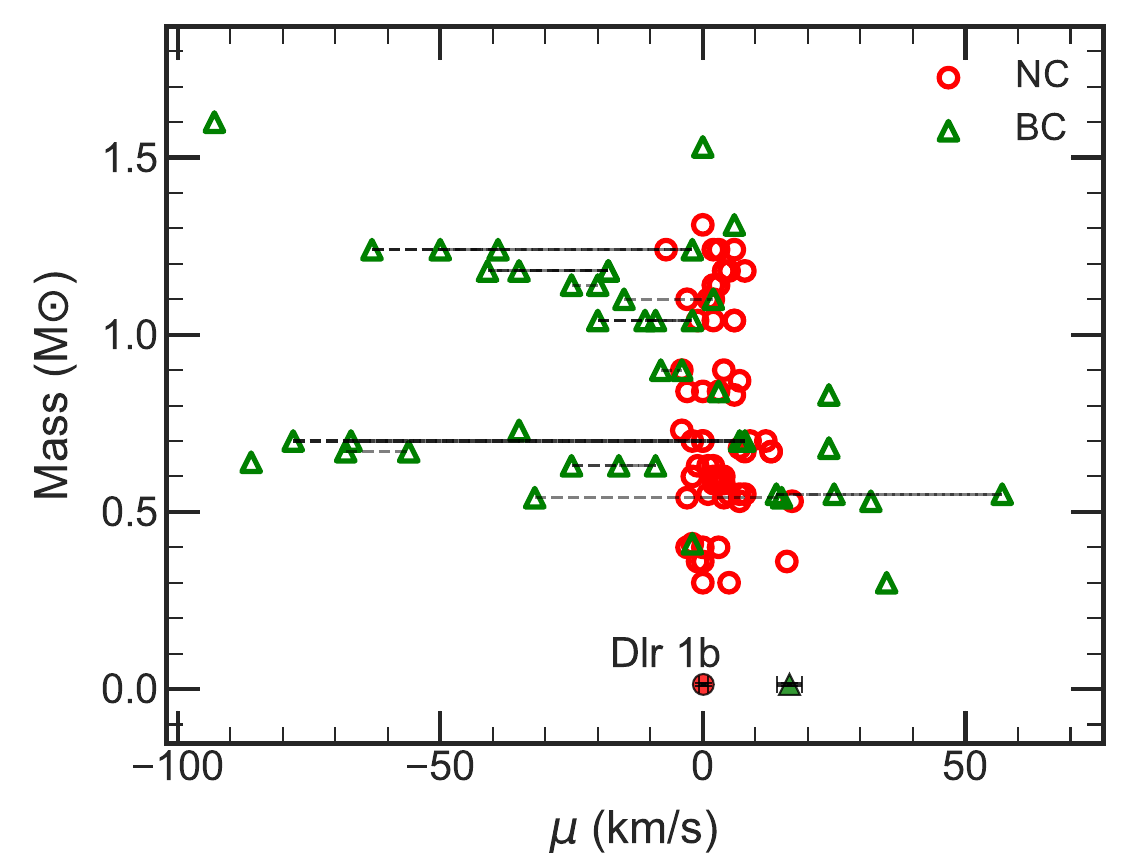}
    \includegraphics[width=0.45\linewidth]{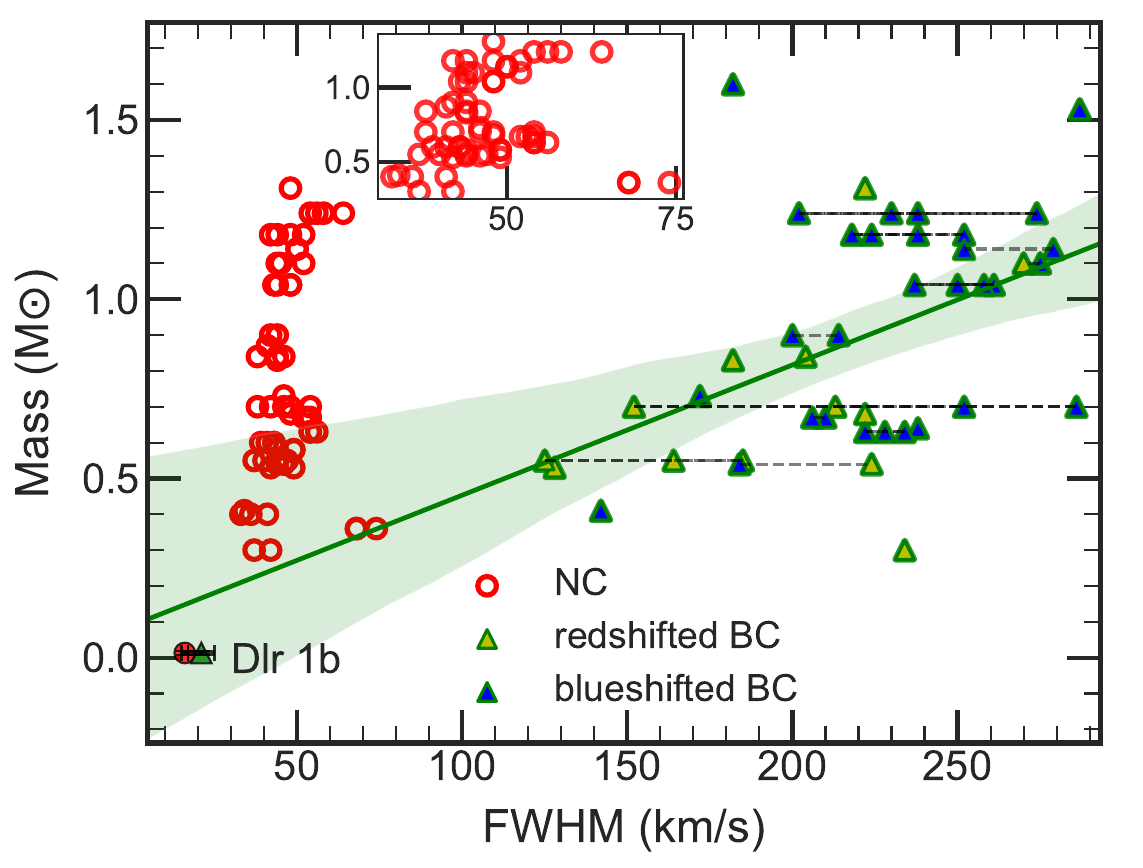}
    \caption{Distribution of (left) the centroid velocities, $\mu$, and (right) FWHM of \hei $\lambda5877$ NC (open circles) and BC (open triangles) of the \citetalias{beristain2001} sample with respect to their mass estimates, with BC colour-coded with respect to centroid velocities in the right panel. The NC and BC measurements of Delorme 1 (AB)b from this work are shown as filled red and green data points in the figure. The inset in the bottom panel zooms in along the FWHM-axis of NC distribution of the CTTS sample. The linear fit to the BC distribution is shown as a green line, along with its $1\sigma$ confidence interval  (shaded green region). Measurements from different observations of the same source are joined by dashed black lines in both plots.}
    \label{fig2}
\end{figure*}

In Fig.~\ref{fig2} we compare the NC and BC kinematics of the \hei~D3 profile in Delorme 1 (AB)b (filled data points) with those of the \citetalias{beristain2001} CTTS sample\footnote{NC and BC kinematics of the CTTS sample are taken from Table 1 of \citetalias{beristain2001}. The respective mass estimates are not from a uniform source.} (unfilled data points). In the left panel, we see the centroid velocity distribution as a function of the mass of the accretor. For NC, the mean velocity ($\mu$) for Delorme 1 (AB)b (see Table~\ref{tab3}) is within the range of NC velocity distribution  among CTTS ($\bar{\mu}_{\mathrm{NC}}=5\pm4$~\kms), although slightly lower. The width distribution as a function of accretor mass (right panel) shows that NC is much narrower in Delorme 1 (AB)b than the typical NC width observed in CTTS (47$\pm$7~\kms). 

Analysing the BC distribution of the \citetalias{beristain2001} sample in Fig.~\ref{fig2}, most of the observed BCs were found in blueshift, while also appearing redshifted in some sources; in fact, five of the sources in the \citetalias{beristain2001} sample (DF Tau, DO Tau, DS Tau, UZ Tau E, and YY Ori) display only redshifted BC in all their observations. In CTTS when the magnetic dipole axis is inclined with respect to the stellar rotation axis (i.e. a non-axisymmetric alignment), the centroid velocity of BC may undergo a phase-dependent velocity modulation with time. Continued monitoring of CTTS emission lines has recorded such velocity shifts with time-resolved observations, including changes between redshift and blueshift \citep[see][]{aguilar2023, armeni2024}. However, the centroid velocity of \hei BC in Delorme 1 (AB)b is always redshifted, closely varying about $\sim+15$~\kms, in our observations spanning days to months. This is within the range of the observed BC redshifts among the CTTS sample.

The width of BC in Delorme 1 (AB)b is not much wider than its NC, unlike what is seen among CTTS. The left panel of Fig.~\ref{fig2} clearly illustrates the very distinct width distribution of NC and BC among CTTS, while in Delorme 1 (AB)b this distinction is quite small, with the BC width much smaller than what is typically observed in CTTS (130--300~\kms); the BC-to-NC width ratio in the CTTS sample is always $>3$, while for the target it is only 1.31$\pm$0.26. The redshifted and blueshifted BCs among CTTS are shown in separate colours (yellow and blue, respectively) in the figure, to highlight that $\sim90\%$ of the blueshifted BCs have FWHM$>$200~\kms. On average, the redshifted BCs among CTTS have widths smaller than the blueshifted ones. A moderate positive correlation exists between mass and FWHM among the CTTS sample with a Pearson's correlation coefficient  $r=0.5$ (at $p\approx10^{-3}$), suggesting that the BC width may be expected to be smaller at planetary masses.
Delorme 1 (AB)b falls within the confidence interval of the linear fit between the BC width and mass of the CTTS sample, albeit larger than what is expected at its mass from the best-fit. This suggests that the smaller BC width for the target could be a consequence of the lower mass of the PMC.   

A comparison with CTTS profiles can also be made for NC in the \hei $\lambda\lambda4473, 6680$ lines. Similarly to $\lambda5877$, these lines also display narrower NCs at smaller redshifts compared to CTTS profiles in \citetalias{beristain2001}, with a mean FWHM $\sim37$~\kms and a mean velocity $\sim6$--$7$~\kms in CTTS for \hei $\lambda6680$ line.

\subsection{Triplet--singlet line ratios}\label{lineratio}
The triplet line \hei $\lambda5877$ and the singlet line \hei $\lambda6680$ share a common upper energy level and are a popular pair in the literature for deciphering the physical conditions of the line emitting region \citep[\citetalias{schneeberger1978, beristain2001};][]{gahm2013}. In particular, the ratio of their line intensities can hint at the temperature and density in these regions and whether they are in local thermodynamic equilibrium \citep[LTE;][]{aguilar2012}. Under nebular conditions, assuming very low electron densities and \hei line origin purely from recombination and cascade, the $\lambda5877$ emission is expected to be stronger than $\lambda6680$ and their line ratio is predicted to be $\sim$3.5 \citep{brocklehurst1972, smith1996}. When collisional excitation contributes at higher densities, this ratio will initially increase but will eventually fall to unity, when thermal equilibrium is reached at very high densities due to coupling between the triplet and singlet levels (\citetalias{beristain2001}). For Delorme 1 (AB)b, the average of the line ratios $\lambda 5877$/$\lambda 6680$ for the respective NCs from individual epochs yields a value of 1.7$\pm$0.4, which is close to unity, indicating very high densities and conditions of near thermal equilibrium in the NC-emitting region. This value is very similar to the reported NC line ratio of 2.0$\pm$0.6 in \citetalias{beristain2001} measured from their CTTS sample. \cite{kwan2011} argue, based on calculations from line excitation models, that the production of optically
thick \hei $\lambda5877$ line emission requires high densities $>10^{11}$~cm$^{-3}$ and high temperatures $>15,000$~K; such physical conditions could be met in the shock generated at the base of accretion columns on the planet surface. \citetalias{beristain2001} also finds that the corresponding BC in CTTS has a line ratio similar to the nebular value, indicating formation in lower density conditions, but we refrain from a direct comparison in our work due to the tentative nature of the singlet BC. 

The main caveat with inferences drawn from such analysis with line ratios is that the triplet 5877~\AA\ emission line is composed of multiple transitions originating from closely spaced energy levels that occur at very similar wavelengths. The bulk of the observed line emission is dominated by transitions from a slightly lower energy level than the 6680~\AA\ transition, since the corresponding transition from the exact same upper energy level has a much lower transition probability. Thus, more accurate estimates of the physical conditions within the emitting region would require dedicated radiative transfer modelling.

\subsection{Accretion luminosity and mass accretion rate}
\cite{ringqvist2023} estimated accretion luminosity (\lacc) for Delorme 1 (AB)b from line luminosities (\lline) of its Balmer emission, using \lline--\lacc scaling relations developed based on 1D planet-surface shock models \citep{aoyama2021}. Similar relations have not been developed for \hei emission lines; alternatively, we used the  empirical scaling relations for \hei lines developed by \cite{fiorellino2025}, updated from those in \cite{alcala2017} based on a sample of Class~\Romannum{2} young stellar objects (YSOs) down to 0.02~\msun, of the form
\begin{equation}
    \log\lacc = a\times\log\lline + b,
\end{equation}
where \lacc, \lline are in units of \lsun. The mean \lacc of the \hei lines (see Table~\ref{tab2} for individual values) is $\log(\lacc/\lsun)=-4.9\pm0.3$. 

In \citetalias{demars2026}, we showed that the wings component of the Balmer emission lines from the target could originate from gas in accretion columns along magnetic field lines. The NC in the target's \hei emission seems to imply physical conditions expected at the accretion shock at the base of these funnels (see Section~\ref{lineratio}). Assuming a magnetospheric accretion geometry from these hints, we derived the corresponding mass accretion rates for the target from the integrated flux of each \hei line using the conventional formula for accreting YSOs \citep{hartmann2016}, 
\begin{equation}
    \mdot = \left(1-\frac{R}{R_{\mathrm{in}}}\right)^{-1} \frac{\lacc R}{G M},
\end{equation}
where $R$, $M$ represents the radius and mass of the accreting body, which are $0.16\pm0.01\,R_{\odot}$ and 13$\pm$5~\mj respectively for Delorme 1 (AB)b \citep{ringqvist2023}. $R_{\mathrm{in}}$, the inner truncation radius, is not well-established for PMOs but is typically set to a standard value of $5R$ for CTTS \citep[see][]{bouvier2007, hartmann2016}. With $R_{\mathrm{in}}=5R$, the mean \mdot obtained from the \hei lines for Delorme 1 (AB)b is $\log(\mdot/\mjyr)=-8.1\pm0.3$ (see Table~\ref{tab2} for individual values), i.e. $\sim7\times10^{-12}\msun\mathrm{yr}^{-1}$. This is comparable to, but slightly higher than the accretion rate derived from its UV excess ($1-3\times10^{-12}\msun\mathrm{yr}^{-1}$; \citetalias{demars2026}) using the hydrogen slab model, as well as the \mdot derived from the \ha line flux in the spectra using \cite{fiorellino2025} scaling relations ($1.4\times10^{-12}\msun\mathrm{yr}^{-1}$). However, recent studies show that the inner disk radius can be as small as $2-3 R$ for T Tauri stars \citep[see][]{pittman2025}. Additionally, \citetalias{demars2026} shows that magnetospheric accretion models \citep{thanathibodee2019} that best fit the Balmer line profiles of Delorme 1 (AB)b predict a smaller $R_{\textrm{in}}$ of $1.5-3R$ for this PMO. If we assume $R_{\textrm{in}}=1.5R$, the mass accretion rate obtained increases to $1.8\times10^{-11}\msun\mathrm{yr}^{-1}$, more than twice the value obtained with $R_{\textrm{in}}=5R$.

We note that the accretion luminosity and mass accretion rate derived from \ha line flux using planetary scaling relations from \cite{aoyama2021} are considerably higher than the corresponding estimates from both the UV excess using the slab model (by $\sim1$~dex) and the \hei lines using empirical stellar scaling relations from \cite{fiorellino2025} (by $\sim0.6$~dex). The latter difference could be partially contributed by the extrapolation of the \lline--\lacc scaling relation for the \hei lines to lower mass regimes. However, the significant difference in these estimates indicates that the \cite{aoyama2018, aoyama2021} shock models may be overestimating the accretion luminosities for wide-orbit planetary companions like Delorme 1 (AB)b, and consequently highlights potential limitations in applying the corresponding scaling relations to protoplanets.

\renewcommand{\arraystretch}{1.2}
\begin{table}[!ht]
\caption{Accretion luminosity of detected \hei lines and corresponding mass accretion rates.}
\centering
\begin{tabular}[c]{c c c c}
\hline\hline
\lmbdrestvac & $\log(\lline/\lsun)$ &  $\log(\lacc/\lsun)$  &  $\log(\mdot/\mjyr)$\\
(\AA) & (dex) & (dex) & (dex) \\
\hline
4027.33 & $-8.2\pm0.2$ & $-5.3\pm1.0$ & $-8.5\pm1.0$ \\
4472.73 & $-7.8\pm0.1$ & $-4.9\pm0.6$ & $-8.2\pm0.7$ \\
4923.31 & $-8.3\pm0.1$ & $-4.8\pm0.6$ & $-8.1\pm0.7$ \\
5017.08 & $-8.3\pm0.1$ & $-4.2\pm0.7$ & $-7.5\pm0.7$ \\
5877.25 & $-7.3\pm0.1$ & $-4.7\pm0.6$ & $-7.9\pm0.6$ \\
6679.99 & $-7.7\pm0.1$ & $-5.3\pm0.8$ & $-8.5\pm0.03$ \\
\hline
\end{tabular}
\tablefoot{We assume no extinction ($A_v=0$) for these calculations.}
\label{tab2}
\end{table}

\subsection{Correlation with UV excess and Balmer emission} \label{correlation}
Accretion hotspots, where magnetospheric funnels hit the surface of the accreting source, can give rise to excess emission in the ultraviolet and optical domain. This excess flux can add to the continuum at these wavelengths, giving rise to the veiling effect seen in T Tauri stars. Hence, both UV excess and continuum veiling observed in young stars have been used as strong proxies for ongoing accretion. 

\begin{figure*}[!ht]
    \centering
    \includegraphics[scale=0.5]{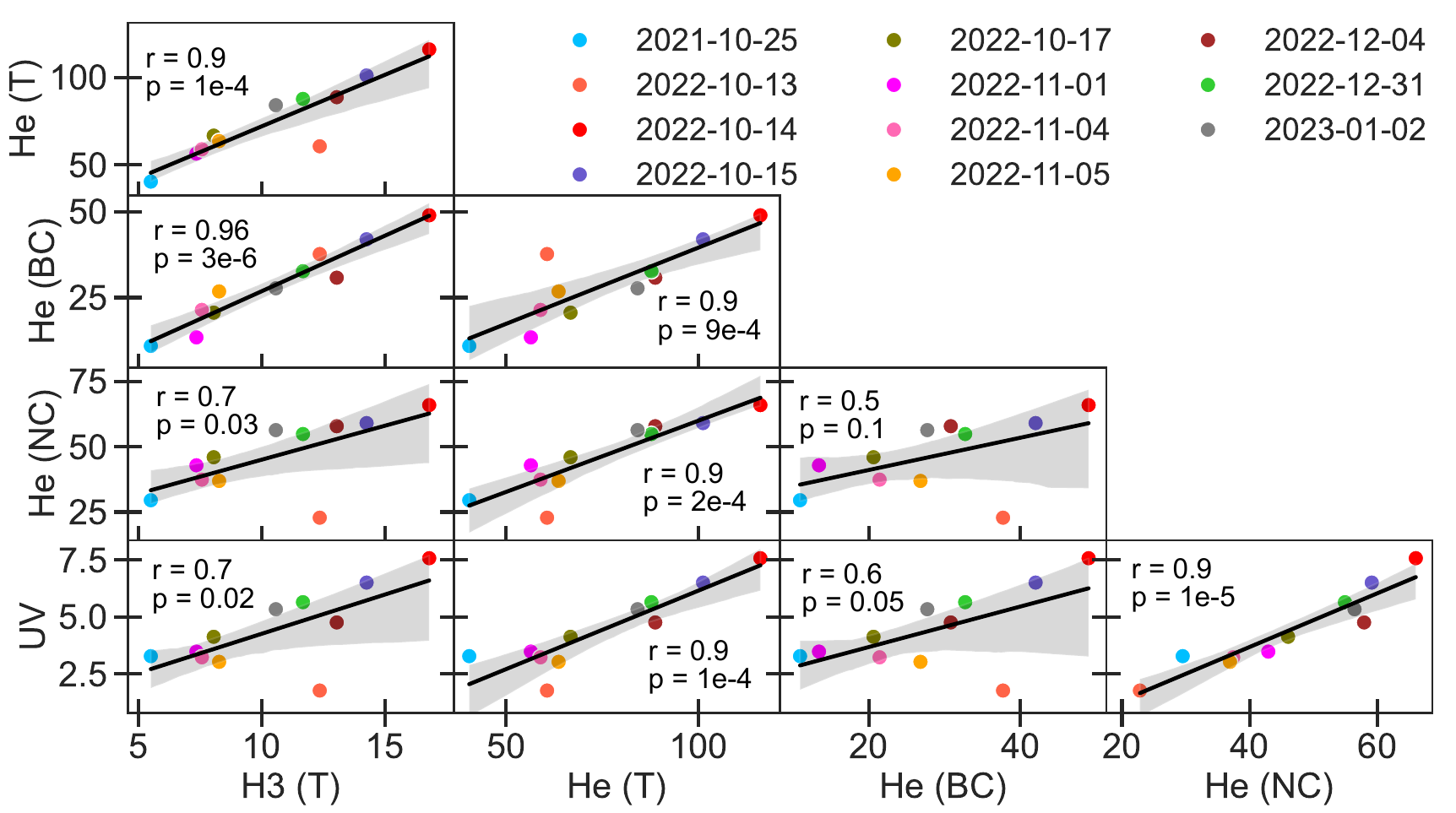}
    \caption{Correlations between \hei $\lambda5877$ line flux measured in this work, and \ha line flux and UV excess reported in \citetalias{demars2026}. The \hei and UV flux are in units of $10^{-17}$ cgs and \ha are in units of  $10^{-15}$ cgs. The Pearson's correlation coefficients ($r$) and the corresponding $p$-values are given in each plot. For both \hei and \ha, `T' denotes the total flux integrated over their entire profile. }
    \label{fig3}
\end{figure*}

The total \ha emission from the target seems to be well correlated with the UV excess (see Fig.~\ref{fig3}), with the correlation driven mostly by the wings component (see \citetalias{demars2026}). In general, the \hei lines detected from the target also display a good correlation with the UV excess similar to \ha. In Fig.~\ref{fig3}, we show the correlation plots for the strongest helium line \hei~D3 with UV excess and \ha. Its correlation is quite strong with the mean UV excess in each epoch (Pearson's $r=0.90$), and it is even stronger for the NC flux (Pearson's $r=0.94$), indicating accretion origin. From Fig.~\ref{fig3}, we see that the NC and BC flux seem uncorrelated with each other, and consequently, there is a relatively weaker correlation between BC flux and UV excess. However, this is mostly driven by a single epoch -- 13 October 2022 -- with strong BC flux and low UV excess. Excluding this epoch, the correlation of BC is strong with both UV excess (Pearson's $r=0.94$, $p=5\times10^{-4}$) and NC flux (Pearson's $r=0.85$, $p=0.02$).

\hei emission lines from the target also seem to closely follow the variation in the \ha flux through the epochs; for \hei~D3 this is illustrated in the figure. Among its two components, BC seems to be strongly correlated with \ha, while the correlation is slightly weaker for NC. This behaviour is in contrast to what we see with the UV excess. However, this is yet again driven by the epoch on 13 October 2022, excluding which the correlation becomes much stronger between NC and \ha (Pearson's $r=0.94$, $p=5\times10^{-5}$). We note here that we also attempted the empirical, non-Gaussian decomposition developed for the target's Balmer profiles in \citetalias{demars2026}, on the helium lines in this work, to compare the respective wings and core components of \hei~D3 and \ha. However, the wings component appears only in a few epochs for \hei~D3 in the resulting profile fits. Moreover, the velocity scales are vastly different for the two lines, as well as the resulting wings components; the entire \hei~D3 profile spans from $-25$ to $+60$~\kms, which is only slightly wider than the FWHM ($\sim70~\kms$) of the Class A \ha profile. Hence, we adhere to the usual Gaussian decomposition and refrain from any comparison with the individual wings and/or core components of \ha.

\begin{figure}[!ht]
    \centering
    \includegraphics[width=\linewidth]{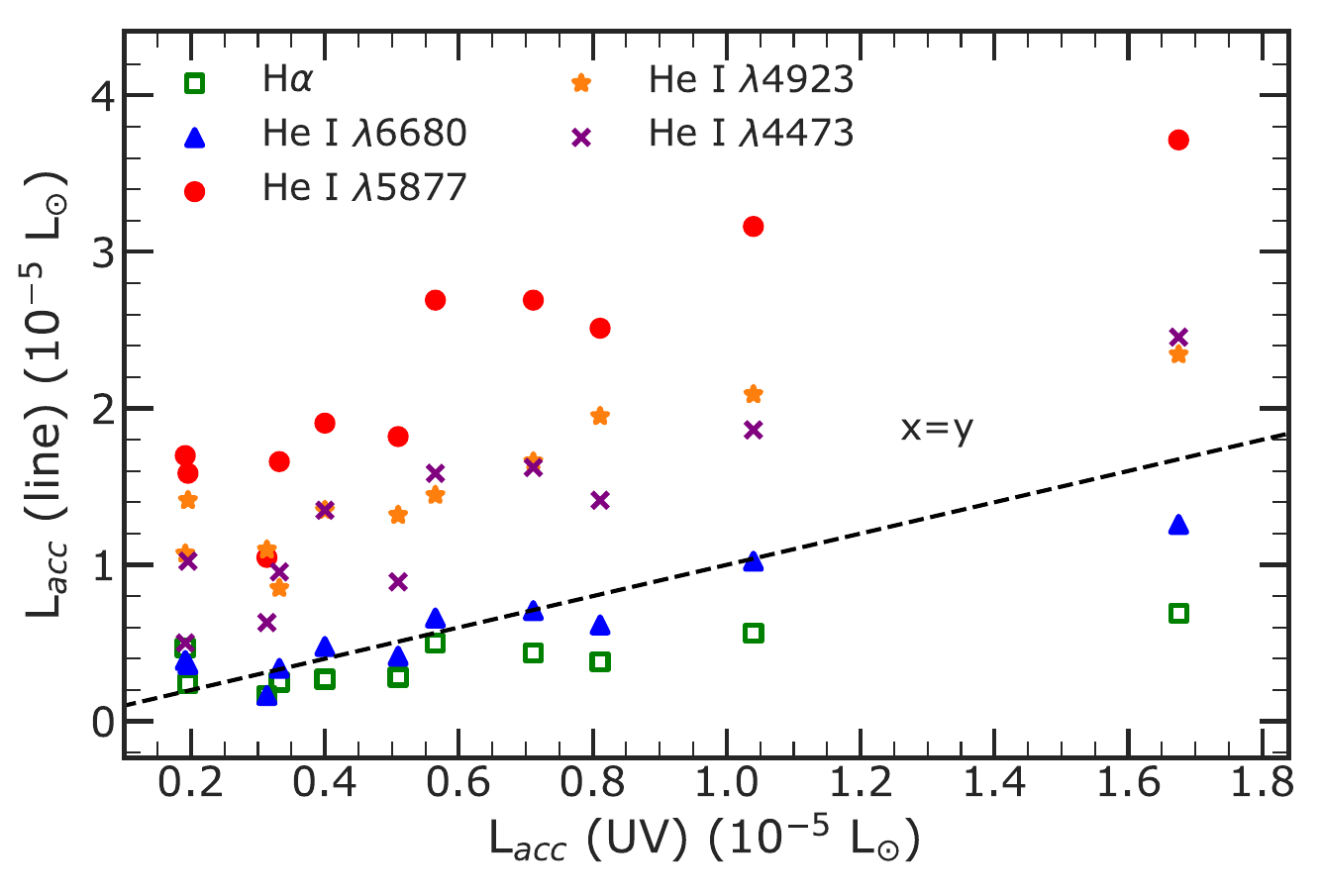}\\
    \caption{Accretion luminosity measurements of Delorme 1 (AB)b derived from \hei and \ha lines from each of the 11 nights using \lline--\lacc scaling relations from \cite{fiorellino2025}, plotted against the respective values measured from UV excess reported in \citetalias{demars2026}. }
    \label{fig4}
\end{figure}

Fig.~\ref{fig4} provides a comparison of accretion luminosity derived from \hei and \ha line flux using \cite{fiorellino2025} scaling relations with the value derived in \citetalias{demars2026} from UV excess emission using hydrogen slab models; the latter is a robust estimation of an upper limit on accretion luminosity as it is independent of the inherent uncertainties of the empirical \lline--\lacc scaling relations. The figure demonstrates that the value of $\lacc$ derived from the integrated \ha flux is relatively consistent with the value derived from the UV excess, although slightly lower. Among the \hei lines,  the \lacc derived from the \hei $\lambda6680$ line flux is highly consistent with the estimate from the UV excess, while those from the asymmetric \hei lines $\lambda\lambda5877,4923,4473$ are generally slightly higher, with the difference being the highest for \hei~D3; the difference between \mdot obtained from \hei~$\lambda6680$ and \hei~D3 is a factor of 4 (see Table.~\ref{tab2}).
The slightly higher values for \lacc estimates from these \hei lines could point to a likely contribution of chromospheric activity. However, the deviation in the \lacc values could also be attributed to uncertainties stemming from the empirical \lline--\lacc scaling relations. These relations are based on accretors of much higher mass than the target; although the sample in \cite{fiorellino2025} goes down to 0.02~\msun, there is only one low-mass target in the entire sample.

\subsection{Comparison to \hei emission in the primary}\label{primary}
The VLT/UVES observations also provide the spectra of the target's M5-type host binary, Delorme 1 AB, which is non-accreting; it does not show any UV excess in its spectra and has no confirmed disk around it. Emission lines in the primary's spectra, attributed solely to chromospheric activity, and taken with the same instrument at the same epochs, serve as a good tool to gauge the contribution to the companion's spectra from activity. 
We detected both the popular activity indicators \ha and \hei~D3 in emission from the primary spectra, the latter at $\sim4\sigma$. The centroid velocity of \hei~D3 is small and redshifted ($2.2\pm0.2$~\kms), but its line profile is symmetric about the centre, unlike the line profile of the companion. \hei emission at $\lambda\lambda4027,4473$ was also detected from the primary at $5\sigma$ and $7\sigma$ respectively, similar to previous detections from M-type dwarfs \citep[e.g][]{maio2020}, but no asymmetry was detected in either of the line profiles. Unlike the Balmer and \hei emission lines from the PMC, which show strong flux variation between the individual epochs as expected from accretion, \ha and \hei lines from its primary show relatively uniform flux levels between the epochs, with the exception on 2 January 2023 where all lines show an overall increase in flux by a factor 1.5 to 2, accompanied by enhanced flux in the blue wing (see Fig.~\ref{figG}). Such blue wing asymmetries are typical during flares from similar M-dwarfs \citep{notsu2024}. This epoch also shows the tentative appearance of \hei emission lines $\lambda\lambda3820, 4923, 6680$ from the primary. \hei $\lambda6680$ emission is usually observed from CTTS and is associated with accretion \citep{muzerolle1998}, but it has been reported from M-type main sequence stars during flare maximum, along with \hei emission at $\lambda4923$ \citep{bleach2002}. The mean \lacc measured from the \hei lines in the primary spectra is $\log(\lacc/\lsun)=-4.2\pm0.1$ and the corresponding accretion rate $\log(\mdot/\msun\mathrm{yr}^{-1})=-11.4\pm0.1$ (assuming $R_{\mathrm{in}}=5R$). Overall, the nature of the primary's \hei line profiles from activity presents a stark contrast to those observed from the companion, further hinting at accretion origin for the latter.

\section{Discussion}  \label{discussion} 

\subsection{Origin of NC and BC} 
Asymmetries in the \hei line profiles are commonly found in the accretion signatures of CTTS. While the NC is interpreted as tracing the accretion shock at the stellar surface in these objects, the exact origin of BC remains inconclusive in the literature, with attributions to both hot-wind origin (for blueshifted BC) and accretion-flow origin \citep[see][]{edwards2003, armeni2024}. In Section~\ref{analysis}, we used a few diagnostic tools to obtain clues about the possible physical origins of the narrow and broad components detected from the \hei line profiles of the target, which we discuss below.

A comparison of the profile characteristics of the target's \hei~D3 emission line to those of CTTS (see Section~\ref{ctts-comparison}) reveals that, while the NC width is much narrower than the typical width seen in CTTS, the slightly redshifted mean NC centroid velocity ($0.1\pm0.8$~\kms) is well within the expected range among CTTS, and could point to emission from a post-shock region. The epoch-to-epoch variation of the NC centroid velocity (see Fig.\ref{vel-variation}), oscillating mostly between $\pm1$~\kms, is also strongly indicative of its formation close to the planet's surface. 
Such velocity oscillations observed in T Tauri stars are consistent with the NC tracing a hot spot localised at high latitudes on the stellar surface and corresponding to the accretion shock at the bottom of an accretion column in a dipolar geometry \citep{mcginnis2020}. The triplet--singlet line ratio derived in Section~\ref{lineratio} for the \hei~D3 NC also implies high densities and near-thermodynamic equilibrium conditions within the line formation regions, which are typical of such post-shock regions. 

The strong correlation of  \hei~D3 NC with the UV excess emission from the target also corroborates an accretion-related origin for the NC emission in general; however, UV excess emission can have contributions from chromospheric activity in stars, which may result in a similar correlation. The near-UV flux from Delorme 1 (AB)b ($F_{\lambda}=2-10\times10^{-17}$~\cgs\AA$^{-1}$ at 3500~\AA; see \citetalias{demars2026}) is in the range of the expected level of chromospheric activity among late M-dwarfs ($M_{\star}>0.08\msun$) older than 10~Myr, as reported in \cite{richey2023}. Their analysis, however, does not distinguish accretors from non-accretors beyond 10~Myr, whereas accretion has been seen to proceed quite late into the formation stage among lower-mass objects like Delorme 1 (AB)b. Additionally, \citetalias{beristain2001} showed that the \hei~D3 NC flux from their CTTS sample, attributed to post-shock origin, is strongly correlated with the continuum veiling produced by UV excess. Furthermore, \cite{kwan2024} showed that for the \hei~D3 profiles among the \citetalias{beristain2001} sample with stronger NC, accretion shock models can reproduce both their NC flux and the UV continuum excess.

The BC in the \hei~D3 emission from the target seems to be of a much lower width compared to those in CTTS, but this may be explained by the much lower mass of the target compared to CTTS. Although the BC is mostly blueshifted among CTTS, its centroid velocity in the target spectra is within the range of the few redshifted BCs in these sources. The detection of similar BCs in the singlet transition \hei~$\lambda4923$, as well as the variation in the BC/NC flux ratio through the epochs, implies that this asymmetry is not simply from the resolution of the triplet components in the emission line. This is slightly corroborated by the lack of asymmetry in the primary's \hei~D3 emission, although there is a possibility in the case of the primary that the line may simply be intrinsically wider than the velocity separation between its triplet components to produce this asymmetry. However, redshifted BCs have been found in CTTS despite large line widths. The strong correlation with UV excess on excluding the 13 October 2022 epoch is also suggestive of an origin from accretion-related processes. 

We saw in Section~\ref{kinematics} that the centroid velocity of the BC shares a strong correlation in time with that of the NC. This hints at a common mechanism of origin for these two components, likely within the post-shock region. This is further corroborated by the fact that we see the BC persistently redshifted in our multi-epoch observations, oscillating closely around $\sim+15$~\kms. Such behaviour is expected from accretion-shock emission from near the planet surface; if the rotational velocity is less than the velocity of the infalling gas, then we may not catch a blueshift but simply observe the BC oscillate around a redshift. The FWHM of the BC tells us that the bulk of its emission is coming from the gas that is moving at $\sim$20--30~\kms, which implies an origin close to the surface of the planet. Instead, if the BC originates within the accretion-flow as proposed for CTTS, its width should be compatible with the free-fall or Keplerian velocity of the gas. For Delorme 1 (AB)b, the free-fall velocity at its surface is 176~\kms and the infall velocity of gas falling from a distance of $R_{\mathrm{in}}=2R$ is 124~\kms. Assuming a distance range  from $R_{\mathrm{in}}=3R$ to the planet surface, the Keplerian velocity will range from 72--124~\kms, with its projection along the line of sight (assuming a typical inclination of $60^{\circ}$) 62--108~\kms. The \hei BC width for the target is much narrower than these estimates, rendering emission from the accretion flow unlikely. 

Alternatively, the asymmetry in the red wing seen in the target's \hei profiles could be a part of the NC line profile. Pure-NC profiles among CTTS have been seen to display some asymmetry in the red part of the line profile due to velocity gradients in the post-shock region, resulting from the incoming gas experiencing a strong deceleration as it approaches the surface of the accretor \citep[see][]{donati2008,donati2012,donati2013}. In fact, some of the pure-NC profiles in \citetalias{beristain2001} show a similar asymmetry, and the extracted NCs from composite profiles in their sample show on average a redwing asymmetry very similar to what is seen from Delorme 1 (AB)b. This is further supported by the slightly asymmetric NC profile of \hei~$\lambda6680$ from Delorme 1 (AB)b. Such an origin may explain the strong correlation seen between NC and BC parameters of these \hei lines from the target. However, asymmetry resulting from such velocity gradients in the \hei~D3 line usually exhibits a smooth profile with an extended red wing and a steep blue wing \citep[see][]{nowacki2023, armeni2024}. Although this may be descriptive of the profile asymmetries in the higher \hei transitions $\lambda\lambda4923, 4473, 4027$ from the target, the asymmetry in the observed \hei~D3 line manifests itself as a clear `bump' in the red wing, which is more suggestive of an additional component in the profile. 

Thus, it is very likely that BC is also formed within the temperature-stratified shock structure, but closer to the shock front than the NC, where the temperature is higher and the gas is still infalling. NC could be formed at a lower temperature region within the shock structure, closer to the planet surface.

\subsection{Accretion or activity?}
Evidence of ongoing accretion at Delorme 1 (AB)b at its age of $\sim$40~Myr emphasizes the challenges in understanding the formation of wide-orbit planetary-mass companions. \cite{malin2025} studied the mid-infrared spectra of the target and showed that its CPD retains a significant amount of gas, despite being older than the typical time scales for gas dispersion around low-mass stars. \cite{stamatellos2015} found that low-mass bound companions, which form via fragmentation of their parent circumstellar disk, can assemble more massive disks around them than previously expected, which evolve slower in time at low viscosities. \cite{betti2023} further deduced that the mass accretion rates for PMCs do not fall as steeply with decreasing mass as in the case of stars and brown dwarfs. An important step in understanding these new trends emerging at planetary masses is to exercise caution and accuracy in determining accretion characteristics, especially distinguishing accretion signatures from those of activity.

Both \ha and \hei~D3 are popular indicators of chromospheric activity among late-type stars and YSOs \citep{saar1997, rachford1997}. In CTTS, such emission from activity can augment the line emission originating in the infalling gas from an accretion disk \citep{petrov2011, manara2013}. It is crucial to gauge this contribution while determining the accretion properties in lower mass regimes, where the accretion levels are weak (\lacc$<10^{-3}\lsun$).  

\begin{figure*}[!ht]
    \centering
    \includegraphics[width=0.45\linewidth]{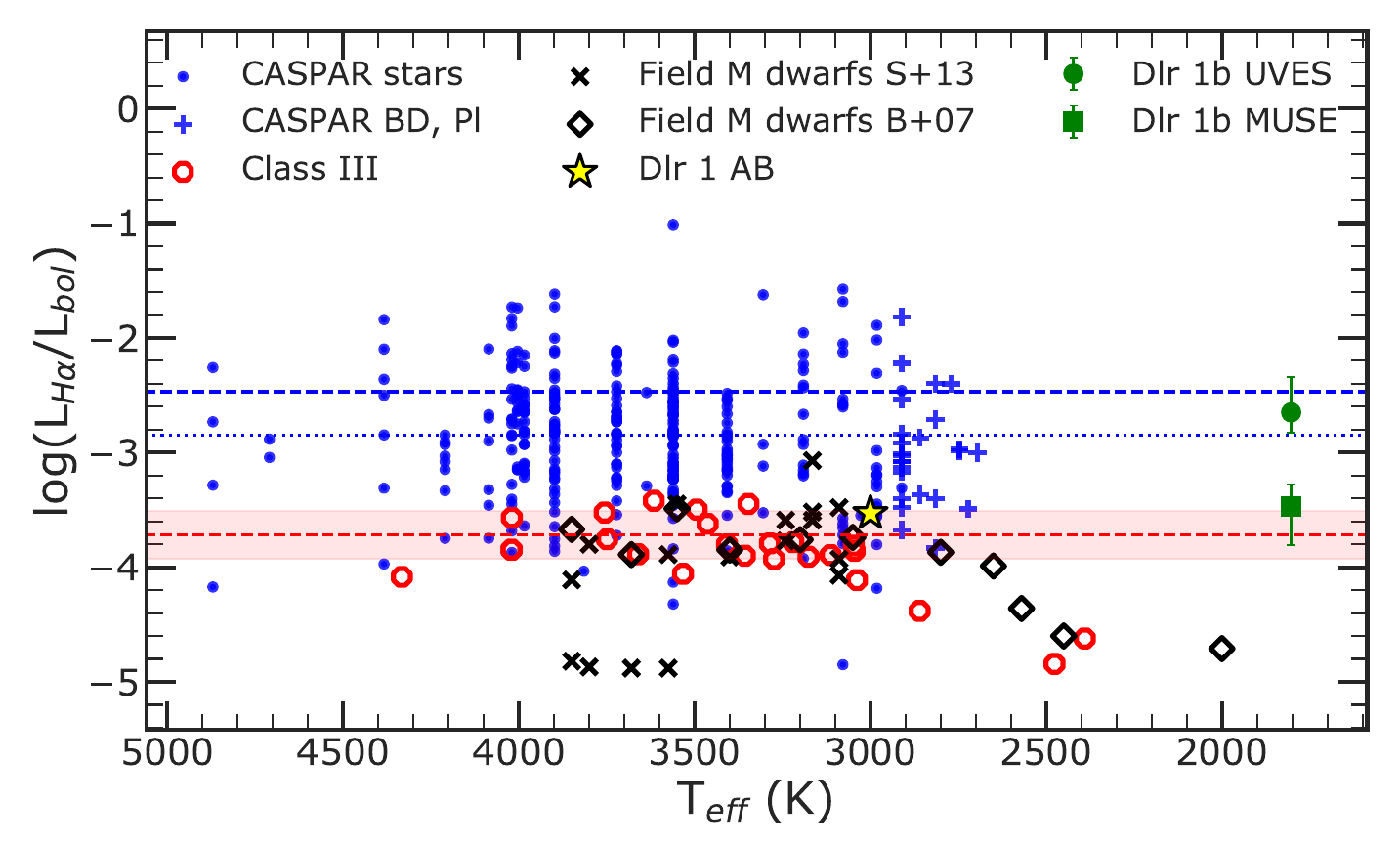}
    \includegraphics[width=0.45\linewidth]{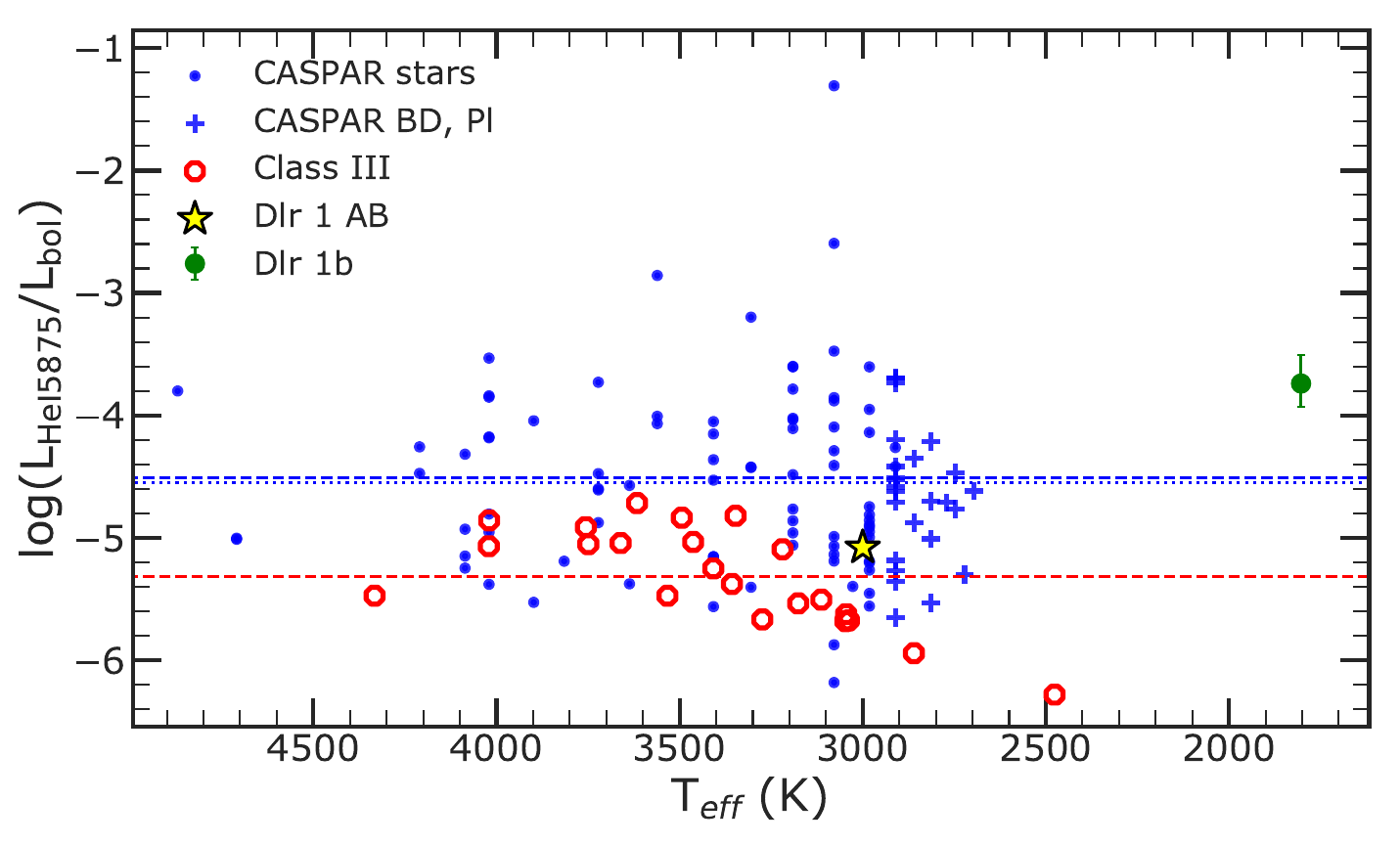}\\
    \includegraphics[scale=0.4]{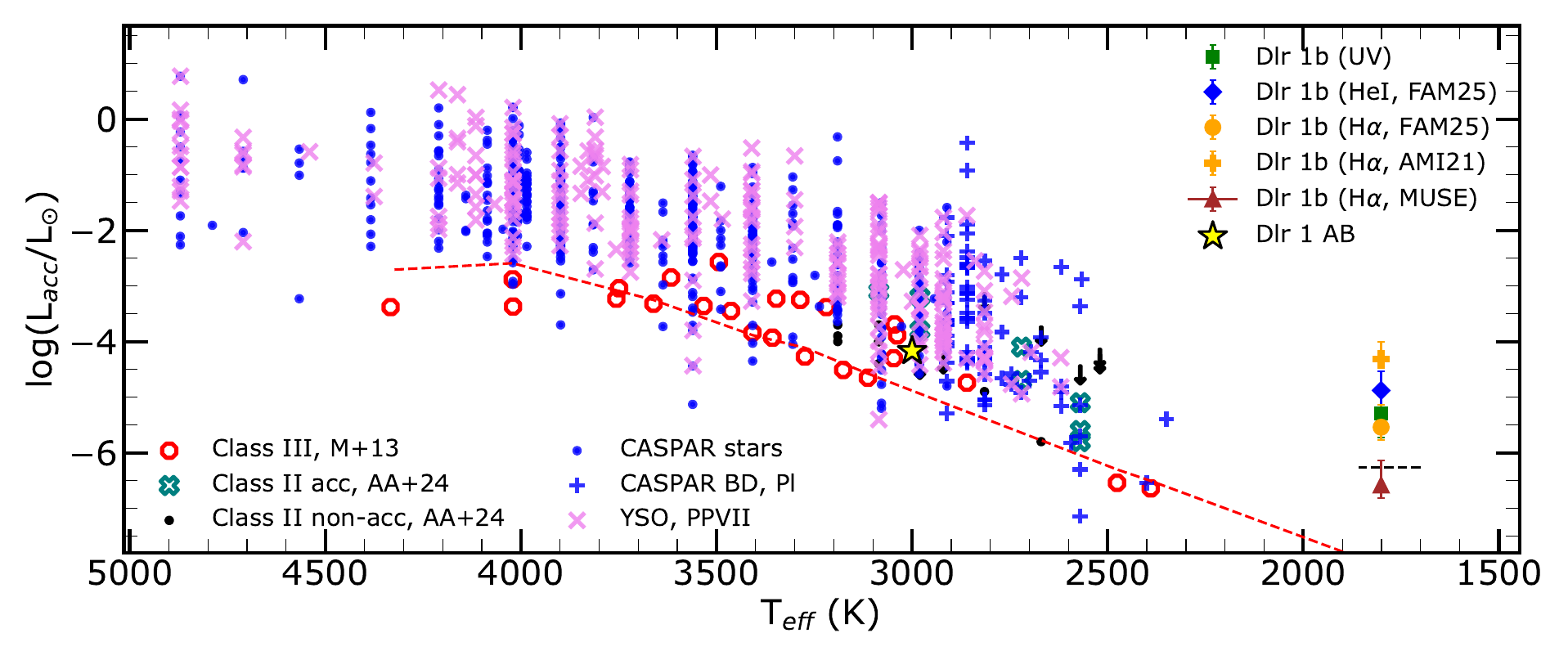}
    \caption{Top: Fractional \ha (left) and \hei $\lambda5877$ (right) luminosity of Delorme 1 (AB)b (filled green circle) shown with respective values from the \citetalias{betti2023} sample of young stars, brown dwarfs, and planets (`BD', `Pl'), sample of active field M-dwarfs from \cite{stelzer2013a} and \cite{bochanski2007}, and sample of Class~\Romannum{3} non-accretors from \cite{manara2013}. The available \ha measurement for the target with MUSE is shown as the filled green square. The dashed red line (shaded region denotes error bars) shows the mean activity index for \ha \citep{stelzer2013b} and \hei (this work). The blue lines denote the median measurements for \citetalias{betti2023} stars (dotted) and brown dwarfs, planets (dashed). Bottom: Accretion luminosity of Delorme 1 (AB)b measured from UV excess, \hei lines, and \ha, along with Class~\Romannum{3} sources from \cite{manara2013} and \citetalias{betti2023} accretors. Also shown are low-mass Class~\Romannum{2} accretors (unfilled teal crosses) and non-accretors (black dots, with upper limits as downward arrows) from \cite{almendros2024} and a sample of accreting YSOs from seven star-forming regions from \cite{manara2023} (pink crosses). \teff$-L_{\mathrm{acc, noise}}$ relation from \cite{manara2013} is shown as the dashed red curve, and the $L_{\mathrm{acc, noise}}$ from \cite{venuti2019} at 20~\mj is shown as the dashed black line. FAM25 and AMI21 denotes \lline--\lacc scaling relations from \cite{fiorellino2025} and \cite{aoyama2021} respectively, and \lacc derived from MUSE epoch is denoted as the brown triangle. In all the panels, measurements from the primary are denoted as a yellow star and fall very close to the measurements from non-accretors. }
    \label{fig5}
\end{figure*}

In Fig.~\ref{fig5}, we compare the emission level measured from the target to those of known accreting objects from \citet[][CASPAR]{betti2023}, as well as non-accreting Class~\Romannum{3} YSOs and field M-dwarfs with activity-induced line emission. The upper panel illustrates the sharply decreasing activity level at late spectral types for the Class~\Romannum{3} sample, for both \ha and \hei~D3 line emission. The fractional \ha luminosity measured from Delorme 1b with UVES, $\log(L_{\ha}/L_{\mathrm{bol}})=-2.65^{+0.31}_{-0.18}$\footnote{The error bars on the fractional luminosity and accretion luminosity measurements correspond to the lowest and  highest values during the observation epochs.}, is more than 1~dex above the mean activity index of the Class~\Romannum{3} sample \citep[$-3.72\pm0.21$;][]{stelzer2013b}, and agrees well with the median measurements of CASPAR accretors. The median \hei fractional luminosity measured in this work, $\log(L_{\mathrm{\hei}}/L_{\mathrm{bol}})=-3.74^{+0.23}_{-0.19}$, is also well above the median Class~\Romannum{3} measurement ($-5.31$), and slightly higher than those of CASPAR accretors ($\sim-4.5$). This indicates that the \ha and \hei line emission in these UVES observations is clearly dominated by accretion. 
However, the target's MUSE observations in September 2018 report a much lower \ha flux with  $\log(L_{\mathrm{\ha}}/L_{\mathrm{bol}})=-3.47^{+0.19}_{-0.34}$, putting it right near the mean activity index. This suggests that Delorme 1 (AB)b, being a weak accretor (\mdot$\sim10^{-12}\msun \mathrm{yr}^{-1}$), might have been activity-dominated in the MUSE epoch; in other words, accretion at the target might be highly episodic. Notably, the MUSE epoch lacks \hei line detections, except for very weak detections of $\lambda6680$ and $\lambda7065$.

For such weak accretors, \cite{manara2013} empirically defines the minimum  accretion luminosity that can be measured from line emission, $L_{\mathrm{acc, noise}}$, below which the line emission is likely to be dominated by chromospheric activity. This relation for `chromospheric noise', based on a sample of Class~\Romannum{3} YSOs up to spectral type M9.5, is given as
\begin{equation}
    \log(L_{\mathrm{acc, noise}}/L_{\star})=(6.17\pm0.53)\log\teff-(24.54\pm1.88),
\end{equation}
where $L_{\star}$ is the bolometric luminosity of the object and \teff~is its effective temperature in kelvin. In line with expectations, the mean \lacc measured from the non-accreting central binary is within $\sim0.6$~dex of the predicted $L_{\mathrm{acc, noise}}$ ($-4.74$) at its temperature \citep[3000~K;][]{ringqvist2023}. By extrapolating this relation to the target's temperature of 1801~K \citep[with $\log(L_{\star}/\lsun)=-3.58$][]{eriksson2020}, we get $\log(L_{\mathrm{acc, noise}}/\lsun)=-8.03$. The \lacc derived for the target from the \ha and \hei lines (see Table~\ref{tab5} for the values) are, respectively, 2.5~dex and 3.2~dex higher than this limit. This is illustrated in the bottom panel of Fig.~\ref{fig5}. We note that the \lacc measured from the target's \ha flux from MUSE observations also falls well above the $L_{\mathrm{acc, noise}}$. As an added measure of extrapolation, we also compared the \lacc from the  \hei lines to the $L_{\mathrm{acc, noise}}$ of the two least massive objects ($\sim20$~\mj) in the Class~\Romannum{3} sample -- M9-type TWA 26 and M9.5-type TWA 29. Both of these objects have $\log(L_{\mathrm{acc, noise}}/\lsun)\sim-6.6$, which is still $1.7$~dex below the target's \lacc, adding confidence to the inference from the extrapolated relation above.

Independently, \cite{venuti2019} estimated the mass-dependent noise threshold, $\dot{M}_{\mathrm{acc, noise}}$, for YSOs down to 20~\mj at 10~Myr based on \lacc measurements of 11 non-accreting stars in the TWA stellar association. At 20~\mj, they predict $\dot{M}_{\mathrm{acc, noise}}\sim3\times10^{-13}\msun\mathrm{yr}^{-1}$, which is $\sim1$~dex lower than the estimated mass accretion rates for Delorme 1 (AB)b with UVES. In Fig.~\ref{fig5} (bottom panel), we show the corresponding $L_{\mathrm{acc, noise}}$; at the MUSE epoch the target clearly falls below this threshold, underlining the earlier inference that the 2018 observations could have been activity-dominated. 

Additionally, for accretion luminosities measured from UV continuum excess, \cite{claes2024} developed a parallel relation for $L_{\mathrm{acc, noise}}$, given by
\begin{equation}
     \log(L_{\mathrm{acc, noise}}/L_{\star})=(5.9\pm0.2)\log(\teff[\mathrm{K}])-(23.3\pm0.7).
\end{equation}
If \lacc falls below this threshold, the observed UV excess alone cannot determine that the target is accreting. At the target's temperature, this threshold value is $-7.67$, which is more than 2~dex below the measured \lacc from the target's UV continuum excess (see Table~\ref{tab5}). Thus, the UV continuum excess and the \ha and \hei line emission detected in these UVES observations are clearly dominated by accretion processes. However, as discussed in Section.~\ref{correlation}, the contribution to the \hei lines from chromospheric activity cannot be dismissed, although not prominent.

\section{Conclusions}  \label{conclusion} 
In this study, we continued our analysis of the high-resolution optical to near-UV spectra of Delorme 1 (AB)b from \citetalias{demars2026}, focusing on helium emission from this planetary-mass object. We presented resolved line profiles of seven \hei emission lines detected with high confidence from the target, including the prominent accretion tracers in CTTS, \hei~$\lambda\lambda5877,6680$. All detected \hei lines show moderate flux variation over time, similar to \hei profiles in CTTS. The median line profiles of \hei~$\lambda\lambda5877,4923,4473$, and 4027 display clear asymmetry with a narrow and broad component; the latter is redshifted, unlike commonly observed in CTTS. Analysis of NC behaviour in \hei~D3 is strongly suggestive of formation within the immediate post-shock region close to the surface of the planet. The BC is most likely of an accretion origin, and its strong correlation in velocity with the NC and the line widths suggest an origin within the shock structure, but closer to the shock front, at higher temperatures. It could also be an asymmetry in the red part of a pure-NC line profile due to a velocity gradient in the post-shock region. 
Comparison of fractional luminosity $L_{\mathrm{\hei}}/L_{\mathrm{bol}}$ and \lacc from \hei flux with those of higher-mass non-accretors indicates that \hei emission is accretion-dominated in these observations, but contribution from activity cannot be completely dismissed.

A key factor in analysing the accretion signatures of such low-mass objects is to gauge the level of contribution from chromospheric activity at these low \teff values or masses. Currently, records of activity indicators from field dwarfs and Class~\Romannum{3} sources extend up to late M-types, with few sources cooler than $\teff=2500$~K, as demonstrated in this work. Although the general trend from higher mass non-accretors indicates a steep decline in activity level at late spectral types, the level of activity at low \teff ranges as Delorme 1 (AB)b remains unclear; the extrapolation of the \cite{manara2013} relation to much lower \teff such as for the target is not really justified since the atmosphere becomes mostly neutral at these temperatures, making activity diagnostics unreliable \citep{mohanty2003b}. Observations of more low-mass Class~\Romannum{3} sources are as equally vital as observations of emission signatures from accretors to understand accretion in the planetary-mass regime. The target sample of the approved VLT/UVES program ENTROPY (PI: Bonnefoy) is designed to tackle exactly this need and will search for prominent accretion indicators, including optical helium line emission, from both accreting and non-accreting sources below 20~\mj.

The \hei $\lambda10833$ emission line is another significant accretion tracer among CTTS in the near-infrared that is sensitive to both accretion and accretion-related outflows such as disk wind or stellar wind \citep{edwards2006, kwan2007, erkal2022}. It has recently been detected in a very low-mass object, TWA 27b \citep[see][]{luhman2023c, marleau2024}, using the JWST Near-Infrared Spectrograph \citep[NIRSpec;][]{jakobsen2022}. The CRyogenic high-resolution InfraRed Echelle Spectrograph \citep[CRIRES;][]{kaufl2004} at VLT can observe the \hei $\lambda10833$ line at a high spectral resolution of $R_{\lambda}\geq\textrm{50,000}$ required to disentangle accretion features in the line profile. The upcoming ArmazoNes high Dispersion Echelle Spectrograph \citep[ANDES;][]{marconi2024} at the Extremely Large Telescope (ELT) that will see its first light early in the next decade has the potential to observe helium emission from protoplanets from near-UV to near-infrared, covering several of the prominent accretion tracers. Such observations will pave the way for population-level studies to understand helium emission from PMOs and can support the development of accretion models, as has been done for hydrogen emission, sculpting the accretion geometry of protoplanets.

\section*{Data availability}
The underlying data used for this work are the flux-calibrated 1D UVES spectra obtained from the ESO observations from 13 October 2022 to 2 January 2023  (programme 0110.C-0203(A)) and on 25 October 2021 (programme 0108.C-0655(A)). Raw science data and calibration products are available through ESO Archive Services. The median stacked He~\textsc{i} line profiles are available through Zenodo \href{https://zenodo.org/records/19006669}{10.5281/zenodo.19006669}.

\begin{acknowledgements}
    We thank Justyn Campbell-White for his valuable inputs and discussions on this work.
      G.V.\ acknowledges support from the Swedish Research Council (\emph{SRC}) via the International Postdoc Grant for the project GENESIS (2024-06609).
      G.-D.M.\ acknowledges the support of the Deutsche Forschungsgemeinschaft (DFG) through grant MA~9185/2-1.
      We acknowledge support in France from the French National Research Agency (ANR) through project grant ANR-20-CE31-0012.
       
\end{acknowledgements}

\bibliographystyle{bibtex/aa_url} 
\bibliography{bibtex/bibliography.bib}

@ARTICLE{alcala2017,
       author = {{Alcal{\'a}}, J.~M. and {Manara}, C.~F. and {Natta}, A. and {Frasca}, A. and {Testi}, L. and {Nisini}, B. and {Stelzer}, B. and {Williams}, J.~P. and {Antoniucci}, S. and {Biazzo}, K. and {Covino}, E. and {Esposito}, M. and {Getman}, F. and {Rigliaco}, E.},
        title = "{X-shooter spectroscopy of young stellar objects in Lupus. Accretion properties of class II and transitional objects}",
      journal = {\aap},
         year = 2017,
        month = apr,
       volume = {600},
          eid = {A20},
        pages = {A20},
          doi = {10.1051/0004-6361/201629929},
archivePrefix = {arXiv},
       eprint = {1612.07054},
 primaryClass = {astro-ph.SR},
       adsurl = {https://ui.adsabs.harvard.edu/abs/2017A&A...600A..20A}
}

@ARTICLE{almendros2024,
       author = {{Almendros-Abad}, V. and {Manara}, C.~F. and {Testi}, L. and {Natta}, A. and {Claes}, R.~A.~B. and {Mu{\v{z}}i{\'c}}, K. and {Sanchis}, E. and {Alcal{\'a}}, J.~M. and {Bayo}, A. and {Scholz}, A.},
        title = "{Evolution of the relation between the mass accretion rate and the stellar and disk mass from brown dwarfs to stars}",
      journal = {\aap},
         year = 2024,
        month = may,
       volume = {685},
          eid = {A118},
        pages = {A118},
          doi = {10.1051/0004-6361/202348649},
archivePrefix = {arXiv},
       eprint = {2402.10523},
 primaryClass = {astro-ph.SR},
       adsurl = {https://ui.adsabs.harvard.edu/abs/2024A&A...685A.118A}
}

@ARTICLE{almendros2025,
       author = {{Almendros-Abad}, Victor and {Scholz}, Aleks and {Damian}, Belinda and {Jayawardhana}, Ray and {Bayo}, Amelia and {Flagg}, Laura and {Mu{\v{z}}i{\'c}}, Koraljka and {Natta}, Antonella and {Pinilla}, Paola and {Testi}, Leonardo},
        title = "{Discovery of an Accretion Burst in a Free-floating Planetary-mass Object}",
      journal = {\apjl},
         year = 2025,
        month = oct,
       volume = {992},
       number = {1},
          eid = {L2},
        pages = {L2},
          doi = {10.3847/2041-8213/ae09a8},
archivePrefix = {arXiv},
       eprint = {2510.01747},
 primaryClass = {astro-ph.SR},
       adsurl = {https://ui.adsabs.harvard.edu/abs/2025ApJ...992L...2A}
}

@ARTICLE{aoyama2018,
       author = {{Aoyama}, Yuhiko and {Ikoma}, Masahiro and {Tanigawa}, Takayuki},
        title = "{Theoretical Model of Hydrogen Line Emission from Accreting Gas Giants}",
      journal = {\apj},
         year = 2018,
        month = oct,
       volume = {866},
       number = {2},
          eid = {84},
        pages = {84},
          doi = {10.3847/1538-4357/aadc11},
archivePrefix = {arXiv},
       eprint = {1808.06776},
 primaryClass = {astro-ph.EP},
       adsurl = {https://ui.adsabs.harvard.edu/abs/2018ApJ...866...84A}
}

@ARTICLE{aoyama2021,
       author = {{Aoyama}, Yuhiko and {Marleau}, Gabriel-Dominique and {Ikoma}, Masahiro and {Mordasini}, Christoph},
        title = "{Comparison of Planetary H{\ensuremath{\alpha}}-emission Models: A New Correlation with Accretion Luminosity}",
      journal = {\apjl},
         year = 2021,
        month = aug,
       volume = {917},
       number = {2},
          eid = {L30},
        pages = {L30},
          doi = {10.3847/2041-8213/ac19bd},
archivePrefix = {arXiv},
       eprint = {2108.01277},
 primaryClass = {astro-ph.EP},
       adsurl = {https://ui.adsabs.harvard.edu/abs/2021ApJ...917L..30A}
}

@ARTICLE{aoyama2024,
       author = {{Aoyama}, Yuhiko and {Marleau}, Gabriel-Dominique and {Hashimoto}, Jun},
        title = "{Analyzing JWST/NIRSpec Hydrogen Line Detections at TWA 27B: Constraining Accretion Properties and Geometry}",
      journal = {\aj},
         year = 2024,
        month = oct,
       volume = {168},
       number = {4},
          eid = {155},
        pages = {155},
          doi = {10.3847/1538-3881/ad67df},
archivePrefix = {arXiv},
       eprint = {2407.15922},
 primaryClass = {astro-ph.EP},
       adsurl = {https://ui.adsabs.harvard.edu/abs/2024AJ....168..155A}
}

@ARTICLE{armeni2024,
       author = {{Armeni}, A. and {Stelzer}, B. and {Frasca}, A. and {Manara}, C.~F. and {Walter}, F.~M. and {Alcal{\'a}}, J.~M. and {Schneider}, P.~C. and {Sicilia-Aguilar}, A. and {Campbell-White}, J. and {Fiorellino}, E. and {Gameiro}, J.~F. and {Gangi}, M.},
        title = "{Evidence for magnetic boundary layer accretion in RU Lup: A spectrophotometric analysis}",
      journal = {\aap},
         year = 2024,
        month = oct,
       volume = {690},
          eid = {A225},
        pages = {A225},
          doi = {10.1051/0004-6361/202451065},
archivePrefix = {arXiv},
       eprint = {2408.14996},
 primaryClass = {astro-ph.SR},
       adsurl = {https://ui.adsabs.harvard.edu/abs/2024A&A...690A.225A}
}

@ARTICLE{arul2023,
       author = {{Arulanantham}, Nicole and {Gronke}, Max and {Fiorellino}, Eleonora and {Gameiro}, Jorge Filipe and {Frasca}, Antonio and {Green}, Joel and {Chang}, Seok-Jun and {Claes}, Rik A.~B. and {Espaillat}, Catherine C. and {France}, Kevin and {Herczeg}, Gregory J. and {Manara}, Carlo F. and {Venuti}, Laura and {{\'A}brah{\'a}m}, P{\'e}ter and {Alexander}, Richard and {Bouvier}, Jerome and {Campbell-White}, Justyn and {Eisl{\"o}ffel}, Jochen and {Fischer}, William J. and {K{\'o}sp{\'a}l}, {\'A}gnes and {Vioque}, Miguel},
        title = "{Ly{\ensuremath{\alpha}} Scattering Models Trace Accretion and Outflow Kinematics in T Tauri Systems}",
      journal = {\apj},
         year = 2023,
        month = feb,
       volume = {944},
       number = {2},
          eid = {185},
        pages = {185},
          doi = {10.3847/1538-4357/acaf70},
archivePrefix = {arXiv},
       eprint = {2301.01761},
 primaryClass = {astro-ph.SR},
       adsurl = {https://ui.adsabs.harvard.edu/abs/2023ApJ...944..185A}
}

@INPROCEEDINGS{bacon2010,
       author = {{Bacon}, R. and {Accardo}, M. and {Adjali}, L. and {Anwand}, H. and {Bauer}, S. and {Biswas}, I. and {Blaizot}, J. and {Boudon}, D. and {Brau-Nogue}, S. and {Brinchmann}, J. and et al.},
        title = "{The MUSE second-generation VLT instrument}",
    booktitle = {Ground-based and Airborne Instrumentation for Astronomy III},
         year = 2010,
       editor = {{McLean}, Ian S. and {Ramsay}, Suzanne K. and {Takami}, Hideki},
       series = {Society of Photo-Optical Instrumentation Engineers (SPIE) Conference Series},
       volume = {7735},
        month = jul,
          eid = {773508},
        pages = {773508},
          doi = {10.1117/12.856027},
archivePrefix = {arXiv},
       eprint = {2211.16795},
 primaryClass = {astro-ph.IM},
       adsurl = {https://ui.adsabs.harvard.edu/abs/2010SPIE.7735E..08B}
}

@ARTICLE{ballester2000,
       author = {{Ballester}, P. and {Modigliani}, A. and {Boitquin}, O. and {Cristiani}, S. and {Hanuschik}, R. and {Kaufer}, A. and {Wolf}, S.},
        title = "{The UVES Data Reduction Pipeline}",
      journal = {The Messenger},
         year = 2000,
        month = sep,
       volume = {101},
        pages = {31-36},
       adsurl = {https://ui.adsabs.harvard.edu/abs/2000Msngr.101...31B}
}

@ARTICLE{beristain2001,
       author = {{Beristain}, Georgina and {Edwards}, Suzan and {Kwan}, John},
        title = "{Helium Emission from Classical T Tauri Stars: Dual Origin in Magnetospheric Infall and Hot Wind}",
      journal = {\apj},
         year = 2001,
        month = apr,
       volume = {551},
       number = {2},
        pages = {1037-1064},
          doi = {10.1086/320233},
       adsurl = {https://ui.adsabs.harvard.edu/abs/2001ApJ...551.1037B}
}

@ARTICLE{betti2022,
       author = {{Betti}, S.~K. and {Follette}, K.~B. and {Ward-Duong}, K. and {Aoyama}, Y. and {Marleau}, G. -D. and {Bary}, J. and {Robinson}, C. and {Janson}, M. and {Balmer}, W. and {Chauvin}, G. and {Palma-Bifani}, P.},
        title = "{Near-infrared Accretion Signatures from the Circumbinary Planetary-mass Companion Delorme 1 (AB)b}",
      journal = {\apjl},
         year = 2022,
        month = aug,
       volume = {935},
       number = {1},
          eid = {L18},
        pages = {L18},
          doi = {10.3847/2041-8213/ac85ef},
archivePrefix = {arXiv},
       eprint = {2208.05016},
 primaryClass = {astro-ph.EP},
       adsurl = {https://ui.adsabs.harvard.edu/abs/2022ApJ...935L..18B}
}

@ARTICLE{betti2023,
       author = {{Betti}, S.~K. and {Follette}, K.~B. and {Ward-Duong}, K. and {Peck}, A.~E. and {Aoyama}, Y. and {Bary}, J. and {Dacus}, B. and {Edwards}, S. and {Marleau}, G. -D. and {Mohamed}, K. and et al.},
        title = "{The Comprehensive Archive of Substellar and Planetary Accretion Rates}",
      journal = {\aj},
         year = 2023,
        month = dec,
       volume = {166},
       number = {6},
          eid = {262},
        pages = {262},
          doi = {10.3847/1538-3881/ad06b8},
archivePrefix = {arXiv},
       eprint = {2310.00072},
 primaryClass = {astro-ph.SR},
       adsurl = {https://ui.adsabs.harvard.edu/abs/2023AJ....166..262B}
}

@ARTICLE{bleach2002,
       author = {{Bleach}, James N. and {Wood}, J.~H. and {Smalley}, B. and {Catal{\'a}n}, M.~S.},
        title = "{Echelle observations of chromospheric activity in post-common-envelope binaries}",
      journal = {\mnras},
         year = 2002,
        month = sep,
       volume = {335},
       number = {3},
        pages = {593-609},
          doi = {10.1046/j.1365-8711.2002.05634.x},
       adsurl = {https://ui.adsabs.harvard.edu/abs/2002MNRAS.335..593B}
}

@ARTICLE{bochanski2007,
       author = {{Bochanski}, John J. and {West}, Andrew A. and {Hawley}, Suzanne L. and {Covey}, Kevin R.},
        title = "{Low-Mass Dwarf Template Spectra from the Sloan Digital Sky Survey}",
      journal = {\aj},
         year = 2007,
        month = feb,
       volume = {133},
       number = {2},
        pages = {531-544},
          doi = {10.1086/510240},
archivePrefix = {arXiv},
       eprint = {astro-ph/0610639},
 primaryClass = {astro-ph},
       adsurl = {https://ui.adsabs.harvard.edu/abs/2007AJ....133..531B}
}

@ARTICLE{brocklehurst1972,
       author = {{Brocklehurst}, M.},
        title = "{The line spectra of helium in gaseous nebulae}",
      journal = {\mnras},
         year = 1972,
        month = jan,
       volume = {157},
        pages = {211},
          doi = {10.1093/mnras/157.2.211},
       adsurl = {https://ui.adsabs.harvard.edu/abs/1972MNRAS.157..211B}
}

@ARTICLE{boucher2016,
       author = {{Boucher}, Anne and {Lafreni{\`e}re}, David and {Gagn{\'e}}, Jonathan and {Malo}, Lison and {Faherty}, Jacqueline K. and {Doyon}, Ren{\'e} and {Chen}, Christine H.},
        title = "{BANYAN. VIII. New Low-mass Stars and Brown Dwarfs with Candidate Circumstellar Disks}",
      journal = {\apj},
         year = 2016,
        month = nov,
       volume = {832},
       number = {1},
          eid = {50},
        pages = {50},
          doi = {10.3847/0004-637X/832/1/50},
archivePrefix = {arXiv},
       eprint = {1608.08259},
 primaryClass = {astro-ph.SR},
       adsurl = {https://ui.adsabs.harvard.edu/abs/2016ApJ...832...50B}
}

@INPROCEEDINGS{bouvier2007,
       author = {{Bouvier}, J. and {Alencar}, S.~H.~P. and {Harries}, T.~J. and {Johns-Krull}, C.~M. and {Romanova}, M.~M.},
        title = "{Magnetospheric Accretion in Classical T Tauri Stars}",
    booktitle = {Protostars and Planets V},
         year = 2007,
       editor = {{Reipurth}, Bo and {Jewitt}, David and {Keil}, Klaus},
        month = jan,
        pages = {479},
          doi = {10.48550/arXiv.astro-ph/0603498},
archivePrefix = {arXiv},
       eprint = {astro-ph/0603498},
 primaryClass = {astro-ph},
       adsurl = {https://ui.adsabs.harvard.edu/abs/2007prpl.conf..479B}
}

@ARTICLE{chinchilla2021,
       author = {{Chinchilla}, P. and {B{\'e}jar}, V.~J.~S. and {Lodieu}, N. and {Zapatero Osorio}, M.~R. and {Gauza}, B.},
        title = "{Strong H{\ensuremath{\alpha}} emission in the young planetary mass companion 2MASS J0249-0557 c}",
      journal = {\aap},
         year = 2021,
        month = jan,
       volume = {645},
          eid = {A17},
        pages = {A17},
          doi = {10.1051/0004-6361/202038731},
archivePrefix = {arXiv},
       eprint = {2011.10002},
 primaryClass = {astro-ph.EP},
       adsurl = {https://ui.adsabs.harvard.edu/abs/2021A&A...645A..17C}
}

@ARTICLE{claes2024,
       author = {{Claes}, R.~A.~B. and {Campbell-White}, J. and {Manara}, C.~F. and {Frasca}, A. and {Natta}, A. and {Alcal{\'a}}, J.~M. and {Armeni}, A. and {Fang}, M. and {Lovell}, J.~B. and {Stelzer}, B. and et al.},
        title = "{FitteR for Accretion ProPErties of T Tauri stars (FRAPPE): A new approach to use class III spectra to derive stellar and accretion properties}",
      journal = {\aap},
         year = 2024,
        month = oct,
       volume = {690},
          eid = {A122},
        pages = {A122},
          doi = {10.1051/0004-6361/202450885},
archivePrefix = {arXiv},
       eprint = {2407.11866},
 primaryClass = {astro-ph.SR},
       adsurl = {https://ui.adsabs.harvard.edu/abs/2024A&A...690A.122C}
}

@ARTICLE{currie2025,
       author = {{Currie}, Thayne and {Hashimoto}, Jun and {Aoyama}, Yuhiko and {Dong}, Ruobing and {Fukagawa}, Misato and {Muto}, Takayuki and {Dykes}, Erica and {El Morsy}, Mona and {Tamura}, Motohide},
        title = "{VLT/MUSE Detection of the AB Aurigae b Protoplanet with H$_{{\ensuremath{\alpha}}}$ Spectroscopy}",
      journal = {\apjl},
         year = 2025,
        month = sep,
       volume = {990},
       number = {2},
          eid = {L42},
        pages = {L42},
          doi = {10.3847/2041-8213/adf7a0},
archivePrefix = {arXiv},
       eprint = {2508.18351},
 primaryClass = {astro-ph.EP},
       adsurl = {https://ui.adsabs.harvard.edu/abs/2025ApJ...990L..42C}
}

@INPROCEEDINGS{dekker2000,
       author = {{Dekker}, Hans and {D'Odorico}, Sandro and {Kaufer}, Andreas and {Delabre}, Bernard and {Kotzlowski}, Heinz},
        title = "{Design, construction, and performance of UVES, the echelle spectrograph for the UT2 Kueyen Telescope at the ESO Paranal Observatory}",
    booktitle = {Optical and IR Telescope Instrumentation and Detectors},
         year = 2000,
       editor = {{Iye}, Masanori and {Moorwood}, Alan F.},
       series = {Society of Photo-Optical Instrumentation Engineers (SPIE) Conference Series},
       volume = {4008},
        month = aug,
        pages = {534-545},
          doi = {10.1117/12.395512},
       adsurl = {https://ui.adsabs.harvard.edu/abs/2000SPIE.4008..534D}
}

@ARTICLE{delorme2013,
       author = {{Delorme}, P. and {Gagn{\'e}}, J. and {Girard}, J.~H. and {Lagrange}, A.~M. and {Chauvin}, G. and {Naud}, M. -E. and {Lafreni{\`e}re}, D. and {Doyon}, R. and {Riedel}, A. and {Bonnefoy}, M. and {Malo}, L.},
        title = "{Direct-imaging discovery of a 12-14 Jupiter-mass object orbiting a young binary system of very low-mass stars}",
      journal = {\aap},
         year = 2013,
        month = may,
       volume = {553},
          eid = {L5},
        pages = {L5},
          doi = {10.1051/0004-6361/201321169},
archivePrefix = {arXiv},
       eprint = {1303.4525},
 primaryClass = {astro-ph.SR},
       adsurl = {https://ui.adsabs.harvard.edu/abs/2013A&A...553L...5D}
}

@ARTICLE{delzanna2020,
       author = {{Del Zanna}, G. and {Storey}, P.~J. and {Badnell}, N.~R. and {Andretta}, V.},
        title = "{Helium Line Emissivities in the Solar Corona}",
      journal = {\apj},
         year = 2020,
        month = jul,
       volume = {898},
       number = {1},
          eid = {72},
        pages = {72},
          doi = {10.3847/1538-4357/ab9d84},
archivePrefix = {arXiv},
       eprint = {2006.08971},
 primaryClass = {physics.atom-ph},
       adsurl = {https://ui.adsabs.harvard.edu/abs/2020ApJ...898...72D}
}

@ARTICLE{demars2023,
       author = {{Demars}, D. and {Bonnefoy}, M. and {Dougados}, C. and {Aoyama}, Y. and {Thanathibodee}, T. and {Marleau}, G.-D. and {Tremblin}, P. and {Delorme}, P. and {Palma-Bifani}, P. and {Petrus}, S. and {Bowler}, B.~P. and {Chauvin}, G. and {Lagrange}, A.-M.},
        title = "{Emission line variability of young 10-30 M$_{Jup}$ companions. I. The case of GQ Lup b and GSC 06214-00210 b}",
      journal = {\aap},
         year = 2023,
        month = aug,
       volume = {676},
          eid = {A123},
        pages = {A123},
          doi = {10.1051/0004-6361/202346221},
archivePrefix = {arXiv},
       eprint = {2305.09460},
 primaryClass = {astro-ph.EP},
       adsurl = {https://ui.adsabs.harvard.edu/abs/2023A&A...676A.123D}
}

@ARTICLE{demars2026,
       author = {{Demars}, Dorian and {Bonnefoy}, Micka{\"e}l and {Dougados}, Catherine and {Viswanath}, Gayathri and {Ringqvist}, Simon C. and {Janson}, Markus and {Aoyama}, Yuhiko and {Thanathibodee}, Thanawuth and {Marleau}, Gabriel-Dominique and {Manara}, Carlo F. and et al.},
        title = "{ExoplaNeT accRetion mOnitoring sPectroscopic surveY (ENTROPY): II. Time series of Balmer line profiles of Delorme 1(AB)b}",
      journal = {\aap},
         year = 2026,
        month = feb,
       volume = {706},
          eid = {A57},
        pages = {A57},
          doi = {10.1051/0004-6361/202554644},
archivePrefix = {arXiv},
       eprint = {2511.01979},
 primaryClass = {astro-ph.EP},
       adsurl = {https://ui.adsabs.harvard.edu/abs/2026A&A...706A..57D}
}

@ARTICLE{donati2008,
       author = {{Donati}, J.-F. and {Jardine}, M.~M. and {Gregory}, S.~G. and {Petit}, P. and {Paletou}, F. and {Bouvier}, J. and {Dougados}, C. and {M{\'e}nard}, F. and {Collier Cameron}, A. and {Harries}, T.~J. and {Hussain}, G.~A.~J. and {Unruh}, Y. and {Morin}, J. and {Marsden}, S.~C. and {Manset}, N. and {Auri{\`e}re}, M. and {Catala}, C. and {Alecian}, E.},
        title = "{Magnetospheric accretion on the T Tauri star BP Tauri}",
      journal = {\mnras},
         year = 2008,
        month = may,
       volume = {386},
       number = {3},
        pages = {1234-1251},
          doi = {10.1111/j.1365-2966.2008.13111.x},
archivePrefix = {arXiv},
       eprint = {0802.2052},
 primaryClass = {astro-ph},
       adsurl = {https://ui.adsabs.harvard.edu/abs/2008MNRAS.386.1234D}
}

@ARTICLE{donati2013,
       author = {{Donati}, J.-F. and {Gregory}, S.~G. and {Alencar}, S.~H.~P. and {Hussain}, G. and {Bouvier}, J. and {Jardine}, M.~M. and {M{\'e}nard}, F. and {Dougados}, C. and {Romanova}, M.~M. and {MaPP Collaboration}},
        title = "{Magnetospheric accretion on the fully convective classical T Tauri star DN Tau}",
      journal = {\mnras},
         year = 2013,
        month = nov,
       volume = {436},
       number = {1},
        pages = {881-897},
          doi = {10.1093/mnras/stt1622},
archivePrefix = {arXiv},
       eprint = {1308.5143},
 primaryClass = {astro-ph.SR},
       adsurl = {https://ui.adsabs.harvard.edu/abs/2013MNRAS.436..881D}
}

@ARTICLE{donati2012,
       author = {{Donati}, J.-F. and {Gregory}, S.~G. and {Alencar}, S.~H.~P. and {Hussain}, G. and {Bouvier}, J. and {Dougados}, C. and {Jardine}, M.~M. and {M{\'e}nard}, F. and {Romanova}, M.~M.},
        title = "{Magnetometry of the classical T Tauri star GQ Lup: non-stationary dynamos and spin evolution of young Suns}",
      journal = {\mnras},
         year = 2012,
        month = oct,
       volume = {425},
       number = {4},
        pages = {2948-2963},
          doi = {10.1111/j.1365-2966.2012.21482.x},
archivePrefix = {arXiv},
       eprint = {1206.1770},
 primaryClass = {astro-ph.SR},
       adsurl = {https://ui.adsabs.harvard.edu/abs/2012MNRAS.425.2948D}
}

@ARTICLE{edwards2003,
       author = {{Edwards}, Susan},
        title = "{Observations of the Star-Disk Interface: Search for Wind Origins}",
      journal = {Astrophysics and Space Science},
         year = 2003,
        month = oct,
       volume = {287},
        pages = {47-57},
          doi = {10.1023/B:ASTR.0000006199.17235.a3}
}

@ARTICLE{edwards2006,
       author = {{Edwards}, Suzan and {Fischer}, William and {Hillenbrand}, Lynne and {Kwan}, John},
        title = "{Probing T Tauri Accretion and Outflow with 1 Micron Spectroscopy}",
      journal = {\apj},
         year = 2006,
        month = jul,
       volume = {646},
       number = {1},
        pages = {319-341},
          doi = {10.1086/504832},
archivePrefix = {arXiv},
       eprint = {astro-ph/0604006},
 primaryClass = {astro-ph},
       adsurl = {https://ui.adsabs.harvard.edu/abs/2006ApJ...646..319E}
}

@ARTICLE{eriksson2020,
       author = {{Eriksson}, Simon C. and {Asensio Torres}, Rub{\'e}n and {Janson}, Markus and {Aoyama}, Yuhiko and {Marleau}, Gabriel-Dominique and {Bonnefoy}, Mickael and {Petrus}, Simon},
        title = "{Strong H{\ensuremath{\alpha}} emission and signs of accretion in a circumbinary planetary mass companion from MUSE}",
      journal = {\aap},
         year = 2020,
        month = jun,
       volume = {638},
          eid = {L6},
        pages = {L6},
          doi = {10.1051/0004-6361/202038131},
archivePrefix = {arXiv},
       eprint = {2005.11725},
 primaryClass = {astro-ph.EP},
       adsurl = {https://ui.adsabs.harvard.edu/abs/2020A&A...638L...6E}
}

@ARTICLE{erkal2022,
       author = {{Erkal}, J. and {Manara}, C.~F. and {Schneider}, P.~C. and {Vincenzi}, M. and {Nisini}, B. and {Coffey}, D. and {Alcal{\'a}}, J.~M. and {Fedele}, D. and {Antoniucci}, S.},
        title = "{The He I {\ensuremath{\lambda}}10830 {\r{A}} line as a probe of winds and accretion in young stars in Lupus and Upper Scorpius}",
      journal = {\aap},
         year = 2022,
        month = oct,
       volume = {666},
          eid = {A188},
        pages = {A188},
          doi = {10.1051/0004-6361/202244254},
archivePrefix = {arXiv},
       eprint = {2208.02940},
 primaryClass = {astro-ph.SR},
       adsurl = {https://ui.adsabs.harvard.edu/abs/2022A&A...666A.188E}
}

@ARTICLE{fiorellino2025,
       author = {{Fiorellino}, E. and {Alcal{\'a}}, J.~M. and {Manara}, C.~F. and {Pittman}, C.~V. and {{\'A}brah{\'a}m}, P. and {Venuti}, L. and {Cabrit}, S. and {Claes}, R. and {Fang}, M. and {K{\'o}sp{\'a}l}, {\'A}. and {Lodato}, G. and {Mauco}, K. and {Tychoniec}, {\L}.},
        title = "{PENELLOPE: VII. Revisiting empirical relations to measure accretion luminosity}",
      journal = {\aap},
         year = 2025,
        month = dec,
       volume = {704},
          eid = {A42},
        pages = {A42},
          doi = {10.1051/0004-6361/202556603},
archivePrefix = {arXiv},
       eprint = {2509.21078},
 primaryClass = {astro-ph.SR},
       adsurl = {https://ui.adsabs.harvard.edu/abs/2025A&A...704A..42F}
}

@ARTICLE{gahm2013,
       author = {{Gahm}, G.~F. and {Stempels}, H.~C. and {Walter}, F.~M. and {Petrov}, P.~P. and {Herczeg}, G.~J.},
        title = "{Face to phase with RU Lupi}",
      journal = {\aap},
         year = 2013,
        month = dec,
       volume = {560},
          eid = {A57},
        pages = {A57},
          doi = {10.1051/0004-6361/201322750},
archivePrefix = {arXiv},
       eprint = {1311.0127},
 primaryClass = {astro-ph.SR},
       adsurl = {https://ui.adsabs.harvard.edu/abs/2013A&A...560A..57G}
}

@ARTICLE{gaiadr3,
       author = {{Gaia Collaboration} and {Vallenari}, A. and {Brown}, A.~G.~A. and {Prusti}, T. and {de Bruijne}, J.~H.~J. and {Arenou}, F. and {Babusiaux}, C. and {Biermann}, M. and {Creevey}, O.~L. and {Ducourant}, C. and {Evans}, D.~W. and {Eyer}, L. and {Guerra}, R. and {Hutton}, A. and {Jordi}, C. and {Klioner}, S.~A. and {Lammers}, U.~L. and {Lindegren}, L. and {Luri}, X. and {Mignard}, F. and {Panem}, C. and {Pourbaix}, D. and {Randich}, S. and {Sartoretti}, P. and {Soubiran}, C. and {Tanga}, P. and {Walton}, N.~A. and {Bailer-Jones}, C.~A.~L. and {Bastian}, U. and {Drimmel}, R. and {Jansen}, F. and {Katz}, D. and {Lattanzi}, M.~G. and {van Leeuwen}, F. and {Bakker}, J. and {Cacciari}, C. and {Casta{\~n}eda}, J. and {De Angeli}, F. and {Fabricius}, C. and {Fouesneau}, M. and {Fr{\'e}mat}, Y. and {Galluccio}, L. and {Guerrier}, A. and {Heiter}, U. and {Masana}, E. and {Messineo}, R. and {Mowlavi}, N. and {Nicolas}, C. and {Nienartowicz}, K. and {Pailler}, F. and {Panuzzo}, P. and {Riclet}, F. and {Roux}, W. and {Seabroke}, G.~M. and {Sordo}, R. and {Th{\'e}venin}, F. and {Gracia-Abril}, G. and {Portell}, J. and {Teyssier}, D. and {Altmann}, M. and {Andrae}, R. and {Audard}, M. and {Bellas-Velidis}, I. and {Benson}, K. and {Berthier}, J. and {Blomme}, R. and {Burgess}, P.~W. and {Busonero}, D. and {Busso}, G. and {C{\'a}novas}, H. and {Carry}, B. and {Cellino}, A. and {Cheek}, N. and {Clementini}, G. and {Damerdji}, Y. and {Davidson}, M. and {de Teodoro}, P. and {Nu{\~n}ez Campos}, M. and {Delchambre}, L. and {Dell'Oro}, A. and {Esquej}, P. and {Fern{\'a}ndez-Hern{\'a}ndez}, J. and {Fraile}, E. and {Garabato}, D. and {Garc{\'\i}a-Lario}, P. and {Gosset}, E. and {Haigron}, R. and {Halbwachs}, J. -L. and {Hambly}, N.~C. and {Harrison}, D.~L. and {Hern{\'a}ndez}, J. and {Hestroffer}, D. and {Hodgkin}, S.~T. and {Holl}, B. and {Jan{\ss}en}, K. and {Jevardat de Fombelle}, G. and {Jordan}, S. and {Krone-Martins}, A. and {Lanzafame}, A.~C. and {L{\"o}ffler}, W. and {Marchal}, O. and {Marrese}, P.~M. and {Moitinho}, A. and {Muinonen}, K. and {Osborne}, P. and {Pancino}, E. and {Pauwels}, T. and {Recio-Blanco}, A. and {Reyl{\'e}}, C. and {Riello}, M. and {Rimoldini}, L. and {Roegiers}, T. and {Rybizki}, J. and {Sarro}, L.~M. and {Siopis}, C. and {Smith}, M. and {Sozzetti}, A. and {Utrilla}, E. and {van Leeuwen}, M. and {Abbas}, U. and {{\'A}brah{\'a}m}, P. and {Abreu Aramburu}, A. and {Aerts}, C. and {Aguado}, J.~J. and {Ajaj}, M. and {Aldea-Montero}, F. and {Altavilla}, G. and {{\'A}lvarez}, M.~A. and {Alves}, J. and {Anders}, F. and {Anderson}, R.~I. and {Anglada Varela}, E. and {Antoja}, T. and {Baines}, D. and {Baker}, S.~G. and {Balaguer-N{\'u}{\~n}ez}, L. and {Balbinot}, E. and {Balog}, Z. and {Barache}, C. and {Barbato}, D. and {Barros}, M. and {Barstow}, M.~A. and {Bartolom{\'e}}, S. and {Bassilana}, J. -L. and {Bauchet}, N. and {Becciani}, U. and {Bellazzini}, M. and {Berihuete}, A. and {Bernet}, M. and {Bertone}, S. and {Bianchi}, L. and {Binnenfeld}, A. and {Blanco-Cuaresma}, S. and {Blazere}, A. and {Boch}, T. and {Bombrun}, A. and {Bossini}, D. and {Bouquillon}, S. and {Bragaglia}, A. and {Bramante}, L. and {Breedt}, E. and {Bressan}, A. and {Brouillet}, N. and {Brugaletta}, E. and {Bucciarelli}, B. and {Burlacu}, A. and {Butkevich}, A.~G. and {Buzzi}, R. and {Caffau}, E. and {Cancelliere}, R. and {Cantat-Gaudin}, T. and {Carballo}, R. and {Carlucci}, T. and {Carnerero}, M.~I. and {Carrasco}, J.~M. and {Casamiquela}, L. and {Castellani}, M. and {Castro-Ginard}, A. and {Chaoul}, L. and {Charlot}, P. and {Chemin}, L. and {Chiaramida}, V. and {Chiavassa}, A. and {Chornay}, N. and {Comoretto}, G. and {Contursi}, G. and {Cooper}, W.~J. and {Cornez}, T. and {Cowell}, S. and {Crifo}, F. and {Cropper}, M. and {Crosta}, M. and {Crowley}, C. and {Dafonte}, C. and {Dapergolas}, A. and {David}, M. and {David}, P. and {de Laverny}, P. and {De Luise}, F. and {De March}, R.},
        title = "{Gaia Data Release 3. Summary of the content and survey properties}",
      journal = {\aap},
         year = 2023,
        month = jun,
       volume = {674},
          eid = {A1},
        pages = {A1},
          doi = {10.1051/0004-6361/202243940},
archivePrefix = {arXiv},
       eprint = {2208.00211},
 primaryClass = {astro-ph.GA},
       adsurl = {https://ui.adsabs.harvard.edu/abs/2023A&A...674A...1G}
}

@ARTICLE{haffert2019,
       author = {{Haffert}, S.~Y. and {Bohn}, A.~J. and {de Boer}, J. and {Snellen}, I.~A.~G. and {Brinchmann}, J. and {Girard}, J.~H. and {Keller}, C.~U. and {Bacon}, R.},
        title = "{Two accreting protoplanets around the young star PDS 70}",
      journal = {Nature Astronomy},
         year = 2019,
        month = jun,
       volume = {3},
        pages = {749-754},
          doi = {10.1038/s41550-019-0780-5},
archivePrefix = {arXiv},
       eprint = {1906.01486},
 primaryClass = {astro-ph.EP},
       adsurl = {https://ui.adsabs.harvard.edu/abs/2019NatAs...3..749H}
}

@ARTICLE{hartmann2016,
       author = {{Hartmann}, Lee and {Herczeg}, Gregory and {Calvet}, Nuria},
        title = "{Accretion onto Pre-Main-Sequence Stars}",
      journal = {\araa},
         year = 2016,
        month = sep,
       volume = {54},
        pages = {135-180},
          doi = {10.1146/annurev-astro-081915-023347},
       adsurl = {https://ui.adsabs.harvard.edu/abs/2016ARA&A..54..135H}
}

@ARTICLE{hashimoto2025,
       author = {{Hashimoto}, Jun and {Aoyama}, Yuhiko},
        title = "{Analyses of Multiple Balmer Emission Lines from Accreting Brown Dwarfs and Very Low Mass Stars}",
      journal = {\aj},
         year = 2025,
        month = feb,
       volume = {169},
       number = {2},
          eid = {93},
        pages = {93},
          doi = {10.3847/1538-3881/ad957e},
archivePrefix = {arXiv},
       eprint = {2411.12133},
 primaryClass = {astro-ph.EP},
       adsurl = {https://ui.adsabs.harvard.edu/abs/2025AJ....169...93H}
}

@ARTICLE{herczeg2023,
       author = {{Herczeg}, Gregory J. and {Chen}, Yuguang and {Donati}, Jean-Francois and {Dupree}, Andrea K. and {Walter}, Frederick M. and {Hillenbrand}, Lynne A. and {Johns-Krull}, Christopher M. and {Manara}, Carlo F. and {G{\"u}nther}, Hans Moritz and {Fang}, Min and {Schneider}, P. Christian and {Valenti}, Jeff A. and {Alencar}, Silvia H.~P. and {Venuti}, Laura and {Alcal{\'a}}, Juan Manuel and {Frasca}, Antonio and {Arulanantham}, Nicole and {Linsky}, Jeffrey L. and {Bouvier}, Jerome and {Brickhouse}, Nancy S. and {Calvet}, Nuria and {Espaillat}, Catherine C. and {Campbell-White}, Justyn and {Carpenter}, John M. and {Chang}, Seok-Jun and {Cruz}, Kelle L. and {Dahm}, S.~E. and {Eisl{\"o}ffel}, Jochen and {Edwards}, Suzan and {Fischer}, William J. and {Guo}, Zhen and {Henning}, Thomas and {Ji}, Tao and {Jose}, Jessy and {Kastner}, Joel H. and {Launhardt}, Ralf and {Principe}, David A. and {Robinson}, Connor E. and {Serna}, Javier and {Siwak}, Michal and {Sterzik}, Michael F. and {Takasao}, Shinsuke},
        title = "{Twenty-five Years of Accretion onto the Classical T Tauri Star TW Hya}",
      journal = {\apj},
         year = 2023,
        month = oct,
       volume = {956},
       number = {2},
          eid = {102},
        pages = {102},
          doi = {10.3847/1538-4357/acf468},
archivePrefix = {arXiv},
       eprint = {2308.14590},
 primaryClass = {astro-ph.SR},
       adsurl = {https://ui.adsabs.harvard.edu/abs/2023ApJ...956..102H}
}

@ARTICLE{hogbom1974,
       author = {{H{\"o}gbom}, J.~A.},
        title = "{Aperture Synthesis with a Non-Regular Distribution of Interferometer Baselines}",
      journal = {\aaps},
         year = 1974,
        month = jun,
       volume = {15},
        pages = {417},
       adsurl = {https://ui.adsabs.harvard.edu/abs/1974A&AS...15..417H}
}

@ARTICLE{jakobsen2022,
       author = {{Jakobsen}, P. and {Ferruit}, P. and {Alves de Oliveira}, C. and {Arribas}, S. and {Bagnasco}, G. and {Barho}, R. and {Beck}, T.~L. and {Birkmann}, S. and {B{\"o}ker}, T. and {Bunker}, A.~J. and et al.},
        title = "{The Near-Infrared Spectrograph (NIRSpec) on the James Webb Space Telescope. I. Overview of the instrument and its capabilities}",
      journal = {\aap},
         year = 2022,
        month = may,
       volume = {661},
          eid = {A80},
        pages = {A80},
          doi = {10.1051/0004-6361/202142663},
archivePrefix = {arXiv},
       eprint = {2202.03305},
 primaryClass = {astro-ph.IM},
       adsurl = {https://ui.adsabs.harvard.edu/abs/2022A&A...661A..80J}
}

@ARTICLE{jayawardhana2003,
       author = {{Jayawardhana}, Ray and {Mohanty}, Subhanjoy and {Basri}, Gibor},
        title = "{Evidence for a T Tauri Phase in Young Brown Dwarfs}",
      journal = {\apj},
         year = 2003,
        month = jul,
       volume = {592},
       number = {1},
        pages = {282-287},
          doi = {10.1086/375573},
archivePrefix = {arXiv},
       eprint = {astro-ph/0303565},
 primaryClass = {astro-ph},
       adsurl = {https://ui.adsabs.harvard.edu/abs/2003ApJ...592..282J}
}

@INPROCEEDINGS{kaufl2004,
       author = {{Kaeufl}, Hans-Ulrich and {Ballester}, Pascal and {Biereichel}, Peter and {Delabre}, Bernard and {Donaldson}, Rob and {Dorn}, Reinhold and {Fedrigo}, Enrico and {Finger}, Gert and {Fischer}, Gerhard and {Franza}, Francis and {Gojak}, Domingo and {Huster}, Gotthard and {Jung}, Yves and {Lizon}, Jean-Louis and {Mehrgan}, Leander and {Meyer}, Manfred and {Moorwood}, Alan and {Pirard}, Jean-Francois and {Paufique}, Jerome and {Pozna}, Esther and {Siebenmorgen}, Ralf and {Silber}, Armin and {Stegmeier}, Joerg and {Wegerer}, Stefan},
        title = "{CRIRES: a high-resolution infrared spectrograph for ESO's VLT}",
    booktitle = {Ground-based Instrumentation for Astronomy},
         year = 2004,
       editor = {{Moorwood}, Alan F.~M. and {Iye}, Masanori},
       series = {Society of Photo-Optical Instrumentation Engineers (SPIE) Conference Series},
       volume = {5492},
        month = sep,
        pages = {1218-1227},
          doi = {10.1117/12.551480},
       adsurl = {https://ui.adsabs.harvard.edu/abs/2004SPIE.5492.1218K}
}

@ARTICLE{kraus2014,
       author = {{Kraus}, Adam L. and {Shkolnik}, Evgenya L. and {Allers}, Katelyn N. and {Liu}, Michael C.},
        title = "{A Stellar Census of the Tucana-Horologium Moving Group}",
      journal = {\aj},
         year = 2014,
        month = jun,
       volume = {147},
       number = {6},
          eid = {146},
        pages = {146},
          doi = {10.1088/0004-6256/147/6/146},
archivePrefix = {arXiv},
       eprint = {1403.0050},
 primaryClass = {astro-ph.SR},
       adsurl = {https://ui.adsabs.harvard.edu/abs/2014AJ....147..146K}
}

@ARTICLE{kwan2007,
       author = {{Kwan}, John and {Edwards}, Suzan and {Fischer}, William},
        title = "{Modeling T Tauri Winds from He I {\ensuremath{\lambda}}10830 Profiles}",
      journal = {\apj},
         year = 2007,
        month = mar,
       volume = {657},
       number = {2},
        pages = {897-915},
          doi = {10.1086/511057},
archivePrefix = {arXiv},
       eprint = {astro-ph/0611585},
 primaryClass = {astro-ph},
       adsurl = {https://ui.adsabs.harvard.edu/abs/2007ApJ...657..897K}
}

@ARTICLE{kwan2011,
       author = {{Kwan}, John and {Fischer}, William},
        title = "{Origins of the H, He I and Ca II line emission in classical T Tauri stars}",
      journal = {\mnras},
         year = 2011,
        month = mar,
       volume = {411},
       number = {4},
        pages = {2383-2425},
          doi = {10.1111/j.1365-2966.2010.17863.x},
archivePrefix = {arXiv},
       eprint = {1010.3265},
 primaryClass = {astro-ph.SR},
       adsurl = {https://ui.adsabs.harvard.edu/abs/2011MNRAS.411.2383K}
}

@ARTICLE{kwan2024,
       author = {{Kwan}, John},
        title = "{Continuum and line emission from accretion shocks at T tauri stars - I. Correlations with shock parameters}",
      journal = {\mnras},
         year = 2024,
        month = jul,
       volume = {531},
       number = {3},
        pages = {3744-3769},
          doi = {10.1093/mnras/stae1385},
archivePrefix = {arXiv},
       eprint = {2406.15618},
 primaryClass = {astro-ph.SR},
       adsurl = {https://ui.adsabs.harvard.edu/abs/2024MNRAS.531.3744K}
}

@ARTICLE{maio2020,
       author = {{Di Maio}, C. and {Argiroffi}, C. and {Micela}, G. and {Benatti}, S. and {Lanza}, A.~F. and {Scandariato}, G. and {Maldonado}, J. and {Maggio}, A. and {Affer}, L. and {Claudi}, R.},
        title = "{The GAPS programme at TNG. XXVI. Magnetic activity in M stars: spectroscopic monitoring of AD Leonis}",
      journal = {\aap},
         year = 2020,
        month = oct,
       volume = {642},
          eid = {A53},
        pages = {A53},
          doi = {10.1051/0004-6361/202038011},
archivePrefix = {arXiv},
       eprint = {2008.04242},
 primaryClass = {astro-ph.SR},
       adsurl = {https://ui.adsabs.harvard.edu/abs/2020A&A...642A..53D}
}

@ARTICLE{malin2025,
       author = {{M{\^a}lin}, Mathilde and {Ward-Duong}, Kimberly and {Grant}, Sierra L. and {Arulanantham}, Nicole and {Tabone}, Beno{\^\i}t and {Pueyo}, Laurent and {Perrin}, Marshall and {Balmer}, William O. and {Betti}, Sarah and {Chen}, Christine H. and {Debes}, John H. and {Girard}, Julien H. and {Hoch}, Kielan K.~W. and {Kammerer}, Jens and {Lu}, Cicero and {Rebollido}, Isabel and {Rickman}, Emily and {Robinson}, Connor and {Worthen}, Kadin and {van der Marel}, Roeland P. and {Lewis}, Nikole K. and {Seager}, Sara and {Valenti}, Jeff A. and {Soummer}, Remi},
        title = "{JWST-TST High Contrast: Medium-resolution spectroscopy reveals a carbon-rich circumplanetary disk around the young accreting exoplanet Delorme 1 AB b}",
      journal = {\aap},
         year = 2025,
        month = dec,
       volume = {704},
          eid = {A181},
        pages = {A181},
          doi = {10.1051/0004-6361/202556792},
archivePrefix = {arXiv},
       eprint = {2510.07253},
 primaryClass = {astro-ph.EP},
       adsurl = {https://ui.adsabs.harvard.edu/abs/2025A&A...704A.181M}
}

@ARTICLE{manara2013,
       author = {{Manara}, C.~F. and {Testi}, L. and {Rigliaco}, E. and {Alcal{\'a}}, J.~M. and {Natta}, A. and {Stelzer}, B. and {Biazzo}, K. and {Covino}, E. and {Covino}, S. and {Cupani}, G. and {D'Elia}, V. and {Randich}, S.},
        title = "{X-shooter spectroscopy of young stellar objects. II. Impact of chromospheric emission on accretion rate estimates}",
      journal = {\aap},
         year = 2013,
        month = mar,
       volume = {551},
          eid = {A107},
        pages = {A107},
          doi = {10.1051/0004-6361/201220921},
archivePrefix = {arXiv},
       eprint = {1301.3058},
 primaryClass = {astro-ph.GA},
       adsurl = {https://ui.adsabs.harvard.edu/abs/2013A&A...551A.107M}
}

@INPROCEEDINGS{manara2023,
       author = {{Manara}, C.~F. and {Ansdell}, M. and {Rosotti}, G.~P. and {Hughes}, A.~M. and {Armitage}, P.~J. and {Lodato}, G. and {Williams}, J.~P.},
        title = "{Demographics of Young Stars and their Protoplanetary Disks: Lessons Learned on Disk Evolution and its Connection to Planet Formation}",
    booktitle = {Protostars and Planets VII},
         year = 2023,
       editor = {{Inutsuka}, S. and {Aikawa}, Y. and {Muto}, T. and {Tomida}, K. and {Tamura}, M.},
       series = {Astronomical Society of the Pacific Conference Series},
       volume = {534},
        month = jul,
        pages = {539},
          doi = {10.48550/arXiv.2203.09930},
archivePrefix = {arXiv},
       eprint = {2203.09930},
 primaryClass = {astro-ph.SR},
       adsurl = {https://ui.adsabs.harvard.edu/abs/2023ASPC..534..539M}
}

@ARTICLE{mcginnis2020,
       author = {{McGinnis}, Pauline and {Bouvier}, J{\'e}r{\^o}me and {Gallet}, Florian},
        title = "{The magnetic obliquity of accreting T Tauri stars}",
      journal = {\mnras},
         year = 2020,
        month = sep,
       volume = {497},
       number = {2},
        pages = {2142-2162},
          doi = {10.1093/mnras/staa2041},
archivePrefix = {arXiv},
       eprint = {2007.06642},
 primaryClass = {astro-ph.SR},
       adsurl = {https://ui.adsabs.harvard.edu/abs/2020MNRAS.497.2142M}
}

@ARTICLE{mohanty2003,
       author = {{Mohanty}, Subhanjoy and {Jayawardhana}, Ray and {Barrado y Navascu{\'e}s}, David},
        title = "{Magellan Echelle Spectroscopy of TW Hydrae Brown Dwarfs}",
      journal = {\apjl},
         year = 2003,
        month = aug,
       volume = {593},
       number = {2},
        pages = {L109-L112},
          doi = {10.1086/378315},
archivePrefix = {arXiv},
       eprint = {astro-ph/0306514},
 primaryClass = {astro-ph},
       adsurl = {https://ui.adsabs.harvard.edu/abs/2003ApJ...593L.109M}
}

@ARTICLE{mohanty2003b,
       author = {{Mohanty}, Subhanjoy and {Basri}, Gibor},
        title = "{Rotation and Activity in Mid-M to L Field Dwarfs}",
      journal = {\apj},
         year = 2003,
        month = jan,
       volume = {583},
       number = {1},
        pages = {451-472},
          doi = {10.1086/345097},
archivePrefix = {arXiv},
       eprint = {astro-ph/0201455},
 primaryClass = {astro-ph},
       adsurl = {https://ui.adsabs.harvard.edu/abs/2003ApJ...583..451M}
}

@ARTICLE{mohanty2005,
       author = {{Mohanty}, Subhanjoy and {Jayawardhana}, Ray and {Basri}, Gibor},
        title = "{The T Tauri Phase Down to Nearly Planetary Masses: Echelle Spectra of 82 Very Low Mass Stars and Brown Dwarfs}",
      journal = {\apj},
         year = 2005,
        month = jun,
       volume = {626},
       number = {1},
        pages = {498-522},
          doi = {10.1086/429794},
archivePrefix = {arXiv},
       eprint = {astro-ph/0502155},
 primaryClass = {astro-ph},
       adsurl = {https://ui.adsabs.harvard.edu/abs/2005ApJ...626..498M}
}

@INPROCEEDINGS{marconi2024,
       author = {{Marconi}, A. and {Abreu}, M. and {Adibekyan}, V. and {Alberti}, V. and {Albrecht}, S. and {Alcaniz}, J. and {Aliverti}, M. and {Allende Prieto}, C. and {Alvarado-Gomez}, J.~D. and {Alves}, C.~S. and {Amado}, P.~J. and {Amate}, M. and {Andersen}, M.~I. and {Antoniucci}, S. and {Artigau}, E. and {Bailet}, C. and {Baker}, C. and {Baldini}, V. and {Balestra}, A. and {Barnes}, S.~A. and {Baron}, F. and {Barros}, S.~C.~C. and {Bauer}, S.~M. and {Beaulieu}, M. and {Bellido-Tirado}, O. and {Benneke}, B. and {Bensby}, T. and {Bergin}, E.~A. and {Berio}, P. and {Biazzo}, K. and {Bigot}, L. and {Bik}, A. and {Birkby}, J.~L. and {Blind}, N. and {Boebion}, O. and {Boisse}, I. and {Bolmont}, E. and {Bolton}, J.~S. and {Bonaglia}, M. and {Bonfils}, X. and {Bonhomme}, L. and {Borsa}, F. and {Bouret}, J.-C. and {Brandeker}, A. and {Brandner}, W. and {Broeg}, C.~H. and {Brogi}, M. and {Brousseau}, D. and {Brucalassi}, A. and {Brynnel}, J. and {Buchhave}, L.~A. and {Buscher}, D.~F. and {Cabona}, L. and {Cabral}, A. and {Calderone}, G. and {Calvo-Ortega}, R. and {Cantalloube}, F. and {Canto Martins}, B.~L. and {Carbonaro}, L. and {Caujolle}, Y. and {Chauvin}, G. and {Chazelas}, B. and {Cheffot}, A.-L. and {Cheng}, Y.~S. and {Chiavassa}, A. and {Christensen}, L. and {Cirami}, R. and {Cirasuolo}, M. and {Cook}, N.~J. and {Cooke}, R.~J. and {Coretti}, I. and {Covino}, S. and {Cowan}, N. and {Cresci}, G. and {Cristiani}, S. and {Cunha Parro}, V. and {Cupani}, G. and {D'Odorico}, V. and {Dadi}, K. and {de Castro Le{\~a}o}, I. and {De Cia}, A. and {De Medeiros}, J.~R. and {Debras}, F. and {Debus}, M. and {Delorme}, A. and {Demangeon}, O. and {Derie}, F. and {Dessauges-Zavadsky}, M. and {Di Marcantonio}, P. and {Di Stefano}, S. and {Dionies}, F. and {Domiciano de Souza}, A. and {Doyon}, R. and {Dunn}, J. and {Egner}, S. and {Ehrenreich}, D. and {Faria}, J.~P. and {Ferruzzi}, D. and {Feruglio}, C. and {Fisher}, M. and {Fontana}, A. and {Frank}, B.~S. and {Fuesslein}, C. and {Fumagalli}, M. and {Fusco}, T. and {Fynbo}, J. and {Gabella}, O. and {Gaessler}, W. and {Gallo}, E. and {Gao}, X. and {Genolet}, L. and {Genoni}, M. and {Giacobbe}, P. and {Giro}, E. and {Gon{\c{c}}alves}, R.~S. and {Gonzalez}, O.~A. and {Gonz{\'a}lez-Hern{\'a}ndez}, J.~I. and {Gouvret}, C. and {Gracia T{\'e}mich}, F. and {Haehnelt}, M.~G. and {Haniff}, C. and {Hatzes}, A. and {Helled}, R. and {Hoeijmakers}, H.~J. and {Hughes}, I. and {Huke}, P. and {Ivanisenko}, Y. and {J{\"a}rvinen}, A.~S. and {J{\"a}rvinen}, S.~P. and {Kaminski}, A. and {Kern}, J. and {Knoche}, J. and {Kordt}, A. and {Korhonen}, H. and {Korn}, A.~J. and {Kouach}, D. and {Kowzan}, G. and {Kreidberg}, L. and {Landoni}, M. and {Lanotte}, A.~A. and {Lavail}, A. and {Lavie}, B. and {Lee}, D. and {Lehmitz}, M. and {Li}, J. and {Li}, W. and {Liske}, J. and {Lovis}, C. and {Lucatello}, S. and {Lunney}, D. and {MacIntosh}, M.~J. and {Madhusudhan}, N. and {Magrini}, L. and {Maiolino}, R. and {Maldonado}, J. and {Malo}, L. and {Man}, A.~W.~S. and {Marquart}, T. and {Marques}, C.~M.~J. and {Marques}, E.~L. and {Martinez}, P. and {Martins}, A. and {Martins}, C.~J.~A.~P. and {Martins}, J.~H.~C. and {Maslowski}, P. and {Mason}, C. and {Mason}, E. and {McCracken}, R.~A. and {Melo e Sousa}, M.~A.~F. and {Mergo}, P. and {Micela}, G. and {Milakovi{\'c}}, D. and {Molli{\`e}re}, P. and {Monteiro}, M.~A. and {Montgomery}, D. and {Mordasini}, C. and {Morin}, J. and {Mucciarelli}, A. and {Murphy}, M.~T. and {N'Diaye}, M. and {Nardetto}, N. and {Neichel}, B. and {Neri}, N. and {Niedzielski}, A.~T. and {Niemczura}, E. and {Nisini}, B. and {Nortmann}, L. and {Noterdaeme}, P. and {Nunes}, N.~J. and {Oggioni}, L. and {Olchewsky}, F. and {Oliva}, E. and {{\"O}nel}, H. and {Origlia}, L. and {{\"O}stlin}, G. and {Ouellette}, N.~N.-Q. and {Pall{\'e}}, E. and {Papaderos}, P. and {Pariani}, G. and {Pasquini}, L.},
        title = "{ANDES, the high resolution spectrograph for the ELT: science goals, project overview, and future developments}",
    booktitle = {Ground-based and Airborne Instrumentation for Astronomy X},
         year = 2024,
       editor = {{Bryant}, Julia J. and {Motohara}, Kentaro and {Vernet}, Jo{\"e}l. R.~D.},
       series = {Society of Photo-Optical Instrumentation Engineers (SPIE) Conference Series},
       volume = {13096},
        month = jul,
          eid = {1309613},
        pages = {1309613},
          doi = {10.1117/12.3017966},
archivePrefix = {arXiv},
       eprint = {2407.14601},
 primaryClass = {astro-ph.IM},
       adsurl = {https://ui.adsabs.harvard.edu/abs/2024SPIE13096E..13M}
}

@ARTICLE{muzerolle1998,
       author = {{Muzerolle}, James and {Hartmann}, Lee and {Calvet}, Nuria},
        title = "{Emission-Line Diagnostics of T Tauri Magnetospheric Accretion. I. Line Profile Observations}",
      journal = {\aj},
         year = 1998,
        month = jul,
       volume = {116},
       number = {1},
        pages = {455-468},
          doi = {10.1086/300428},
       adsurl = {https://ui.adsabs.harvard.edu/abs/1998AJ....116..455M}
}

@ARTICLE{muzerolle2005,
       author = {{Muzerolle}, James and {Luhman}, Kevin L. and {Brice{\~n}o}, C{\'e}sar and {Hartmann}, Lee and {Calvet}, Nuria},
        title = "{Measuring Accretion in Young Substellar Objects: Approaching the Planetary Mass Regime}",
      journal = {\apj},
         year = 2005,
        month = jun,
       volume = {625},
       number = {2},
        pages = {906-912},
          doi = {10.1086/429483},
archivePrefix = {arXiv},
       eprint = {astro-ph/0502023},
 primaryClass = {astro-ph},
       adsurl = {https://ui.adsabs.harvard.edu/abs/2005ApJ...625..906M}
}

@ARTICLE{notsu2024,
       author = {{Notsu}, Yuta and {Kowalski}, Adam F. and {Maehara}, Hiroyuki and {Namekata}, Kosuke and {Hamaguchi}, Kenji and {Enoto}, Teruaki and {Tristan}, Isaiah I. and {Hawley}, Suzanne L. and {Davenport}, James R.~A. and {Honda}, Satoshi and {Ikuta}, Kai and {Inoue}, Shun and {Namizaki}, Keiichi and {Nogami}, Daisaku and {Shibata}, Kazunari},
        title = "{Apache Point Observatory (APO)/SMARTS Flare Star Campaign Observations. I. Blue Wing Asymmetries in Chromospheric Lines during Mid-M-Dwarf Flares from Simultaneous Spectroscopic and Photometric Observation Data}",
      journal = {\apj},
         year = 2024,
        month = feb,
       volume = {961},
       number = {2},
          eid = {189},
        pages = {189},
          doi = {10.3847/1538-4357/ad062f},
archivePrefix = {arXiv},
       eprint = {2310.02450},
 primaryClass = {astro-ph.SR},
       adsurl = {https://ui.adsabs.harvard.edu/abs/2024ApJ...961..189N}
}

@ARTICLE{nowacki2023,
       author = {{Nowacki}, H. and {Alecian}, E. and {Perraut}, K. and {Zaire}, B. and {Folsom}, C.~P. and {Pouilly}, K. and {Bouvier}, J. and {Manick}, R. and {Pantolmos}, G. and {Sousa}, A.~P. and et al.},
        title = "{Star-disk interactions in the strongly accreting T Tauri star S CrA N}",
      journal = {\aap},
         year = 2023,
        month = oct,
       volume = {678},
          eid = {A86},
        pages = {A86},
          doi = {10.1051/0004-6361/202347145},
archivePrefix = {arXiv},
       eprint = {2308.03442},
 primaryClass = {astro-ph.SR},
       adsurl = {https://ui.adsabs.harvard.edu/abs/2023A&A...678A..86N}
}

@ARTICLE{petrov2011,
       author = {{Petrov}, P.~P. and {Gahm}, G.~F. and {Stempels}, H.~C. and {Walter}, F.~M. and {Artemenko}, S.~A.},
        title = "{Accretion-powered chromospheres in classical T Tauri stars}",
      journal = {\aap},
         year = 2011,
        month = nov,
       volume = {535},
          eid = {A6},
        pages = {A6},
          doi = {10.1051/0004-6361/201116721},
archivePrefix = {arXiv},
       eprint = {1109.1266},
 primaryClass = {astro-ph.SR},
       adsurl = {https://ui.adsabs.harvard.edu/abs/2011A&A...535A...6P}
}

@ARTICLE{pittman2025,
       author = {{Pittman}, Caeley V. and {Espaillat}, Catherine C. and {Zhu}, Zhaohuan and {Thanathibodee}, Thanawuth and {Robinson}, Connor E. and {Calvet}, Nuria and {K{\'o}sp{\'a}l}, {\'A}gnes},
        title = "{The ODYSSEUS Survey. Using Accretion and Stellar Rotation to Reveal the Star─Disk Connection in T Tauri Stars}",
      journal = {\apj},
         year = 2025,
        month = nov,
       volume = {993},
       number = {2},
          eid = {181},
        pages = {181},
          doi = {10.3847/1538-4357/ae03b3},
archivePrefix = {arXiv},
       eprint = {2509.03767},
 primaryClass = {astro-ph.SR},
       adsurl = {https://ui.adsabs.harvard.edu/abs/2025ApJ...993..181P}
}

@ARTICLE{rachford1997,
       author = {{Rachford}, Brian L.},
        title = "{Chromospheric Activity in Dwarf and Evolved Late A- and Early F-Type Stars}",
      journal = {\apj},
         year = 1997,
        month = sep,
       volume = {486},
       number = {2},
        pages = {994-999},
          doi = {10.1086/304570},
       adsurl = {https://ui.adsabs.harvard.edu/abs/1997ApJ...486..994R}
}

@ARTICLE{richey2023,
       author = {{Richey-Yowell}, Tyler and {Shkolnik}, Evgenya L. and {Schneider}, Adam C. and {Peacock}, Sarah and {Huseby}, Lori A. and {Jackman}, James A.~G. and {Barman}, Travis and {Osby}, Ella and {Meadows}, Victoria S.},
        title = "{HAZMAT. IX. An Analysis of the UV and X-Ray Evolution of Low-mass Stars in the Era of Gaia}",
      journal = {\apj},
         year = 2023,
        month = jul,
       volume = {951},
       number = {1},
          eid = {44},
        pages = {44},
          doi = {10.3847/1538-4357/acd2dc},
archivePrefix = {arXiv},
       eprint = {2305.06561},
 primaryClass = {astro-ph.SR},
       adsurl = {https://ui.adsabs.harvard.edu/abs/2023ApJ...951...44R}
}

@ARTICLE{ringqvist2023,
       author = {{Ringqvist}, Simon C. and {Viswanath}, Gayathri and {Aoyama}, Yuhiko and {Janson}, Markus and {Marleau}, Gabriel-Dominique and {Brandeker}, Alexis},
        title = "{Resolved near-UV hydrogen emission lines at 40-Myr super-Jovian protoplanet Delorme 1 (AB)b. Indications of magnetospheric accretion}",
      journal = {\aap},
         year = 2023,
        month = jan,
       volume = {669},
          eid = {L12},
        pages = {L12},
          doi = {10.1051/0004-6361/202245424},
archivePrefix = {arXiv},
       eprint = {2212.03207},
 primaryClass = {astro-ph.EP},
       adsurl = {https://ui.adsabs.harvard.edu/abs/2023A&A...669L..12R}
}

@ARTICLE{saar1997,
       author = {{Saar}, S.~H. and {Huovelin}, J. and {Osten}, R.~A. and {Shcherbakov}, A.~G.},
        title = "{HeI D3 absorption and its relation to rotation and activity in G and K dwarfs.}",
      journal = {\aap},
         year = 1997,
        month = oct,
       volume = {326},
        pages = {741-750},
       adsurl = {https://ui.adsabs.harvard.edu/abs/1997A&A...326..741S}
}

@ARTICLE{santamaria2018,
       author = {{Santamar{\'\i}a-Miranda}, Alejandro and {C{\'a}ceres}, Claudio and {Schreiber}, Matthias R. and {Hardy}, Adam and {Bayo}, Amelia and {Parsons}, Steven G. and {Gromadzki}, Mariusz and {Aguayo Villegas}, Aurora Bel{\'e}n},
        title = "{Accretion signatures in the X-shooter spectrum of the substellar companion to SR12}",
      journal = {\mnras},
         year = 2018,
        month = apr,
       volume = {475},
       number = {3},
        pages = {2994-3003},
          doi = {10.1093/mnras/stx3325},
archivePrefix = {arXiv},
       eprint = {1712.09297},
 primaryClass = {astro-ph.SR},
       adsurl = {https://ui.adsabs.harvard.edu/abs/2018MNRAS.475.2994S}
}

@ARTICLE{santamaria2019,
       author = {{Santamar{\'\i}a-Miranda}, Alejandro and {C{\'a}ceres}, Claudio and {Schreiber}, Matthias R. and {Hardy}, Adam and {Bayo}, Amelia and {Parsons}, Steven G. and {Gromadzki}, Mariusz and {Aguayo Villegas}, Aurora Bel{\'e}n},
        title = "{Erratum: Accretion signatures in the X-shooter spectrum of the substellar companion to SR12}",
      journal = {\mnras},
         year = 2019,
        month = oct,
       volume = {488},
       number = {4},
        pages = {5852-5853},
          doi = {10.1093/mnras/stz2173},
       adsurl = {https://ui.adsabs.harvard.edu/abs/2019MNRAS.488.5852S}
}

@ARTICLE{schneeberger1978,
       author = {{Schneeberger}, T.~J. and {Linsky}, J.~L. and {Worden}, S.~P.},
        title = "{The helium triplet-to-singlet ratio in T Tauri stars.}",
      journal = {\aap},
         year = 1978,
        month = jan,
       volume = {62},
       number = {3},
        pages = {447},
       adsurl = {https://ui.adsabs.harvard.edu/abs/1978A&A....62..447S}
}

@ARTICLE{aguilar2012,
       author = {{Sicilia-Aguilar}, A. and {K{\'o}sp{\'a}l}, {\'A}. and {Setiawan}, J. and {{\'A}brah{\'a}m}, P. and {Dullemond}, C. and {Eiroa}, C. and {Goto}, M. and {Henning}, Th. and {Juh{\'a}sz}, A.},
        title = "{Optical spectroscopy of EX Lupi during quiescence and outburst. Infall, wind, and dynamics in the accretion flow}",
      journal = {\aap},
         year = 2012,
        month = aug,
       volume = {544},
          eid = {A93},
        pages = {A93},
          doi = {10.1051/0004-6361/201118555},
archivePrefix = {arXiv},
       eprint = {1206.3081},
 primaryClass = {astro-ph.SR},
       adsurl = {https://ui.adsabs.harvard.edu/abs/2012A&A...544A..93S}
}

@ARTICLE{aguilar2023,
       author = {{Sicilia-Aguilar}, A. and {Campbell-White}, J. and {Roccatagliata}, V. and {Desira}, J. and {Gregory}, S.~G. and {Scholz}, A. and {Fang}, M. and {Cruz-Saenz de Miera}, F. and {K{\'o}sp{\'a}l}, {\'A}. and {Matsumura}, S. and {{\'A}brah{\'a}m}, P.},
        title = "{Stable accretion in young stars: the cases of EX Lupi and TW Hya}",
      journal = {\mnras},
         year = 2023,
        month = dec,
       volume = {526},
       number = {4},
        pages = {4885-4907},
          doi = {10.1093/mnras/stad3029},
archivePrefix = {arXiv},
       eprint = {2310.02681},
 primaryClass = {astro-ph.SR},
       adsurl = {https://ui.adsabs.harvard.edu/abs/2023MNRAS.526.4885S}
}

@ARTICLE{smith1996,
       author = {{Smits}, Derck P.},
        title = "{Theoretical HeI line intensities in low-density plasmas}",
      journal = {\mnras},
         year = 1996,
        month = feb,
       volume = {278},
       number = {3},
        pages = {683-687},
          doi = {10.1093/mnras/278.3.683},
       adsurl = {https://ui.adsabs.harvard.edu/abs/1996MNRAS.278..683S}
}

@ARTICLE{stamatellos2015,
       author = {{Stamatellos}, Dimitris and {Herczeg}, Gregory J.},
        title = "{The properties of discs around planets and brown dwarfs as evidence for disc fragmentation}",
      journal = {\mnras},
         year = 2015,
        month = jun,
       volume = {449},
       number = {4},
        pages = {3432-3440},
          doi = {10.1093/mnras/stv526},
archivePrefix = {arXiv},
       eprint = {1503.05209},
 primaryClass = {astro-ph.GA},
       adsurl = {https://ui.adsabs.harvard.edu/abs/2015MNRAS.449.3432S}
}

@ARTICLE{stelzer2013a,
       author = {{Stelzer}, B. and {Marino}, A. and {Micela}, G. and {L{\'o}pez-Santiago}, J. and {Liefke}, C.},
        title = "{The UV and X-ray activity of the M dwarfs within 10 pc of the Sun}",
      journal = {\mnras},
         year = 2013,
        month = may,
       volume = {431},
       number = {3},
        pages = {2063-2079},
          doi = {10.1093/mnras/stt225},
archivePrefix = {arXiv},
       eprint = {1302.1061},
 primaryClass = {astro-ph.SR},
       adsurl = {https://ui.adsabs.harvard.edu/abs/2013MNRAS.431.2063S}
}

@ARTICLE{stelzer2013b,
       author = {{Stelzer}, B. and {Frasca}, A. and {Alcal{\'a}}, J.~M. and {Manara}, C.~F. and {Biazzo}, K. and {Covino}, E. and {Rigliaco}, E. and {Testi}, L. and {Covino}, S. and {D'Elia}, V.},
        title = "{X-shooter spectroscopy of young stellar objects. III. Photospheric and chromospheric properties of Class III objects}",
      journal = {\aap},
         year = 2013,
        month = oct,
       volume = {558},
          eid = {A141},
        pages = {A141},
          doi = {10.1051/0004-6361/201321979},
archivePrefix = {arXiv},
       eprint = {1308.5563},
 primaryClass = {astro-ph.SR},
       adsurl = {https://ui.adsabs.harvard.edu/abs/2013A&A...558A.141S}
}

@ARTICLE{szulagyi2020,
       author = {{Szul{\'a}gyi}, Judit and {Ercolano}, Barbara},
        title = "{Hydrogen Recombination Line Luminosities and Variability from Forming Planets}",
      journal = {\apj},
         year = 2020,
        month = oct,
       volume = {902},
       number = {2},
          eid = {126},
        pages = {126},
          doi = {10.3847/1538-4357/abb5a2},
archivePrefix = {arXiv},
       eprint = {2002.09918},
 primaryClass = {astro-ph.EP},
       adsurl = {https://ui.adsabs.harvard.edu/abs/2020ApJ...902..126S}
}

@ARTICLE{temmink2024,
       author = {{Temmink}, Milou and {van Dishoeck}, Ewine F. and {Grant}, Sierra L. and {Tabone}, Beno{\^\i}t and {Gasman}, Danny and {Christiaens}, Valentin and {Samland}, Matthias and {Argyriou}, Ioannis and {Perotti}, Giulia and {G{\"u}del}, Manuel and {Henning}, Thomas and {Lagage}, Pierre-Olivier and {Abergel}, Alain and {Absil}, Olivier and {Barrado}, David and {Caratti o Garatti}, Alessio and {Glauser}, Adrian M. and {Kamp}, Inga and {Lahuis}, Fred and {Olofsson}, G{\"o}ran and {Ray}, Tom P. and {Scheithauer}, Silvia and {Vandenbussche}, Bart and {Waters}, L.~B.~F.~M. and {Arabhavi}, Aditya M. and {Jang}, Hyerin and {Kanwar}, Jayatee and {Morales-Calder{\'o}n}, Maria and {Rodgers-Lee}, Donna and {Schreiber}, J{\"u}rgen and {Schwarz}, Kamber and {Colina}, Luis},
        title = "{MINDS: The DR Tau disk. I. Combining JWST-MIRI data with high-resolution CO spectra to characterise the hot gas}",
      journal = {\aap},
         year = 2024,
        month = jun,
       volume = {686},
          eid = {A117},
        pages = {A117},
          doi = {10.1051/0004-6361/202348911},
archivePrefix = {arXiv},
       eprint = {2403.13591},
 primaryClass = {astro-ph.EP},
       adsurl = {https://ui.adsabs.harvard.edu/abs/2024A&A...686A.117T}
}

@ARTICLE{thanathibodee2019,
       author = {{Thanathibodee}, Thanawuth and {Calvet}, Nuria and {Bae}, Jaehan and {Muzerolle}, James and {Hern{\'a}ndez}, Ramiro Franco},
        title = "{Magnetospheric Accretion as a Source of H{\ensuremath{\alpha}} Emission from Protoplanets around PDS 70}",
      journal = {\apj},
         year = 2019,
        month = nov,
       volume = {885},
       number = {1},
          eid = {94},
        pages = {94},
          doi = {10.3847/1538-4357/ab44c1},
archivePrefix = {arXiv},
       eprint = {1909.06450},
 primaryClass = {astro-ph.EP},
       adsurl = {https://ui.adsabs.harvard.edu/abs/2019ApJ...885...94T}
}

@ARTICLE{venuti2019,
       author = {{Venuti}, L. and {Stelzer}, B. and {Alcal{\'a}}, J.~M. and {Manara}, C.~F. and {Frasca}, A. and {Jayawardhana}, R. and {Antoniucci}, S. and {Argiroffi}, C. and {Natta}, A. and {Nisini}, B. and et al.},
        title = "{X-shooter spectroscopy of young stars with disks. The TW Hydrae association as a probe of the final stages of disk accretion}",
      journal = {\aap},
         year = 2019,
        month = dec,
       volume = {632},
          eid = {A46},
        pages = {A46},
          doi = {10.1051/0004-6361/201935745},
archivePrefix = {arXiv},
       eprint = {1909.06699},
 primaryClass = {astro-ph.SR},
       adsurl = {https://ui.adsabs.harvard.edu/abs/2019A&A...632A..46V}
}

@ARTICLE{viswanath2024,
       author = {{Viswanath}, Gayathri and {Ringqvist}, Simon C. and {Demars}, Dorian and {Janson}, Markus and {Bonnefoy}, Micka{\"e}l and {Aoyama}, Yuhiko and {Marleau}, Gabriel-Dominique and {Dougados}, Catherine and {Szul{\'a}gyi}, Judit and {Thanathibodee}, Thanawuth},
        title = "{ExoplaNeT accRetion mOnitoring sPectroscopic surveY (ENTROPY): I. Evidence for magnetospheric accretion in the young isolated planetary-mass object 2MASS J11151597+1937266}",
      journal = {\aap},
         year = 2024,
        month = nov,
       volume = {691},
          eid = {A64},
        pages = {A64},
          doi = {10.1051/0004-6361/202450881},
archivePrefix = {arXiv},
       eprint = {2409.12187},
 primaryClass = {astro-ph.EP},
       adsurl = {https://ui.adsabs.harvard.edu/abs/2024A&A...691A..64V}
}

@ARTICLE{ulrich1981,
       author = {{Ulrich}, R.~K. and {Wood}, B.~C.},
        title = "{Observations and analysis of the helium I recombination lines lam 5876 and lam 10830 in eight T Tau stars.}",
      journal = {\apj},
         year = 1981,
        month = feb,
       volume = {244},
        pages = {147-156},
          doi = {10.1086/158692},
       adsurl = {https://ui.adsabs.harvard.edu/abs/1981ApJ...244..147U}
}

@ARTICLE{wells2015,
       author = {{Wells}, Martyn and {Pel}, J.-W. and {Glasse}, Alistair and {Wright}, G.~S. and {Aitink-Kroes}, Gabby and {Azzollini}, Ruym{\'a}n and {Beard}, Steven and {Brandl}, B.~R. and {Gallie}, Angus and {Geers}, V.~C. and {Glauser}, A.~M. and {Hastings}, Peter and {Henning}, Th. and {Jager}, Rieks and {Justtanont}, K. and {Kruizinga}, Bob and {Lahuis}, Fred and {Lee}, David and {Martinez-Delgado}, I. and {Mart{\'\i}nez-Galarza}, J.~R. and {Meijers}, M. and {Morrison}, Jane E. and {M{\"u}ller}, Friedrich and {Nakos}, Thodori and {O'Sullivan}, Brian and {Oudenhuysen}, Ad and {Parr-Burman}, P. and {Pauwels}, Evert and {Rohloff}, R.-R. and {Schmalzl}, Eva and {Sykes}, Jon and {Thelen}, M.~P. and {van Dishoeck}, E.~F. and {Vandenbussche}, Bart and {Venema}, Lars B. and {Visser}, Huib and {Waters}, L.~B.~F.~M. and {Wright}, David},
        title = "{The Mid-Infrared Instrument for the James Webb Space Telescope, VI: The Medium Resolution Spectrometer}",
      journal = {\pasp},
         year = 2015,
        month = jul,
       volume = {127},
       number = {953},
        pages = {646},
          doi = {10.1086/682281},
archivePrefix = {arXiv},
       eprint = {1508.03070},
 primaryClass = {astro-ph.IM},
       adsurl = {https://ui.adsabs.harvard.edu/abs/2015PASP..127..646W}
}

@ARTICLE{white2021,
       author = {{Campbell-White}, Justyn and {Sicilia-Aguilar}, Aurora and {Manara}, Carlo F. and {Matsumura}, Soko and {Fang}, Min and {Frasca}, Antonio and {Roccatagliata}, Veronica},
        title = "{The STAR-MELT PYTHON package for emission-line analysis of YSOs}",
      journal = {\mnras},
         year = 2021,
        month = nov,
       volume = {507},
       number = {3},
        pages = {3331-3350},
          doi = {10.1093/mnras/stab2300},
archivePrefix = {arXiv},
       eprint = {2108.02552},
 primaryClass = {astro-ph.SR},
       adsurl = {https://ui.adsabs.harvard.edu/abs/2021MNRAS.507.3331C}
}

@ARTICLE{white2023,
       author = {{Campbell-White}, Justyn and {Manara}, Carlo F. and {Sicilia-Aguilar}, Aurora and {Frasca}, Antonio and {Nielsen}, Louise D. and {Christian Schneider}, P. and {Nisini}, Brunella and {Bayo}, Amelia and {Ercolano}, Barbara and {{\'A}brah{\'a}m}, P{\'e}ter and {Claes}, Rik and {Fang}, Min and {Fedele}, Davide and {Gameiro}, Jorge Filipe and {Gangi}, Manuele and {K{\'o}sp{\'a}l}, {\'A}gnes and {Mauc{\'o}}, Karina and {Petr-Gotzens}, Monika G. and {Rigliaco}, Elisabetta and {Robinson}, Connor and {Siwak}, Michal and {Tychoniec}, Lukasz and {Venuti}, Laura},
        title = "{Empirical determination of the lithium 6707.856 {\r{A}} wavelength in young stars}",
      journal = {\aap},
         year = 2023,
        month = may,
       volume = {673},
          eid = {A80},
        pages = {A80},
          doi = {10.1051/0004-6361/202245696},
archivePrefix = {arXiv},
       eprint = {2303.03843},
 primaryClass = {astro-ph.SR},
       adsurl = {https://ui.adsabs.harvard.edu/abs/2023A&A...673A..80C}
}

@ARTICLE{white2023b,
       author = {{Campbell-White}, Justyn and {Manara}, Carlo F. and {Benisty}, Myriam and {Natta}, Antonella and {Claes}, Rik A.~B. and {Frasca}, Antonio and {Bae}, Jaehan and {Facchini}, Stefano and {Isella}, Andrea and {P{\'e}rez}, Laura and {Pinilla}, Paola and {Sicilia-Aguilar}, Aurora and {Teague}, Richard},
        title = "{A Magnetically Driven Disk Wind in the Inner Disk of PDS 70}",
      journal = {\apj},
         year = 2023,
        month = oct,
       volume = {956},
       number = {1},
          eid = {25},
        pages = {25},
          doi = {10.3847/1538-4357/acf0c0},
archivePrefix = {arXiv},
       eprint = {2308.09554},
 primaryClass = {astro-ph.SR},
       adsurl = {https://ui.adsabs.harvard.edu/abs/2023ApJ...956...25C}
}

@ARTICLE{luhman2023c,
       author = {{Luhman}, K.~L. and {Tremblin}, P. and {Birkmann}, S.~M. and {Manjavacas}, E. and {Valenti}, J. and {Alves de Oliveira}, C. and {Beck}, T.~L. and {Giardino}, G. and {L{\"u}tzgendorf}, N. and {Rauscher}, B.~J. and {Sirianni}, M.},
        title = "{JWST/NIRSpec Observations of the Planetary Mass Companion TWA 27B}",
      journal = {\apjl},
         year = 2023,
        month = jun,
       volume = {949},
       number = {2},
          eid = {L36},
        pages = {L36},
          doi = {10.3847/2041-8213/acd635},
archivePrefix = {arXiv},
       eprint = {2305.18603},
 primaryClass = {astro-ph.EP},
       adsurl = {https://ui.adsabs.harvard.edu/abs/2023ApJ...949L..36L}
}

@ARTICLE{marleau2024,
       author = {{Marleau}, Gabriel-Dominique and {Aoyama}, Yuhiko and {Hashimoto}, Jun and {Zhou}, Yifan},
        title = "{Revisiting the Helium and Hydrogen Accretion Indicators at TWA 27B: Weak Mass Flow at Near-freefall Velocity}",
      journal = {\apj},
         year = 2024,
        month = mar,
       volume = {964},
       number = {1},
          eid = {70},
        pages = {70},
          doi = {10.3847/1538-4357/ad1ee9},
archivePrefix = {arXiv},
       eprint = {2401.04763},
 primaryClass = {astro-ph.EP},
       adsurl = {https://ui.adsabs.harvard.edu/abs/2024ApJ...964...70M}
}

\begin{appendix}

\section{Variation in the UVES spectra of Delorme 1 (AB)b}\label{app-a}
\begin{figure*}
    \centering
    \includegraphics[width=1\linewidth]{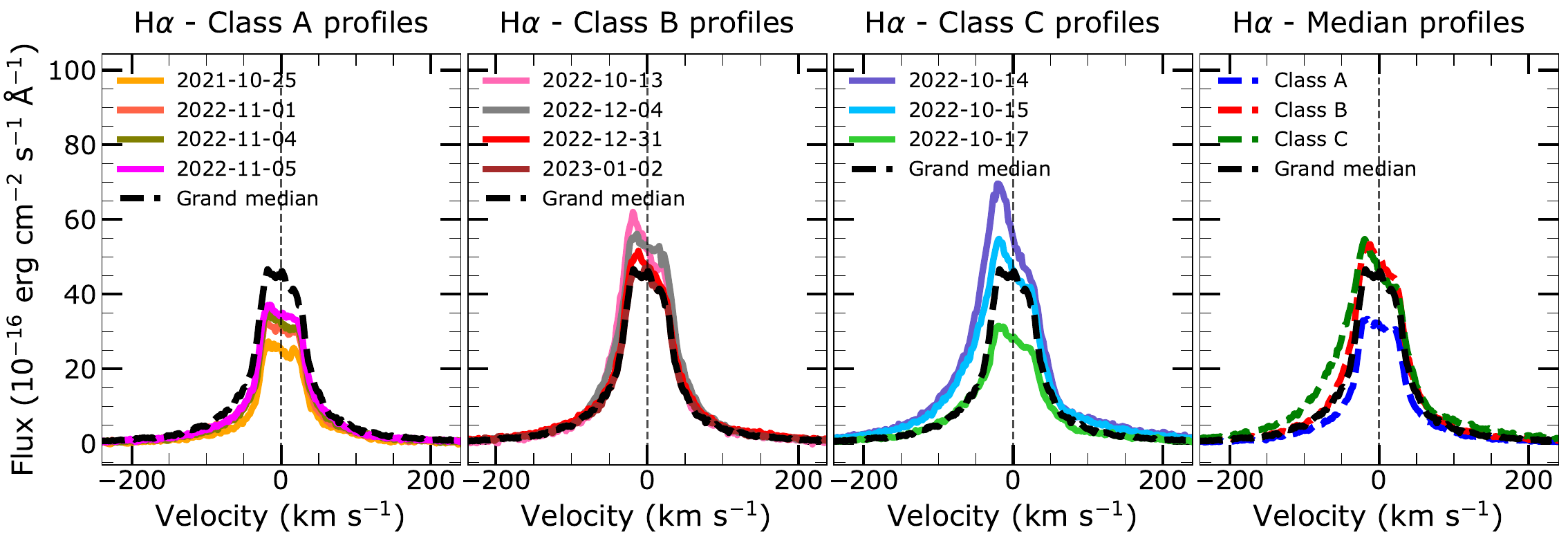}\\
    \includegraphics[width=1\linewidth]{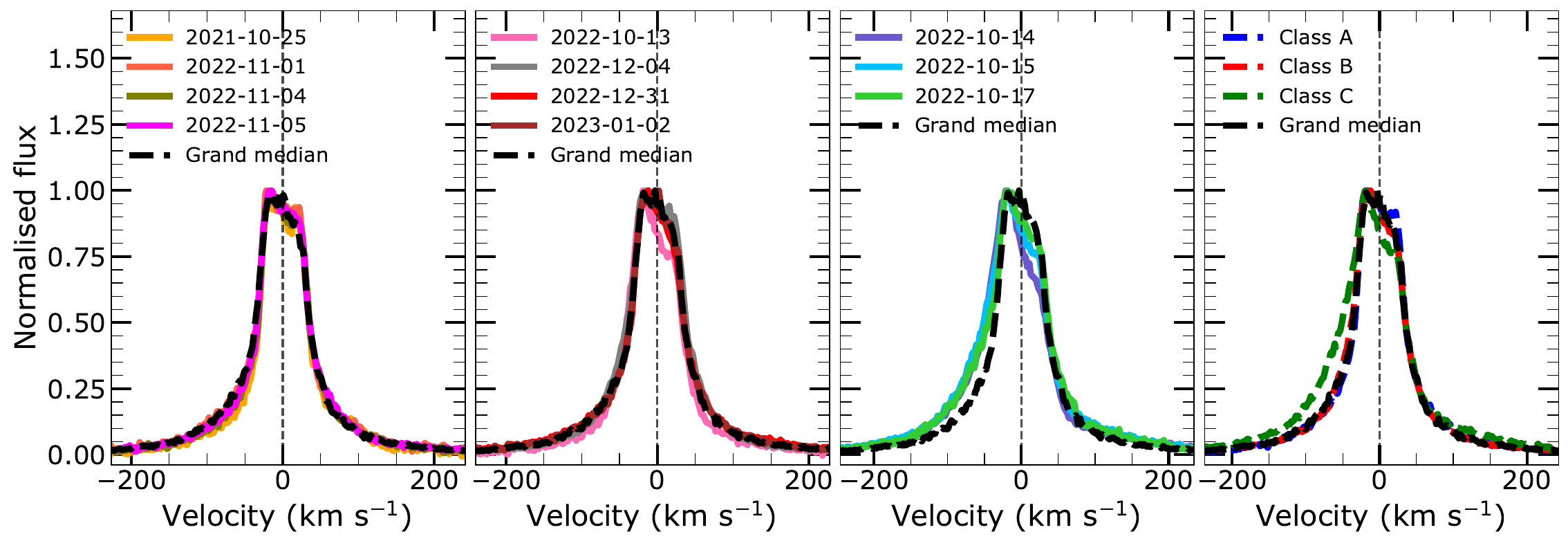}
    \caption{Class A, class B, and class C profiles of the \ha emission of Delorme 1 (AB)b. Right panels show the respective median profiles along with the grand median profile. Top panels illustrate the flux variation between individual epochs (in units of $10^{-16}$~erg~cm$^{-2}$~s$^{-1}$~\AA$^{-1}$) and bottom panels (with flux normalised to \ha peak intensity in the y-axis) demonstrate the comparison of profile shapes between the three classes.}
    \label{figA1}
\end{figure*}

\begin{figure*}
    \centering
    \includegraphics[scale=0.5]{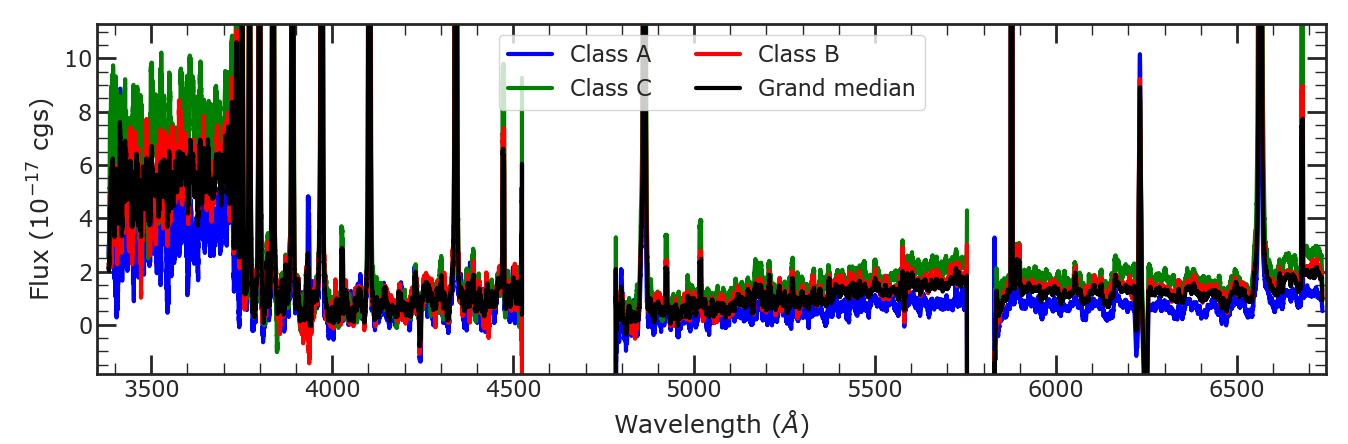}
    \caption{UVES spectrum of Delorme 1 (AB)b over the whole wavelength range, smoothened with a box size of 500 pixels. The rise in the UV excess and continuum level beyond $\sim5300~$\AA\ can be clearly seen in the class C median compared to class A and class B. `cgs' denotes \cgs\AA$^{-1}$.}
    \label{figA2}
\end{figure*}

A by-eye classification of the shape and intensity of the \ha emission line reveals three distinct classes of profiles as mentioned in Section~\ref{results}. Class A profiles displayed during the 2021 epoch as well on 1, 4, and 5 November 2022 are marked by low intensities, with plateau-like cores and no notable deviation in the wings from that of a Gaussian profile. Class A epochs span a total of $\sim$6.2~hr of total integration time over the observations. There is very little flux variability between these individual class A epochs in 2022, with a maximum of $\sim$40\% increase on 5 November 2022 compared to 25 October 2021. In comparison, there is notable flux increase on the nights of 13 October, 4 December, and 31 December 2022 and 2 January 2023, with shallow sloping core profiles -- we classify these as class B (4.1~hr of total integration time). On 14, 15, and 17 October 2022 (3.3~hr of total integration time), the \ha profile displays a significant increase in the blue wing of the profile compared to both class A and class B profiles. We call these the class C profiles. Top panel of Fig.~\ref{figA1} illustrates the difference between the three profile classes in terms of their respective median profiles. The bottom panel shows these line profiles normalised with respect to their respective peak intensities, and demonstrate that class A and class B profiles mostly differ in flux levels but share similar profile shapes. Compared to these classes, class C profiles vary largely in flux between individual epochs. The significant rise in Balmer emission during these epochs is also reflected in the respective UV excess and the continuum level beyond $\sim5300$~\AA\ (see Fig.~\ref{figA2}), possibly tracing an accretion burst or chromospheric activity burst. The peak outburst is expected to be around 14 October 2022, showing $\sim$130\% increase in \ha intensity compared to 17 October 2022.  

\section{Tentative \hei detection}\label{app-c}
\begin{figure}
    \centering
    \includegraphics[scale=0.4]{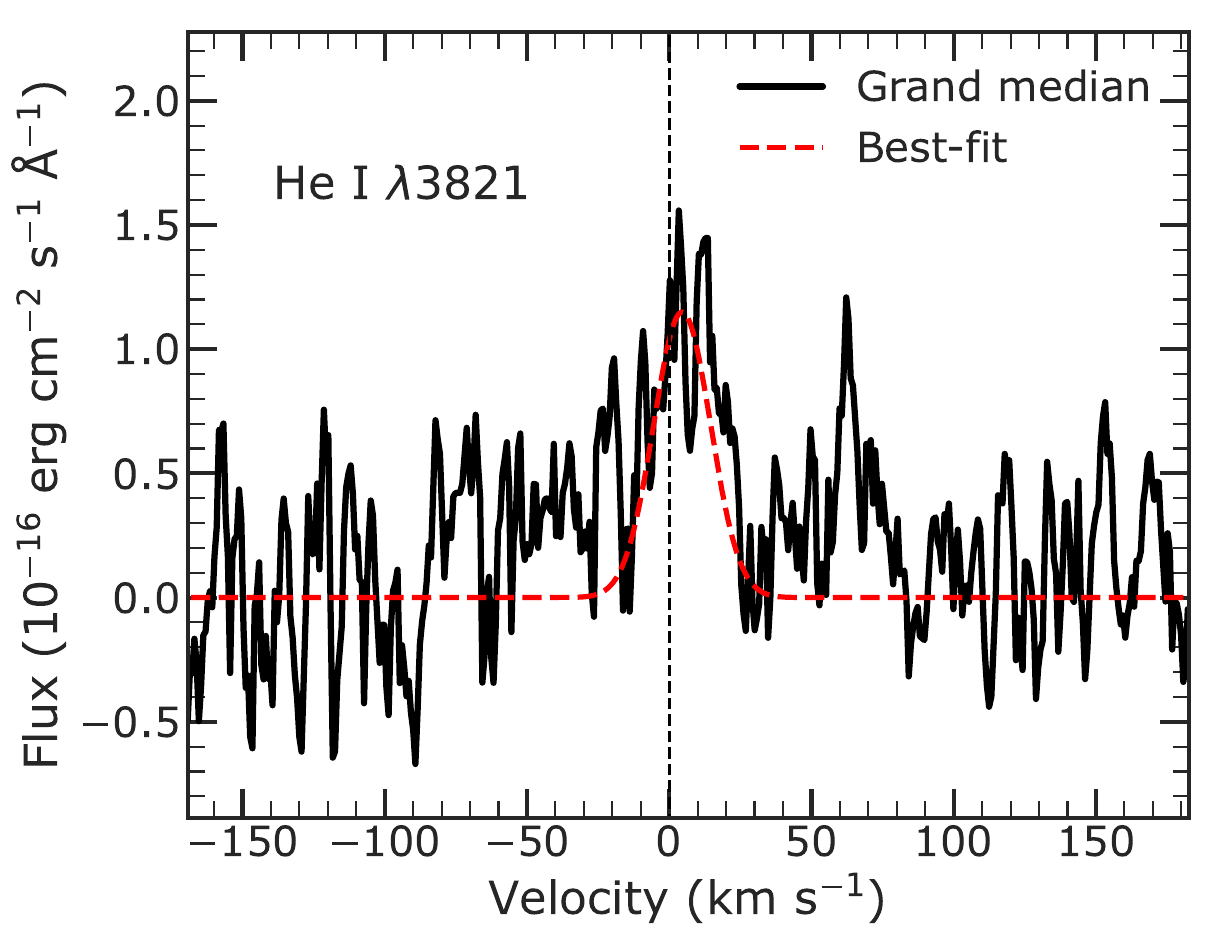}
    \caption{Tentative detection of \hei $\lambda$3821 emission in the grand median UVES spectrum of Delorme 1 (AB)b. The least-$\chi^2$ fit to the line profile is shown in red.}
    \label{figC}
\end{figure}

From a slightly higher energy level ($E_{k}=24.21$ eV) than the rest of the \hei detections in the spectra, \hei $\lambda$3821 emission is detected tentatively in the grand median at a confidence of 2$\sigma$ (see Fig.~\ref{figC}). It is barely present in the class A epochs (0.5$\sigma$), but appears in the class B and class C epochs (1.4$\sigma$), with maximum intensity in class C profiles (2.2$\sigma$). The emission is red-shifted with a centroid velocity of $4.5\pm1.0$~\kms, similar to the red-shifted profiles of the confirmed \hei detections in this work. The FWHM (23.5$\pm$0.6~\kms) also agrees with those of the other \hei lines in Table~\ref{tab1}.

\section{Disentangling \hei \texorpdfstring{$\lambda$}{lambda}3890 from H8}\label{app-d}
\begin{figure}[H]
    \centering
    \includegraphics[scale=0.4]{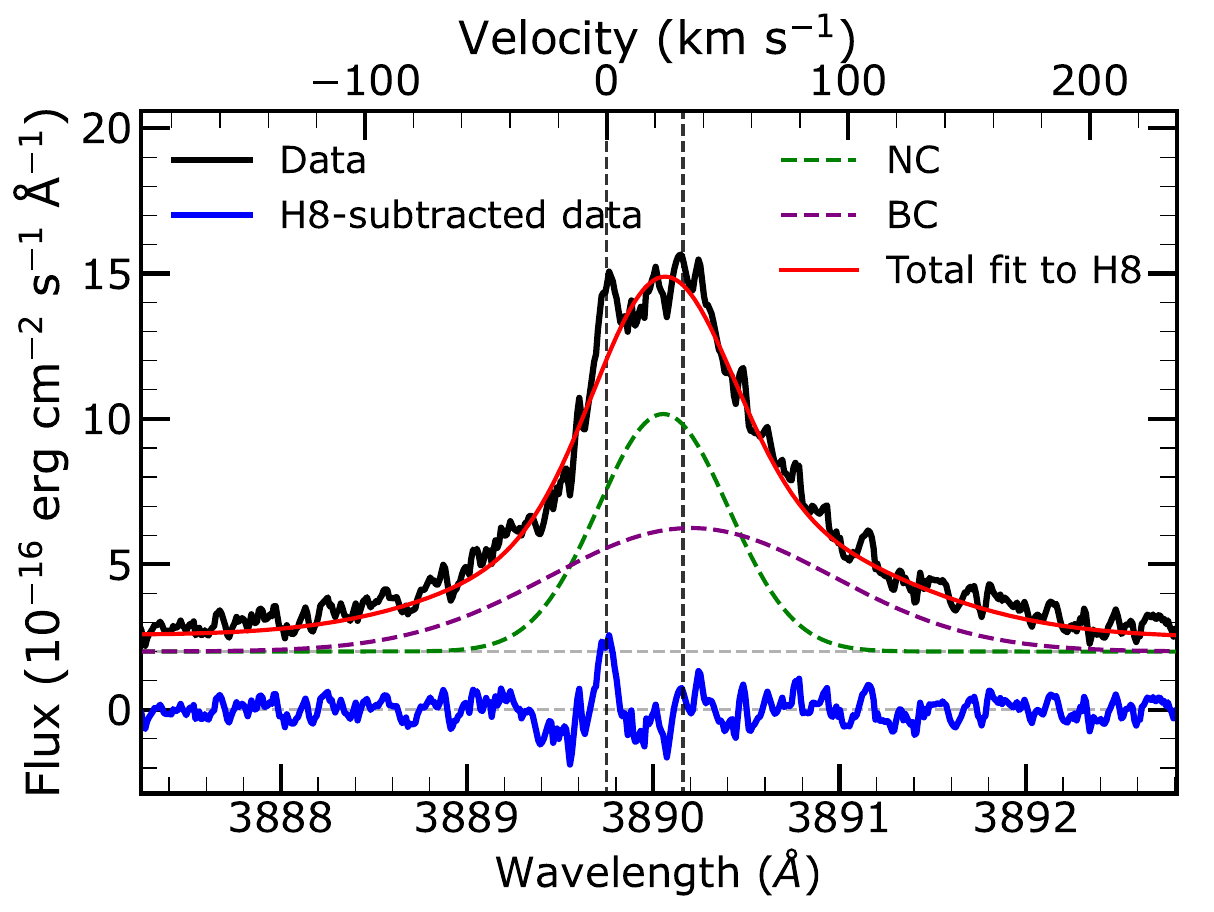}
    \caption{Least-$\chi^2$ fit (red) to H8 in the spectrum (black) composed of NC (green) and BC (purple), plotted with a vertical offset of $2\times10^{-16}$~erg~s$^{-1}$~cm$^{-2}$~\AA$^{-1}$. The data after subtracting this composite fit is shown in blue, revealing the \hei emission at 3889.75~\AA.}
    \label{figD}
\end{figure}

Fig.~\ref{figD} illustrates the \hei $\lambda$3890 emission after Balmer emission at 3890.16~\AA\ (H8) was modelled and subtracted from the spectrum. H8 emission has a wide, asymmetric profile spanning $\sim$200~\kms on each wing. The least-$\chi^2$ fit to its profile yields an NC (FWHM$=66$~\kms) at $-8.2$~\kms and a BC (FWHM$=165$~\kms) at $3.3$~\kms. The composite fit to H8 was then subtracted from the spectrum, revealing the \hei emission at 3889.75~\AA.

\section{Evolution of NC and BC from class A to class C profiles}\label{app-e}
\begin{figure*}[!ht]
    \centering
    \includegraphics[scale=0.5]{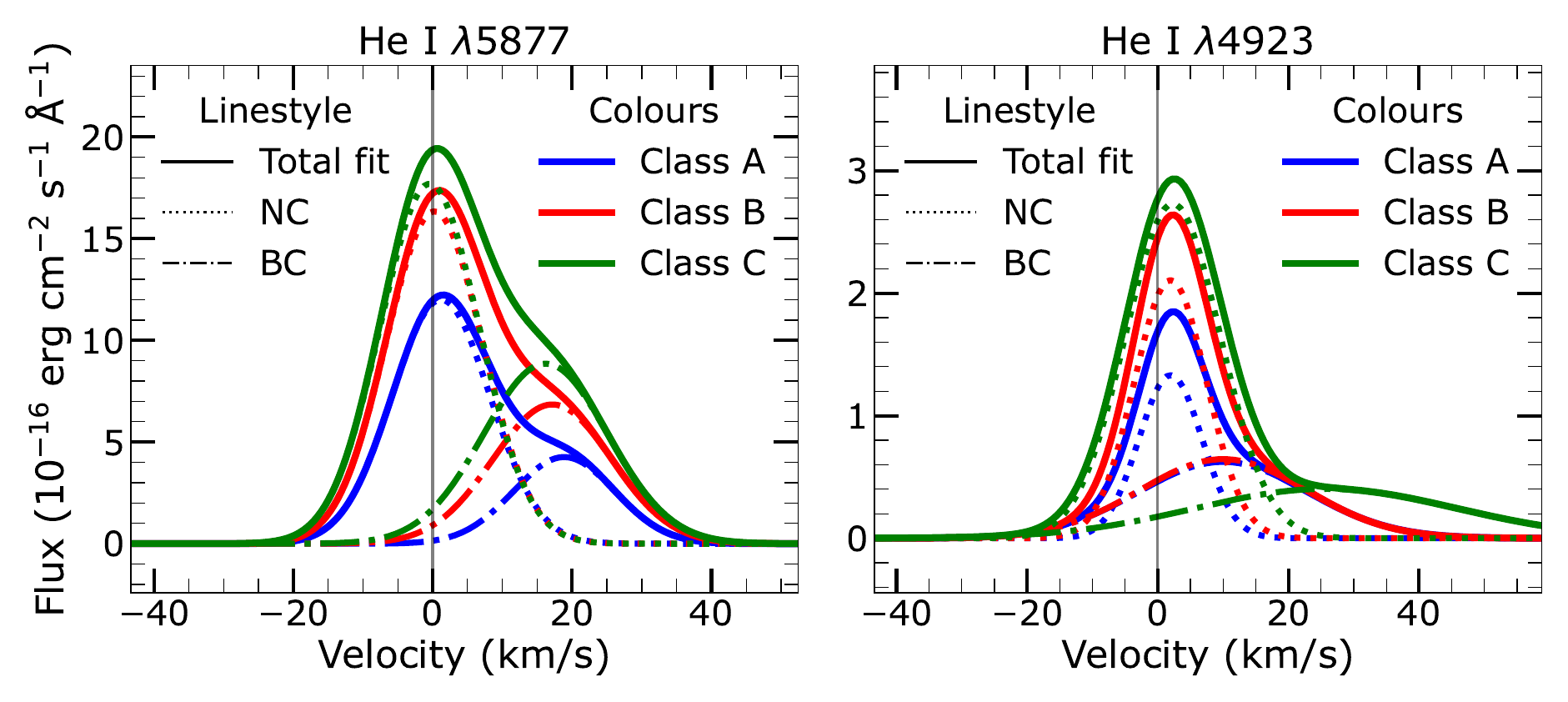}
    \caption{Decomposition into NC and BC for the median profiles of the \hei lines $\lambda5877$ and $\lambda4923$ from class A, class B, and class C epochs of Delorme 1 (AB)b. The line-styles represent the NC (dotted), BC (dashed) and total fits (solid) to the profiles, while the colours represent the median class A (blue), class B (red) and class C (green) epochs.}
    \label{figE}
\end{figure*}

\begin{figure*}[!ht]
    \centering
    \includegraphics[width=0.4\linewidth]{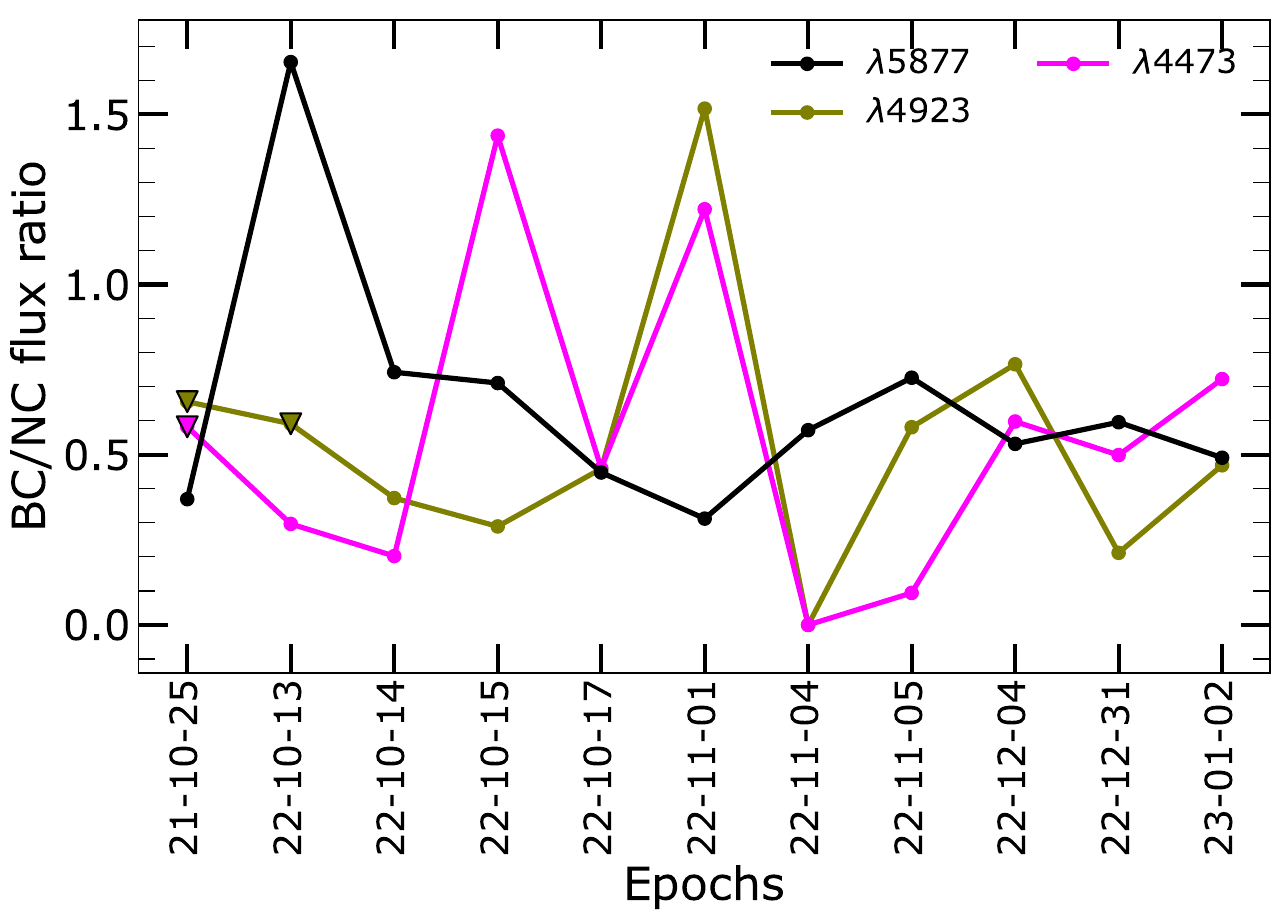}
    \includegraphics[width=0.4\linewidth]{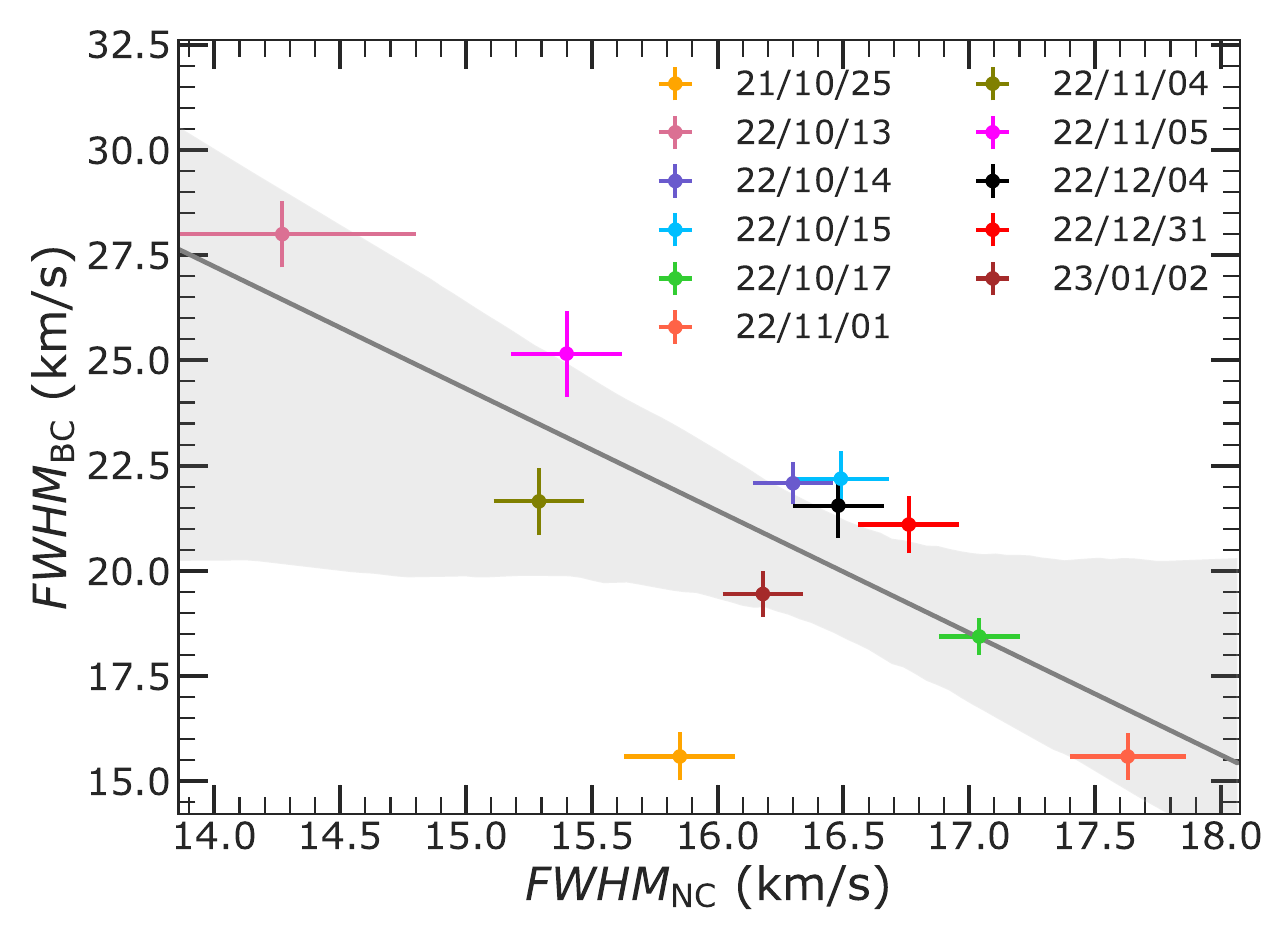}
    \caption{(Left) Variation in the BC/NC flux ratio of \hei $\lambda\lambda5877,4923,4473$ emission lines with time. \hei $\lambda4027$ is not shown in the figure since it is of much lower S/N with its profiles from individual nights relatively noisy. (Right) Variation in the FWHM of NC and BC through the epochs for \hei~D3 line. The widths of both the components are anti-correlated (grey line and shaded region denotes the linear fit and its uncertainty), with the overall profile shape remaining the same with time.}
    \label{figE2}
\end{figure*}

The NC and BC profiles of the asymmetric \hei lines in the Delorme 1 (AB)b spectra show an increasing trend in intensities and widths from class A to class C. This is illustrated in Fig.~\ref{figE} for the triplet \hei $\lambda5877$ and the singlet \hei $\lambda4923$. The BC in the \hei $\lambda4923$ profile is broader and more redshifted in class C epochs. Fig.~\ref{figE2} shows the variation in the BC/NC flux ratio of the \hei $\lambda\lambda5877,4923,4473$ line profiles from epoch to epoch. The line profiles of \hei $\lambda4923$ and $\lambda4473$ vary from BC-dominated (BC/NC$>$1) to NC-dominated  (BC/NC$<$1). \hei $\lambda5877$ line profile remains mostly NC-dominated except on 13 October 2022.

\section{Line flux from individual epochs}
\begin{figure*}
    \centering
        \includegraphics[scale=0.5]{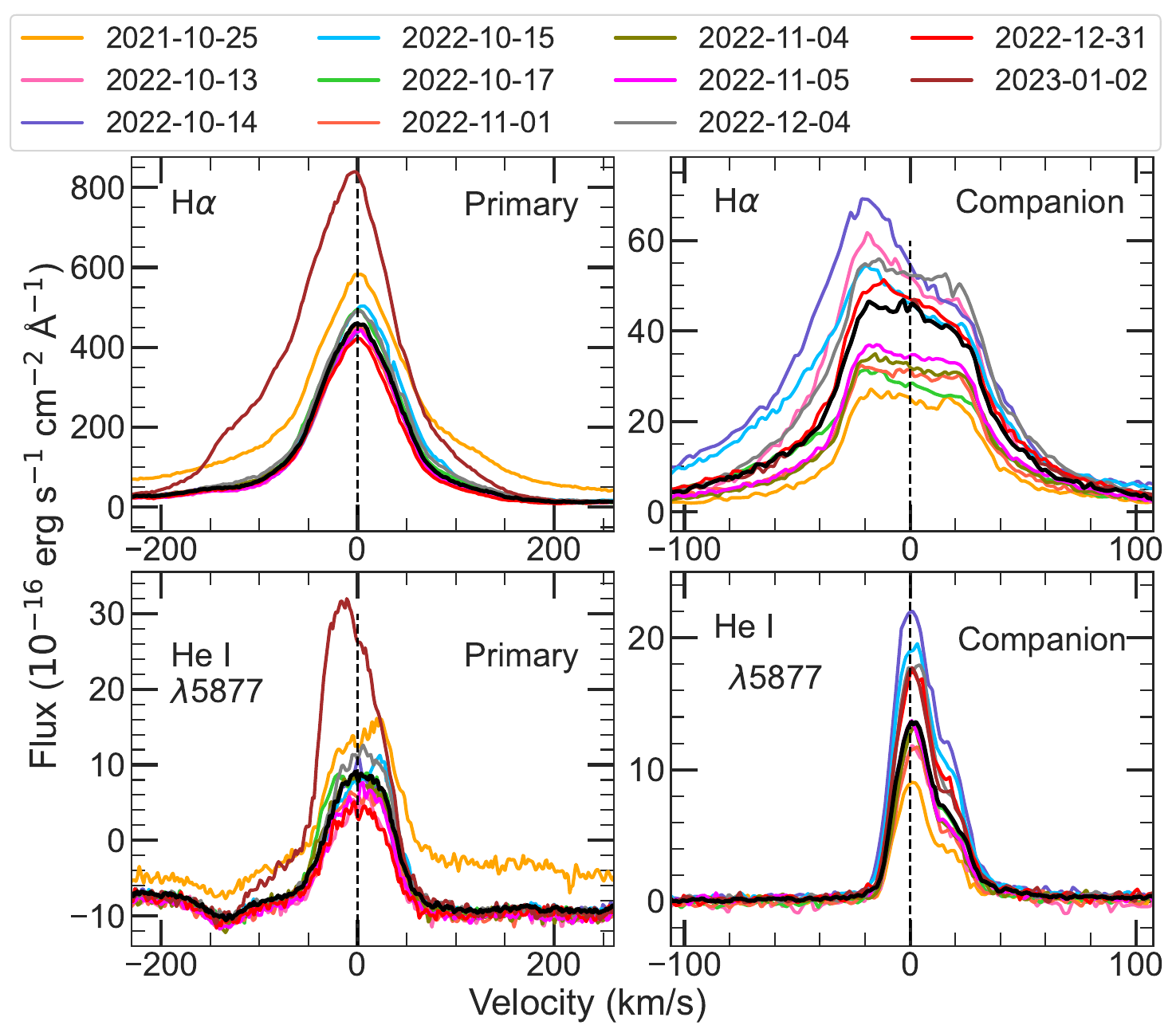}
    \caption{\ha (top panel) and \hei $\lambda$5877 (bottom panel) emission lines detected from the primary (left) and the companion (right) during each observation night. The grand median profile is represented by a thick black curve in each panel.}
    \label{figG}
\end{figure*}

\begin{figure*}[hb]
    \centering
    \includegraphics[width=1.\linewidth]{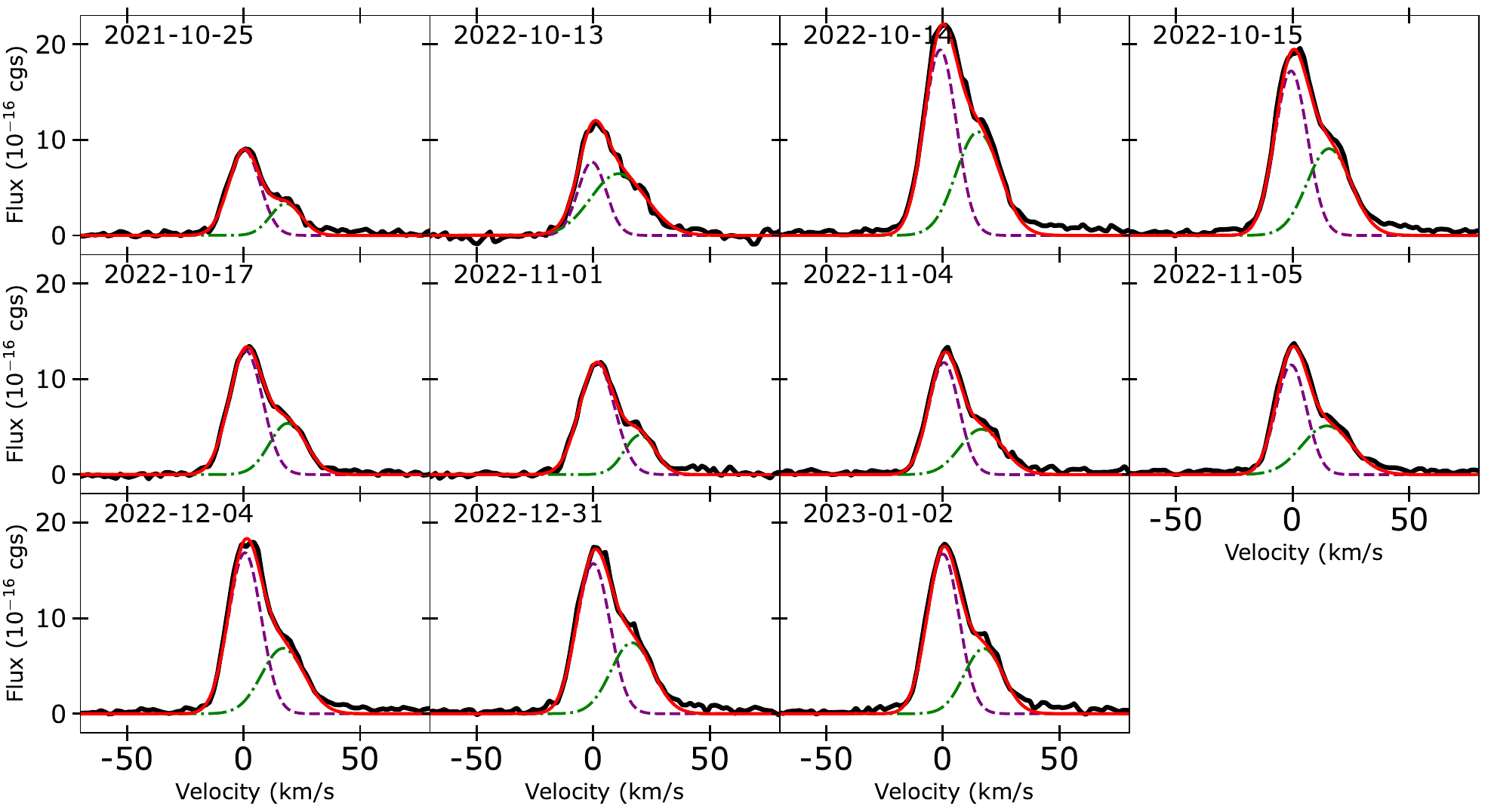}
    \caption{\hei $\lambda5877$ emission from the median of each night of observation, with the respective least-$\chi^2$ fit to its profile. Colours and symbols hold the same meaning as in Fig.~\ref{fig1}. cgs stands for \cgs\AA$^{-1}$.}
    \label{figG2}
\end{figure*}

Table~\ref{tabg1} lists the integrated flux of \hei lines detected from Delorme 1 (AB)b in this work from the median spectra from each observation night. Fig~\ref{figG} compares the variation of \ha and \hei $\lambda5877$ emission lines from both the target and its primary with time. As evident from the plots, the primary shows very little variation compared to the companion, as is expected from chromospheric activity, except likely during a flare on 2 January 2023. The emission profile from the primary also lacks any asymmetry unlike in the companion spectra. Fig~\ref{figG2} shows the line profile of \hei $\lambda5877$ from the median of each night of observation, along with the respective NC and BC from the least-$\chi^2$ fit to the profile. The characteristics of the respective profile fits from each night is given in Table~\ref{tabe2}.

\begin{table*}
    \centering
    \caption{Integrated line fluxes for \hei lines from individual nights, in units of $10^{-17}$~\cgs.}
    \begin{tabular}{cccccccc}
        \hline
         Epoch & $\lambda3890$ & $\lambda4027$ & $\lambda4473$ & $\lambda4923$ & $\lambda5017$ & $\lambda5877$ & $\lambda6680$\\\hline
         2021-10-25 & $\phantom{1}1.3\pm8.4$\tablefootmark{a} & $\phantom{1}\phantom{1}5.7\pm8.9$\tablefootmark{a} & $12.2\pm9.5$\tablefootmark{a} & $\phantom{1}\phantom{1}5.8\pm4.7$\tablefootmark{a} & $\phantom{1}\phantom{1}3.6\pm4.2$\tablefootmark{a} & $40.4\pm1.9$ & $13.6\pm2.9$\\
         2022-10-13 & $\phantom{1}\phantom{1}1.6\pm14.0$\tablefootmark{a} & $\phantom{1}\phantom{1}\phantom{1}6.2\pm11.8$\tablefootmark{a} & $10.1\pm14.3$ & $\phantom{1}\phantom{1}5.7\pm7.3$\tablefootmark{a} & $\phantom{1}5.2\pm6.8$ & $60.6\pm3.7$ & $24.4\pm4.3$ \\
         2022-10-14 & $5.9\pm9.2$ & $17.1\pm7.3$ & $39.1\pm8.8$ & $12.1\pm3.7$ & $\phantom{1}5.2\pm3.8$ & $116.2\pm2.5$ & $54.6\pm5.3$\\
         2022-10-15 & $7.0\pm6.5$ & $13.3\pm5.0$ & $31.2\pm6.2$ & $10.7\pm3.1$ & $13.4\pm3.8$ & $101.1\pm2.0$ & $46.8\pm2.8$ \\
         2022-10-17 & $3.0\pm7.0$ & $\phantom{1}9.8\pm5.6$ & $23.4\pm7.2$ & $\phantom{1}7.1\pm3.8$ & $\phantom{1}6.6\pm3.2$ & $66.7\pm2.0$ & $28.2\pm2.3$ \\
         2022-11-01 & $\phantom{1}4.1\pm10.5$ & $\phantom{1}\phantom{1}3.8\pm8.4$\tablefootmark{a} & $18.5\pm9.4$ & $\phantom{1}7.5\pm4.4$ &  $\phantom{1}3.7\pm4.6$ & $56.4\pm3.2$ & $23.2\pm8.9$ \\
         2022-11-04 & $3.8\pm6.9$ & $\phantom{1}\phantom{1}7.6\pm5.7$\tablefootmark{a} & $20.0\pm7.1$ & $\phantom{1}4.6\pm4.1$ &  $\phantom{1}4.1\pm3.8$ & $58.9\pm2.1$ & $22.3\pm2.1$ \\
         2022-11-05 & $4.3\pm7.3$ & $\phantom{1}3.7\pm4.6$ & $16.5\pm8.0$& $\phantom{1}6.9\pm3.7$ & $\phantom{1}4.9\pm3.4$ & $63.6\pm2.3$ & $25.5\pm2.3$ \\
         2022-12-04 & $7.9\pm6.0$ & $11.6\pm6.0$ & $27.1\pm7.2$ & $\phantom{1}4.1\pm3.8$ & $\phantom{1}7.4\pm3.0$ & $88.7\pm1.9$ & $35.2\pm2.1$\\
         2022-12-31 & $9.6\pm8.4$ & $\phantom{1}9.6\pm6.4$ & $27.3\pm7.8$ & $\phantom{1}8.6\pm3.6$ & $\phantom{1}8.2\pm3.7$ & $87.7\pm1.3$ & $36.8\pm2.2$\\
         2023-01-02 & $7.8\pm9.1$ & $\phantom{1}7.9\pm7.2$ & $24.6\pm9.4$ & $10.2\pm4.6$ & $\phantom{1}7.9\pm5.6$ & $84.1\pm2.3$ & $33.3\pm2.4$\\\hline
    \end{tabular}
    \tablefoot{\\ \tablefoottext{a}{Tentative Detections with S/N $1-3\sigma$, where $\sigma$ denotes the standard deviation of the flux within $\pm150$~\kms .}\\
    \tablefoottext{b}{Error bars on the flux values indicate the rms of the local continuum for each night.}}
    \label{tabg1}
\end{table*}

\begin{table*}[!ht]
    \centering
    \caption{Centroid velocities and FWHM, in units of \kms, of NC and BC from the least-$\chi^2$ fit to the \hei $\lambda5877$ emission line profile from each observation night.} 
    \begin{tabular}{cccccc}
        \hline
         Epoch & $\mu_{\mathrm{NC}}$ & FWHM$_{\mathrm{NC}}$ & $\mu_{\mathrm{BC}}$ & FWHM$_{\mathrm{BC}}$ & Flux$_{\mathrm{BC}}$/Flux$_{\mathrm{NC}}$\\\hline
         2021-10-25 & $\phantom{1}0.4\pm0.1$ & $15.8\pm0.2$ & $18.1\pm0.3$ & $15.6\pm0.6$ & 0.37 \\
         2022-10-13 & $-0.5\pm0.1$ & $14.3\pm0.5$ & $10.8\pm0.8$ & $28.0\pm0.8$ & 1.65 \\
         2022-10-14 & $-1.1\pm0.1$ & $16.3\pm0.2$ & $15.1\pm0.4$ & $22.1\pm0.5$ & 0.76 \\
         2022-10-15 & $-0.6\pm0.1$ & $16.5\pm0.2$ & $15.6\pm0.5$ & $22.2\pm0.6$ & 0.71 \\
         2022-10-17 & $\phantom{1}1.0\pm0.1$ & $17.0\pm0.2$ & $19.0\pm0.3$ & $18.4\pm0.4$ & 0.45 \\
         2022-11-01 & $\phantom{1}1.6\pm0.1$ & $17.6\pm0.2$ & $20.0\pm0.1$ & $15.6\pm0.6$ & 0.31 \\
         2022-11-04 & $\phantom{1}0.4\pm0.1$ & $15.3\pm0.2$ & $16.5\pm0.5$ & $21.6\pm0.8$ & 0.57 \\
         2022-11-05 & $-0.6\pm0.1$ & $15.4\pm0.2$ & $14.7\pm0.8$ & $25.2\pm1.0$ & 0.73\\
         2022-12-04 & $\phantom{1}0.6\pm0.1$ & $16.5\pm0.2$ & $17.0\pm0.5$ & $21.6\pm0.8$ & 0.53 \\
         2022-12-31 & $\phantom{1}0.1\pm0.2$ & $16.8\pm0.2$ & $16.6\pm0.5$ & $21.1\pm0.7$ & 0.60 \\
         2023-01-02 & $\phantom{1}0.1\pm0.1$ & $16.2\pm0.2$ & $17.5\pm0.3$ & $19.4\pm0.5$ & 0.49 \\\hline
    \end{tabular}
    \label{tabe2}
\end{table*}

\section{\lacc and \mdot measurements from different accretion proxies}
Table~\ref{tab5} shows the \lacc and \mdot measurements for the target obtained from UV excess using hydrogen slab models \citetalias{demars2026} and using \ha line luminosity and \hei line luminosities based on \lline--\lacc scaling relations. For the latter, we use both the \cite{fiorellino2025} empirical scaling relations based on low-mass stars, as well as the \cite{aoyama2021} planetary scaling relations developed based on shock emission models.

\renewcommand{\arraystretch}{1.2}
\begin{table}
\caption{\lacc and \mdot measured from Delorme 1 (AB)b using different parameters.}
\centering
\begin{tabular}[c]{c c c c}
\hline\hline
Epoch & Measured from & $\log(L_{\mathrm{acc}}/\lsun)$ &  $\mdot\,(\msun \mathrm{yr}^{-1})$\\
\hline
UVES & UV excess & $-5.3^{+0.5}_{-0.4}$ & $1-3\times10^{-12}$ \\
UVES & \hei (FAM25) & $-4.9^{+0.3}_{-0.3}$ & $7.4\times10^{-12}$ \\
UVES & \ha (FAM25) & $-5.5^{+0.4}_{-0.2}$ & $1.4\times10^{-12}$ \\
UVES & \ha (AMI21) & $-4.3^{+0.3}_{-0.2}$ & $2.6\times10^{-11}$ \\
MUSE & \ha (FAM25) & $-6.6^{+0.4}_{-0.2}$ & $1.3\times10^{-13}$\\
 \hline
\end{tabular}
\tablefoot{\\ FAM25 and AMI21 denotes \lline--\lacc scaling relations from \cite{fiorellino2025} and \cite{aoyama2021} respectively. The uncertainties indicated on the \lacc measurements are based on the lowest and highest values over the observations epochs.}
\label{tab5}
\end{table}

\FloatBarrier

\end{appendix}
\end{document}